\documentclass[twocolumn,aps,nofootinbib,pra]{revtex4-2}

\usepackage{amsmath,amsfonts}
\usepackage{amssymb}
\usepackage{enumerate}
\usepackage{mathrsfs}
\usepackage{amsthm}
\usepackage{textcomp}
\usepackage{babel}
\usepackage[colorlinks,citecolor=blue,linkcolor=blue]{hyperref}
\usepackage{babel}

\newcommand{\be}{\begin{equation}}
\newcommand{\ee}{\end{equation}}
\newcommand{\bey}{\begin{eqnarray}}
\newcommand{\eey}{\end{eqnarray}}
\newcommand{\ba}{\begin{array}}
\newcommand{\ea}{\end{array}}
\newcommand{\bi}{\begin{itemize}}
\newcommand{\ei}{\end{itemize}}
\newcommand{\bem}{\begin{enumerate}}
\newcommand{\eem}{\end{enumerate}}
\newcommand{\bw}{\begin{widetext}}
\newcommand{\ew}{\end{widetext}}

\newcommand{\ra}{\rangle}
\newcommand{\la}{\langle}

\newcommand{\ov}{\overline}

\newcommand{\wh}{\widehat}
\newcommand{\ww}{\widetilde}

\newcommand{\bk}{{\bf k}}

\newcommand{\bp}{{\bf p}}

\newcommand{\bq}{{\bf q}}

\newcommand{\bx}{{\bf x}}

\newcommand{\C}{{\mathscr{C}}}
\newcommand{\cs}{\mathcal{S}}

\newcommand{\D}{{\mathcal{D}}}
\newcommand{\E}{{\cal E}}

\newcommand{\G}{{\cal G}}
\newcommand{\HH}{{\mathscr{H}}}

\newcommand{\LL}{{\mathcal{L}}}
\newcommand{\N}{{\mathscr{N}}}

\newcommand{\OO}{\mathcal{O}}

\newcommand{\WW}{{\mathscr{W}}}

\newcommand{\VV}{{\mathscr{V}}}

\begin{document}
%\draft

 \title{ 
 Models  for elementary particles under a unified framework without making use of gauge symmetry
 }
 
\author{Wen-ge Wang}
\affiliation{
 Department of Modern Physics, Anhui Center for fundamental sciences in theoretical physics,
 and CAS Key Laboratory of Microscale Magnetic Resonance,
 University of Science and Technology of China,
 Hefei 230026, China
 }
%\affiliation{ CAS Key Laboratory of Microscale Magnetic Resonance,
%University of Science and Technology of China, Hefei 230026, China}

 \date{\today}
   
\begin{abstract}
 This paper consists of two parts.
 In the first part, a concise reformulation is derived for that part of the Glashow-Weinberg-Salam (GWS) 
 electroweak interaction Hamiltonian, which describes interactions
 between leptons (in the first generation) and vector bosons.
 In this reformulation, different particle species are associated with 
 different types of spinor spaces for the description of their spin states and, 
 as a consequence, the former are characterizable by configurations of the latter.  
 In the second part, based on results of the first part,
 a framework is proposed which may be useful for constructing models
 of elementary particles and their interactions without making use of gauge symmetry
 at the fundamental level. 
 Three models within this framework are constructed and studied.
 The first model predicts the above mentioned part of the GWS electroweak interactions,
 while, it suffers a flaw. 
 Remove of the flaw by a small modification gives rise to a second model;
 and, as a byproduct of the modification, this model
 contains only one coupling constant, with the Weinberg angle fixed at 
 $\sin^2 \theta_{W} = 0.25$.
 The third model predicts interactions that show certain similarities to strong interactions,
 with the quantum number of color related to certain type of spinor-space configuration. 
\end{abstract}

%\keywords{quantum electrodynamics, Lorentz covariance, spinor spaces, interaction Lagrangian}
 \maketitle

%\tableofcontents
%\begin{multicols}{2}

\section{Introduction}

%\subsection{Motivation}

 One of the ultimate goals of Physics is to unify descriptions of fundamental interactions observed in Nature.
 This goal is partially achieved in 
 the so-called standard model (SM), the most successful quantum field theory (QFT)
 that has ever been developed for the description of the electrodynamic, weak, and strong interactions 
 (see, e.g., textbooks \cite{Weinberg-book,Peskin,Itzy});
 meanwhile, the gravitational interaction is described by the theory of general relativity. 
 In the SM, the electrodynamic and weak interactions are successfully unified
 by making use of the gauge symmetry of $U(1) \otimes SU(2)$,
 known as the  Glashow-Weinberg-Salam (GWS) electroweak theory.
 However, a further unification of the electroweak interaction and strong interaction, 
 the latter of which is independently related to the gauge symmetry of $SU(3)$,
 sometimes referred to as grand unification, 
 meets serious difficulties in the gauge-symmetry approach.
 In fact, going beyond the SM as a topic has attracted lots of attention in recent decades
 (see, e.g., Refs.\cite{IWW-RMP24,FR-RMP21,BB-RMP21,FADOSV-RMP21,DeGrand-MP16,
 GGS99,KO01,Polch-book,Ross84,Raby93,book-fundam-QFT,GN03,SV06}).

 The main purpose of this paper is to propose a framework for the description of elementary particles,
 which may be useful in pursuing the grand unification of interactions. 
 This will be done, based on 
 a concise reformulation to be derived for a major part of the GWS interaction Hamiltonian.
 The framework to be proposed is directly founded on a quantum-mechanical ground,
 not making use of any gauge-symmetry assumption
 which has to be introduced firstly for classical fields.

 As well known, the GWS electroweak interaction Hamiltonian denoted by $H^{\rm gws}_{\rm int}$, 
 which is obtained after the Higgs mechanism has been used to introduce particle masses,
 consists of two parts,
 \begin{align}\label{Hint-gws}
  H^{\rm gws}_{\rm int} =  H^{\rm gws}_{\rm int,I} + H^{\rm gws}_{\rm int,II}, 
 \end{align}
 where $H^{\rm gws}_{\rm int,I}$ includes all those interactions that are between leptons
 and vector bosons,
 while, $H^{\rm gws}_{\rm int,II}$ includes interactions merely among vector bosons.
 In this paper, we mainly discuss interactions described by $H^{\rm gws}_{\rm int,I}$ and 
 are to call them \emph{lepton-vector-boson interactions}.
 And, we are to discuss the first generation of leptons only.

 The paper mainly consists of two parts. 
 The first part gives a reformulation of $H^{\rm gws}_{\rm int,I}$.
 To achieve this goal, we are to utilize mathematical tools supplied by the theory of spinors,
 which is based on the so-called $SL(2,C)$ group 
 --- a covering group of the proper orthochronous Lorentz group
 \cite{Penrose-book,Kim-group,CM-book,Corson,pra16-commu}.
 \footnote{As is known (see, e.g., Ref.\cite{Penrose-book}),  for the purpose of 
 addressing spin states of elementary particles, 
 the $SL(2,C)$ group is more powerful than the Clifford algebra.
 This is at least for two reasons:
 (i) The Clifford algebra is for Dirac fermions, while, 
 the $SL(2,C)$ group is for both  fermions and bosons.
 And, (ii) in the Clifford algebra, to make a distinction between 
 the LH and RH parts of Dirac spinors, 
 one must discuss within a special representation, the so-called
 chiral representation of $\gamma^\mu$-matrices;
 while, this distinction comes out in a natural way in the theory of spinors.
 }
 In particular, the reformulation is based on an isomorphic relationship between 
 the direct product of two Weyl-spinor spaces
 (for the left-handed (LH) and right-handed (RH) parts of Dirac spinors, respectively)
 and a four-component-vector space,
 through the so-called  Enfeld-van der Waerden symbols, in short, \emph{EvdW symbols}.
 Moreover, in the reformulation to be given, 
 different particle species are associated with 
 different types of spinor spaces for the description of their spin states, 
 such that the former are characterizable by configurations of the latter.

 In the second part, based results obtained in the first part,
 a framework is proposed for the description of elementary particles and their interactions.
 It includes five basic assumptions
 and certain generic aspects of four model-dependent assumptions. 
 Three concrete models are to be discussed within the framework. 
 The first model predicts the same interaction Hamiltonian as $H^{\rm gws}_{\rm int,I}$ of the GWS theory, 
 while, it shows certain flaw in view of the proposed framework.
 Overcoming the flaw gives rise to the second model,
 which retains only one coupling constant, with the Weinberg angle fixed at  $\sin^2 \theta_{W} = 0.25$.
 The third model shows a possibility of supplying an alternative description for strong interaction,
 with the quantum number of color related to certain 
 type of configuration of spinor space.

 The paper is organized as follows. 
 Major notations are given and discussed in Sec.\ref{sect-main-notation}.
 Basic results of the 
 reformulation of $H^{\rm gws}_{\rm int,I}$ are presented in Sec.\ref{sect-main-results},
 whose detailed derivation is given in Secs.\ref{sect-reformulate-Hint}
 and \ref{sect-EvdW-electroweak}.
 The framework for introducing elementary particles and their interactions
 is proposed in Sec.\ref{sect-framework}.
 Applications of the framework as three models are discussed in Sec.\ref{sect-models}. 
 Finally, conclusions and discussions are given in Sec.\ref{sect-conclusion}.

\section{Notation system}\label{sect-main-notation}

 In this section, we discuss basic notations to be used in later sections.
 It proves convenient to write spinors in the abstract ket-bra form, 
 when expressing some of the main results to be derived in later sections. 
 \footnote{ 
 For detailed discussions on the abstract ket-bra notation for spinors, see Appendix \ref{sect-2-spinor}.
 Brief discussions about $SL(2,C)$ transformations are given in Appendix \ref{sect-SL2C-transf}.
 }
 Most of the notations are explained in Sec.\ref{sect-some-basic-notation}
 and those for bras are discussed in Secs.\ref{sect-scalar-bra} and \ref{sect-single-p-notation}.

\subsection{Basic notations not including spinor bras}\label{sect-some-basic-notation}

% We first explain conventions to be adopted for overline and repeated index,
% then, discuss basic notations to be employed.

\vspace{0.2cm}

 \noindent (a) \emph{Overline}. 
 \\ {Overlines} are to be used in three different situations, as explained below. 
 \\ (i) An {overline} above a spinor indicates its complex conjugate
 and similar for symbols related to spinors.
 \footnote{This is a convention usually adopted in the theory of spinors.
 We adopt it here, for the sake of convenience it brings when dealing with spinors. 
 To avoid confusion, in this paper, say, $U^\dag \gamma^0$  for a Dirac spinor $U$
 is written explicitly, but not in the usual abbreviation of $\ov U$.
 }
 \\ (ii) For a fermion species denoted by $f$, we use $\ov f$ to indicate the antiparticle of $f$,
 with  $\ov {\ov f} \equiv f$ implied. 
 In particular, electron and positron are indicated as $e$ and $\ov e$, respectively;
 meanwhile, electron neutrino and electron antineutrino as $\nu$ and $\ov \nu$, respectively. 
 \footnote{
 As will become clear in later discussions [see, e.g., Eq.(\ref{single-p-s})],
 the first and second usages of overline are in fact consistent.}
 \\ (iii) An overline above a three-momentum $\bp$ 
 indicates the opposite momentum, i.e., 
\begin{align}\label{}
 \ov\bp := - \bp.
\end{align}

\vspace{0.2cm}
\noindent (b) Sign of $p^0$.
 \\  A crucial point in the approach to be studied is to utilize
 negative-$p^0$ solutions of the Dirac equation for free fermion, 
 where $p^0$ is the first component of a four-momentum $p^\mu$ of $\mu=0,1,2,3$.
 \footnote{See Appendix \ref{app-neg-E-UV} for detailed properties of the
 negative-$p^0$ solutions of the free Dirac equation, 
 as well as their relationship to the ordinarily used positive-$p^0$ solutions.}
 A label $\varrho$, with $\varrho = \pm$, is used to represent the sign of $p^0$, i.e., 
 \begin{align}\label{p0-varrho}
  p^0 = \varrho |p^0|.
\end{align}
 
\vspace{0.2cm}
\noindent (c) Convention of summation.
\\ The convention of repeated index implying summation  is obeyed,
except for the label $\varrho$. 

\vspace{0.2cm}
\noindent (d) Spaces of Weyl spinors.
\\ The two smallest nontrivial representation spaces of the $SL(2,C)$ group
 are denoted by $\WW$ and $\ov \WW$,
 which are spanned by two-component Weyl spinors.
 \footnote{In the theory of spinors, 
 each nontrivial representation space of the $SL(2,C)$ group 
 is either one of $\WW$ and $\ov\WW$, or built from them.
 }
 The two spaces are the complex conjugate of each other.
\footnote{  Weyl spinors in $\WW$ and $\ov \WW$ are called
 the LH and RH parts of Diract spinors in the ordinary formulation. }
 An arbitrary spinor in $\WW$ possesses two components
 written as, say, $\kappa_A$,  where $A$ is the spinor index with $A=0,1$. 
 The complex conjugate of $\kappa_A$ gives a Weyl spinor in the space $\ov\WW$,
 written as $\ov \kappa_{A'}$ with a primed index $A' = 0', 1'$.

\vspace{0.2cm}
\noindent (e) Some other spinor spaces.
\\ We use $\VV$ to indicate the spinor space of four-component vectors,
which are useful in the description of spin states of vector bosons. 
The space $\VV$ is isomorphic to $\WW \otimes \ov\WW$, as known in the theory of spinors.
Spaces that are isomorphic to $\WW \otimes \WW$ and $\ov \WW \otimes \ov\WW$
 are denoted by $\C$ and $\ov \C$, respectively.

\vspace{0.2cm}
\noindent (f) Raising and lowering Weyl-spinor indices. 
 \\ Two symbols of $\epsilon^{AB}$ and $\epsilon_{AB}$ are used 
 for raising and lowering indices of spinors in $\WW$, respectively, which have the matrix expression of 
\begin{align}\label{}
 [\epsilon^{AB}] = [\epsilon_{AB}] =  \left( \begin{array}{cc} 0 & 1 \\ -1 & 0 \end{array} \right).
\end{align}
 For example, 
\begin{align}\label{kappa-raise-lower}
 \kappa^A = \epsilon^{AB} \kappa_B, \quad \kappa_A = \kappa^B \epsilon_{BA}. 
\end{align}
 The corresponding symbols for $\ov\WW$ are written as $\epsilon^{A'B'}$ and $\epsilon_{A'B'}$, respectively,
 described by the same matrix.

\vspace{0.2cm}
\noindent (g) Kets of spinors in the spaces of $\WW, \ov\WW$, and $\VV$.
 \\ We use a ket $|S^A\ra$ $(A=0,1)$ to indicate a basis in the space $\WW$.
 Thus, a Weyl spinor $\kappa_A \in \WW$ is written as
\begin{align}\label{kappa-ket-expan}
 |\kappa\ra = \kappa_A|S^A\ra.
\end{align}
 Correspondingly, a basis in $\ov\WW$, as the complex conjugate of $|S^A\ra$,  is written as $|\ov S^{A'}\ra$.
 The complex conjugate of $|\kappa\ra$, namely $|\ov \kappa \ra \in \ov\WW$, 
 is written as ${| \ov \kappa \ra} = \ov\kappa_{A'}|\ov S^{A'}\ra$.
\\
 A basis in  a four-component vector space $\VV$ is written as $|T^\mu\ra$ ($\mu =0,1,2,3$). 
 On this basis, an arbitrary ket $|K\ra \in \VV$ is expanded as $|K\ra = K_\mu |T^\mu\ra$.
 For example, polarization vectors $\varepsilon^{\lambda}_{ \mu}(\bk)$ are written as 
 $ | \varepsilon^{\lambda}(\bk)\ra = \varepsilon^{\lambda}_{ \mu}(\bk) |T^\mu\ra $.
 The index $\mu$ is raised and lowered by the Minkovski metric $g^{\mu \nu}$
 and $g_{\mu\nu}$, respectively [cf.~Eq.(\ref{g-munu}) in Appendix \ref{sect-recall-vector}].
% with $[g^{\mu \nu}] = [g_{\mu\nu}]$.

\vspace{0.2cm}
\noindent (h) 
 Three-momentum states.
 \\ States of three-momentum are written in the ordinary ket-bra form, say, $|\bp\ra$ 
 and $\la \bp |$.
 They possess the following inner product,
\begin{gather}\label{<bp|bq>}
  \la \bq |\bp \ra   = |p^0| \delta^3(\bp-\bq),
\end{gather}
 which is Lorentz invariant and, hence, is a scalar product of the $SL(2,C)$ group. 

\vspace{0.2cm}
\noindent (i) Kets for single particle states. 
 \\ For a free fermion $f$, which possesses a three-momentum $\bp$
 and a sign $\varrho$ of $p^0$ in Eq.(\ref{p0-varrho}), 
 the ket form of its state is written as $|{f}_{\bp r \varrho} \ra$, 
 where $r$  ($r=0,1$) is  the ordinary  spin label
 which is raised by the  Kroneck symbol $\delta^{rs}$ and lowered by $\delta_{rs}$.
 The space, which is spanned by those vectors $|f_{\bp  r \varrho }\ra$
 that possess a given sign $\varrho$, is denoted by $\E_{f \varrho}$,
\begin{align}\label{}
 \label{Ef} & \E_{f \varrho} = \bigoplus_{\bp,r} |f_{\bp r \varrho}\ra.
\end{align}
 The ket form of a single-particle state of a vector boson $B$,
 with a three-momentum $\bk$ and an ordinary polarization index $\lambda$ ($\lambda =0,1,2,3$),
 is written as $|B_{\bk \lambda}\ra$.
 The space, which is spanned by $|B_{\bk \lambda}\ra$, is indicated as $\E_{B}$,
  \begin{align}\label{}
  \label{E1-boson} & \E_{B}  = \bigoplus_{\bk, \lambda} |B_{\bk \lambda}\ra.
 %  = \bigoplus_{\bk} |\bk\ra \otimes \VV,
\end{align}
 Fermionic and bosonic states satisfy the usual anticommutation and commutation relations,
\begin{subequations}
\begin{align}\label{}
 \label{ff'-f'f} &  |f_{\bp r \varrho}\ra |f'_{\bp' r' \varrho'}\ra = - |f'_{\bp' r' \varrho'}\ra |f_{\bp r \varrho}\ra ,
 \\ & |B_{\bk \lambda}\ra|B'_{\bk' \lambda'}\ra = |B'_{\bk' \lambda'}\ra|B_{\bk \lambda}\ra.
\end{align}
\end{subequations}

\vspace{0.2cm}
\noindent (j) Direct product form of single-particle states. 
 \\ The above single-particle states are direct products of momentum states and spin states.
 In the ordinary notation, the spin states are not written in the ket-bra form;
 more exactly, fermionic spin states are written as $U^{r}(\bp)$ and $V^{r}(\bp)$, as 
 positive-$p^0$ stationary solutions of the free Dirac equation 
 [see Eqs.(\ref{Up-uv-app}) and (\ref{Vp-uv-app}) in Appendix \ref{sect-Dirac-spinor}]
 (e.g., for electron and positron, respectively).
 But, for reasons to be explained later around Eq.(\ref{ovU-delta-rs}), 
 positron's spin states are to be described by $\ov U^{r}(\bp)$, the complex conjugate of $U^{r}(\bp)$. 
 Meanwhile, bosonic spin states are written as polarization vectors $\varepsilon^{\lambda}_\mu(\bk)$.
 In the ket form, with the sign $\varrho$ indicated explicitly,
 single-particle states of electron, positron, and photon are to be written as follows, 
\begin{subequations}\label{single-p-s}
\begin{align}
 \label{|b>} & |e_{\bp r \varrho}\ra=  |\bp\ra |U_{r \varrho}(\bp)\ra,
 \\  \label{|d>} &  |\ov e_{\bp r \varrho}\ra = |\bp\ra |\ov U_{r \varrho}(\bp)\ra ,
 \\ \label{|a>} & |A_{\bk \lambda}\ra = |\bk\ra |\varepsilon_\lambda(\bk)\ra.
\end{align}
\end{subequations}

\vspace{0.2cm}
\noindent (k)
 The direct-product space $\WW \otimes\ov\WW$.
 \\ Basis spinors in this space are written as $|S_{AB'}\ra \equiv |S_A\ra | \ov S_{B'}\ra $.
 The following anticommutation relation is to be used, 
\begin{align}\label{SAB-commu-main}
 |S_A\ra | \ov S_{B'}\ra = - |\ov S_{B'}\ra |S_A\ra |,
\end{align}
 in computations related to interaction amplitudes that will be carried out later.  
 \footnote{We note that the same final results as to be given this paper are obtainable without imposing 
 the anticommutation relation in Eq.(\ref{SAB-commu}).
 But, this would require more complicated definitions for some quantities to be used later. 
 }

\subsection{Bras introduced for scalar products of spinors}\label{sect-scalar-bra}

 One main purpose of introducing bras in quantum mechanics is to express inner product
 in a convenient way.
 However, in the theory of spinors,
 of primary importance is the concept of scalar product under $SL(2,C)$ transformations, 
 but not inner product.
 This requires that bras should be first of all introduced for scalar products, 
 as discussed in Ref.\cite{pra16-commu}.
 Below, we call such a bra a \emph{scalar-product-based bra}
 and use the symbol of ``$\la\la \cdot |$'' to indicate it. 
 For example, such a bra for $|S^A\ra $ is written as $\la\la S^A|$.
 \footnote{
 In discussions given in Ref.\cite{pra16-commu}, which are relatively simple compared with
 those to be given later in this paper, 
 the ordinary symbol of $\la \cdot |$ was also used for scalar-product-based bras.
 But, here, in order to avoid confusion, we use the specific notation of ``$\la\la \cdot |$'' for them. 
 }

 In the theory of spinors, 
 the most fundamental scalar product is that of Weyl spinors.
 In the space $\WW$, it is
 written as $ \kappa_A \chi^A$ for two arbitrary Weyl spinors $\kappa^A$ and $\chi^A$. 
 (This scalar product is not an inner product.)
 In the ket-bra form, this product is written as 
\begin{equation}\label{<chi|kappa>-main}
 \la\la\kappa|\chi\ra \equiv \kappa_A \chi^A.
\end{equation}
 For a ket $|\kappa\ra$ expanded in Eq.(\ref{kappa-ket-expan}), this requires 
 the following scalar-product-based bra $\la\la \kappa|$,
\begin{align}\label{kappa-bra-expan}
 \la\la \kappa| = \la\la S^A| \kappa_A.
\end{align}
 Making use of Eq.(\ref{kappa-raise-lower}), it is straightforward to find that
\begin{equation}\label{SA-SB-main}
  \la\la S^{A}|S^{B}\ra = \epsilon^{A B}.
\end{equation}
 The antisymmetry of $\epsilon^{A B}$ implies that
\begin{equation}\label{ck=-kc-main}
  \la\la \chi |\kappa\ra \equiv \chi_A \kappa^A = -\la\la\kappa |\chi\ra
  \equiv -\kappa_A \chi^A .
\end{equation}
 It is easy to check that  [cf.~Eq.(\ref{kappa-A})]
\begin{equation}\label{kappa-A-main}
  \kappa^A = \la\la S^{A}|\kappa\ra, \quad \kappa_A = \la\la S_{A}|\kappa\ra.
\end{equation}
 The above discussions are also applicable to spinors in the space $\ov\WW$.

 For a space $\VV$, the scalar product of two four-component vectors
 $K^\mu$ and $J^\mu$, which has the well known form of $K_\mu J^\mu$,
 is written as $ \la\la K|J\ra$ in the abstract notation.
 Similar to Eq.(\ref{kappa-bra-expan}), 
 the scalar-product-based bra of $|K\ra$ is expanded as
\begin{align}\label{<<K|}
 \la\la K| := \la\la T^\mu|  K_\mu,
\end{align}
 where $\la\la T^\mu|$ is a basis for the scalar-product-based bras, satisfying
\begin{eqnarray}\label{Tmu-Tnu-main}
  \la\la T^\mu|T^\nu\ra =  g^{\mu\nu}.
\end{eqnarray}
 It is easy to check that $ \la\la K|J\ra = \la\la J|K\ra$ and 
%  [cf.~Eqs.(\ref{K-mu}) and (\ref{<K|J>=<J|K>})]
\begin{align}  \label{K-mu-main}
 & K^\mu = \la\la T^\mu |K\ra , \quad K_\mu = \la\la T_\mu |K\ra .
\end{align}

\subsection{Bras for single-particle states}\label{sect-single-p-notation}

 In this section, making use of the above-discussed scalar-product-based bras,
 we discuss construction of bras that are useful for expressing single-particle states.

 As is well known, transferring kets for single-particle states to the corresponding bras 
 should involve a procedure of complex conjugation. 
 However, the right-hand sides (rhs) of Eqs.(\ref{kappa-bra-expan}) and (\ref{<<K|})
 contain the components ($\kappa_A$ and $K_\mu$), but, not their complex conjugates.
 This implies that the scalar-product-based bras should not be directly used
 in the construction of bras for single-particle states.
 \footnote{ 
 In fact, it is impossible to directly use scalar-product-based bras of spinors and the momentum bras $\la \bp|$
 to construct a linear space dual to that of kets.
 For example, $\la\la S^A| \la\bp|$ do not span a dual space. 
 The problem comes from 
 the fact that the construction of scalar-product-based bras of spinors
 does not involve any procedure of complex conjugation
 [see Eqs.(\ref{kappa-bra-expan}) and (\ref{<<K|})],
 while, the ket-bra transformation for momentum states takes the form of $c|\bp\ra \to \la \bp | c^*$.
 In fact, when trying to write a bra for a ket of $|\psi\ra = \int d^3 p \ c_A(\bp)|S^A\ra |\bp\ra$,
 $\la \la S^A|$ requires expansion coefficients $c_A(\bp)$,
 while, $\la \bp|$  requires $c^*_A(\bp)$. 
 }
 In other words, to construct bras for single-particle states, 
 certain changes involving complex conjugation are needed.

 Let us first discuss fermions whose spin states are described by Dirac spinors. 
 As is well known, Dirac spinors possess inner product, which involves complex conjugation.
 To write the inner product in the abstract notation, 
 one may make use of the so-called \emph{hat-bras} of Dirac spinors
 as discussed in Ref.\cite{pra16-commu},
 which are briefly recalled in Appendix \ref{sect-Dirac-spinor}. 
 For example, the hat-bra of a ket $|U(\bp)\ra$ is indicated as $\la \wh U(\bp)|$
 [see Eq.(\ref{wh-U-app}) for its expression].
 Then, the bra for the single-particle ket $|{f}_{\bp r \varrho}\ra$ of $f=e$
 [Eq.(\ref{|b>})], which is to be indicated as $\la {f}_{\bp r \varrho}|$, is written as
\begin{align}\label{}
  \la {e}_{\bp r \varrho}| = \la \wh U_{r\varrho}(\bp)| \la \bp|.
\end{align}
 (See Appendix \ref{app-hatbra-spinors} for explicit expressions of various hat-bras.)

 Next, we discuss bosons whose spin states are described by four-component vectors.
 For them, one may take advantage of 
 the fact that the space $\VV$ coincides with its complex-conjugate space.
 More exactly, one may use, say, $\la \la \ov K|$ in the construction of bras for single-particle states. 
 To be consistent with the ordinary notation, we indicate $\la \la \ov K|$ as $\la K|$, 
\begin{align}\label{<K|=<<ov-K|}
 \la K| := \la \la \ov K|.
\end{align}
 Thus, the bra of $|K\ra |\bp\ra$ is written as $\la \bp| \la K|$. 
 By this method, one simply gets the bra of a ket $|B_{\bk \lambda}\ra$ for a bosonic state,
 indicated as $\la B_{\bk \lambda}|$.

 Scalar and inner product of fermionic states is written as follows,
\begin{align}\label{<f'-varrho|f>=delta}
  \la {f'}_{\bp' r' \varrho'} |{f}_{\bp r \varrho} \ra 
 = \varrho |p^0| \delta^3(\bp-\bp') \delta_{rr'} \delta_{ff'} \delta_{\varrho \varrho'},
\end{align}
 where Eq.(\ref{p0-varrho}) has been used.
 It is easy to check the relation in Eq.(\ref{<f'-varrho|f>=delta}) in the case of $\varrho = \varrho'$,
 by making use of properties of spinors discussed in Appendix \ref{sect-2-spinor}.
 While, the term $\delta_{\varrho \varrho'}$ is imposed as an assumption for \textit{essential difference 
 between fermions with positive and negative $p^0$}
 \footnote{This assumption is in consistency with a rule to be introduced in Sec.\ref{sect-main-results-2}
 (rule for negative-$p^0$ states). 
 }.
 Making use of Eq.(\ref{<f'-varrho|f>=delta}), it is straightfoward to 
 check that $\varrho \varrho'  I_{f\varrho, f' \varrho'} $
 is an identity operator that acts on the space of 
 $\E_{f\varrho} \otimes \E_{f' \varrho'}$, where
\begin{align}\label{} 
 & I_{f\varrho, f' \varrho'} := \int d\ww p d\ww p'
 | f_{\bp \varrho}^r\ra | {f'}_{\bp' \varrho'}^{r'}\ra \la f'_{\bp' r'\varrho'}| \la  f_{\bp r \varrho}|.
\label{I-ff'} 
\end{align}
 Here and hereafter, $d\ww p$ indicates the following abbreviation (similar for $d\ww k$),
\begin{gather}\label{dwwp}
 d\ww p := \frac{1}{|p^0|} d^3p.
% quad  \& \quad d\ww k = \frac{1}{k^0} d^3k,
\end{gather}

 For bosonic states, we consider four-momenta $k^\mu$ with positive $k^0$ only.
 Normalized single-boson states possess the ordinary scalar products, written as 
\begin{gather}\label{<B|B>}
  \la B_{\bk \lambda}|B'_{\bk' \lambda'}\ra =   k^0 \delta^3(\bk-\bk') g_{\lambda \lambda'}
  \delta_{BB'},
\end{gather}
 where $g_{\lambda\lambda'}$ is the Minkovski metric.
 It is easy to check that the identity operator on the space $\E_B$ is written as
\begin{gather}\label{IB-main} 
 I_{B} =  \int d\ww k  |B_{\bk \lambda}\ra  \la B_{\bk}^{ \lambda}|.
% = \int d\ww k |\bk\ra |\varepsilon_{\lambda}(\bk)\ra  \la  \varepsilon^{\lambda}(\bk)| \la \bk| .
\end{gather}
% with $k_0=|\bk|$ in $d\ww k$.

\section{Main results of the first part}\label{sect-main-results}

 In this section, we present main results of the first part of this paper,
 which are for the GWS lepton-vector-boson interaction Hamiltonian $H^{\rm gws}_{\rm int,I}$
 in the mass representation obtained by the Higgs mechanism.
 The major achievement is a concise reformulation of $H^{\rm gws}_{\rm int,I}$,
 expressed by means of two concepts to be introduced below ---
 fundamental interaction operator (FIO) and fundamental vacuum fluctuation (FVF).
 To achieve this goal, 
 it proves convenient to discuss in the Schr\"{o}dinger picture, which is to be adopted.
 Meanwhile, we are to write creation and annihilation operators
 in their equivalent forms of kets and bras, respectively.
% \footnote{\label{note-creation-operator-as-ket}
% The ket-bra forms of fermionic creation and annihilation operators possess positive $\varrho$,
% but not negative $\varrho$, due to a rule for negative-$p^0$ states to be introduced 
% in Sec.\ref{sect-main-results-2}. }.

 Specifically, spinor spaces for spin states of particles are discussed in Sec.\ref{sect-two-layer}.
 The two concepts of FIO and FVF, with their mathematical expressions, are discussed 
 in Secs.\ref{sect-FIO} and \ref{sect-FVF}, respectively. 
 Finally, the main result is presented in Sec.\ref{sect-main-results-2}.

\subsection{Particle species and their spinor spaces}\label{sect-two-layer}

 Only eight particle species are involved in discussions to be given below
 for the above mentioned reformulation of $H^{\rm gws}_{\rm int,I}$;
 that is, four leptons of the first generation (electron $e$, positron $\ov e$, 
 electron neutrino $\nu$, and electron antineutrino $\ov \nu$)
 and four vector bosons mediating electroweak interactions
 (photon, $Z^0$ boson, and $W^\pm$ bosons). 
 In this section, we discuss basic structures to be employed for spin spaces of these particle species,
 leaving their exact descriptions to a later section (Sec.\ref{sect-GWS-two-layer}).

 The spin space for each of the leptonic species is assumed to be 
 a direct sum of two Weyl-spinor spaces (namely, two of $\WW / \ov\WW$).  
 Clearly, there are only four possibilities for the spinor-space configuration. 
 We assume that they are related to the four leptonic species as given below (for kets), 
\begin{subequations}\label{config-WovW}
 \begin{align}\label{config-WovW-e}
 e: \left( \begin{array}{c} \WW \\ \ov\WW  \end{array} \right), \ \quad
  \ov e: \left( \begin{array}{c} \ov \WW \\ \WW  \end{array} \right), \ 
  \\  \nu: \left( \begin{array}{c} \WW \\ \WW  \end{array} \right), \  \quad
   \ \ov \nu: \left( \begin{array}{c} \ov \WW \\ \ov\WW  \end{array} \right). \label{config-WovW-nu}
 \end{align}
\end{subequations}
 One notes that the electron's spinor configuration given above
 is the same as the ordinarily-used one, while, others' are not. 
 The spinor-space configurations in (\ref{config-WovW}) show a \emph{two-layer} structure,
 with each layer as a Weyl-spinor space. 
 The upper layer will be referred to as the \emph{first} layer and the lower one as the second layer.

 Spin states of vector bosons are assumed to possess a two-layer structure, too.
 For them, the spinor space for each layer is assumed to be isomorphic to 
 the direct product of two Weyl-spinor spaces.
 There are three possibilities of the direct product, i.e., $\WW \otimes \ov\WW$,
 $\WW \otimes \WW$, and $\ov \WW \otimes \ov\WW$,
 and they correspond to the spinor spaces of $\VV$, $\C$, and $\ov \C$, respectively,
 which have been discussed in ``(e)'' of Sec.\ref{sect-some-basic-notation}.
 Since a vector boson should possess at least one $\VV$-type layer,
 we are to set the first layer always as a $\VV$-type layer.

 Although the space $\VV$ is equal to its complex-conjugate space $\ov\VV$, 
 most vectors in $\VV$ are not equal to their complex conjugates.
% \footnote{See Appendix \ref{sect-vector-abstract} for detailed discussions.} 
 For this reason, 
 the configuration of $ \left( \begin{array}{c} \VV  \\ \VV  \end{array} \right)$
 may be regarded as different from that of $ \left( \begin{array}{c} \VV  \\ \ov\VV  \end{array} \right)$
\footnote{ Meanwhile, there is no essential difference between 
 $ \left( \begin{array}{c} \VV  \\ \VV  \end{array} \right)$ and 
 $ \left( \begin{array}{c} \ov\VV  \\ \ov\VV  \end{array} \right)$,
 and similar for $ \left( \begin{array}{c} \VV  \\ \ov\VV  \end{array} \right)$ and 
 $ \left( \begin{array}{c} \ov\VV  \\ \VV  \end{array} \right)$.}.
 Then, there are totally four independent two-layer configurations
 built from $\VV$, $\C$, and $\ov \C$ for vector bosons,
 which are to be related to photon, $Z^0$ boson, and $W^\pm$ bosons, i.e.,
\begin{subequations}\label{config-VCC}
  \begin{align}\label{photon-Z0-VV}
  & \text{photon} :  \left( \begin{array}{c} \VV
     \\ \ov\VV  \end{array} \right), \quad
  Z^0 : \left( \begin{array}{c} \VV
     \\ \VV  \end{array} \right), \ 
  \\ & W^+_\mu: \left( \begin{array}{c} \VV 
     \\ \C  \end{array} \right), \quad 
  W^-_\mu:   \left( \begin{array}{c} \VV 
       \\ \ov\C  \end{array} \right). \ \label{config-W-otimes-ovW}
\end{align}
\end{subequations}

 To summarize, there are eight two-layer configurations of spinor spaces, 
 which satisfy requirements discussed above, and they are used for the 
 eight particle species, respectively. 
 In this sense, the eight particle species are characterizable by their spinor-space configurations.

 More specifically, kets for spin states of the leptons $f$ and bosons $B$ are written 
 in the following column form, 
\begin{subequations}\label{S-fB-gen}
  \begin{align}\label{S-f-gen}
   &  |\cs_{fr\varrho}(\bp)\ra = 
    \left( \begin{array}{c} |\LL^f_1 \ra \\  | \LL^f_2 \ra 
     \end{array} \right)
     \quad \text{($|\LL^{f}_{1,2} \ra \in \WW$ or $\ov\WW$)},
 \\ &    \label{S-B-gen}   |\cs_{B \lambda}(\bk) \ra = 
     \left( \begin{array}{c} |\LL^B_1 \ra \\ |\LL^B_2 \ra 
     \end{array} \right)
     \quad \text{($|\LL^{B}_2 \ra \in \VV$, or $\C$, or $\ov\C$)},
   \end{align}  
\end{subequations}
 with $|\LL^{B}_1 \ra \in \VV$;
 meanwhile, bras are written in the row form. 
 It will turn out that, when computing the interaction amplitudes
 of the reformulated $H^{\rm gws}_{\rm int,I}$,
 explicit expressions only of $|\cs_{f r \varrho}(\bp)\ra$ and 
 $\la \cs_{B \lambda}(\bk) |$ are to be used,
 which will be given in Sec.\ref{sect-GWS-two-layer}.
 \footnote{ In fact, $|\cs_{B \lambda}(\bk) \ra$ may be gotten directly from $\la \cs_{B \lambda}(\bk) |$
 by taking complex conjugation.
 But, the situation with the bras $\la \cs_{f r \varrho}(\bp)|$ is more complicated.
 Their explicit expressions are discussed in Appendix \ref{app-hatbra-spinors}.
 Making use of the explicit expressions given there, one may check that
 the relations in  Eqs.(\ref{<f'-varrho|f>=delta}) and (\ref{<B|B>}) still hold.
 }
 Kets for single-particle states in the two-layer description are written as 
\begin{subequations}\label{single-p-s-gen-2L}
      \begin{align}\label{}
       \label{|f>2L} & |f_{\bp r \varrho}\ra=  |\bp\ra |\cs_{f r \varrho}(\bp)\ra,
       \\ \label{|B>2L} & |B_{\bk \lambda}\ra = |\bk\ra |\cs_{B \lambda}(\bk) \ra;
      \end{align}
\end{subequations}
 and, the corresponding bras are written as
\begin{subequations}\label{single-p-s-gen-2L-bra}
      \begin{align}\label{}
       \label{<f|2L} & \la f_{\bp r \varrho}| =  \la \bp| \la \cs_{f r \varrho}(\bp)|,
       \\ \label{<B|2L} & \la B_{\bk \lambda}| = \la \bk| \la \cs_{B \lambda}(\bk) |.
      \end{align}
\end{subequations}
 The single-particle states obey the ordinary commutation rules
 (anticommutability for fermionic states and commutability for bosonic states).

 One remark. Although most of the descriptions of single-particle states in Eq.(\ref{single-p-s-gen-2L})
 use spinor spaces differing from the ordinarily-used ones, 
 they should possess the same scalar products 
 as those given in Eqs.(\ref{<f'-varrho|f>=delta}) and (\ref{<B|B>})
 and should finally produce the same interaction amplitudes for $H^{\rm gws}_{\rm int,I}$.
 These points will be shown in later sections, when their detailed descriptions are given.

\subsection{Fundamental interaction operator}
\label{sect-FIO}

 The concept of \textit{fundamental interaction operator} (FIO) 
 is introduced, based on the following feature of the operator of $H^{\rm gws}_{\rm int,I}$, 
 particularly, when expanded by means of creation and annihilation operators.
 That is, each interaction term in it in fact represents a map between two state spaces,
 e.g., a map from the state space of an electron-positron pair to that of a photon.
 A crucial observation, which will be explained later [see, e.g., Eq.(\ref{H1-8-Vf})], 
 is that all the maps may be traced back to two types of simple maps.
 This gives rise to two types of FIOs explained below.

 One type of FIO is for the change of a pair of fermions $(f,f')$ to one vector boson $B$
 and is to be referred to as {FIO1}. 
 For two fermions $f$ and $f'$, which possess signs $\varrho$ and $\varrho'$ 
 [Eq.(\ref{p0-varrho})], respectively, 
 the FIO1 is denoted as $H^{{\rm FIO1} }_{\varrho \varrho', \eta}$, where 
 \begin{align}\label{eta-define}
 \eta := (f,f', B).
 \end{align} 
 The other type of FIO is for the reverse map, namely $B \to ff'$, and is called FIO2, 
 denoted by $H^{{\rm FIO2} }_{\varrho \varrho', \eta}$.
 Each FIO2 is assumed to be the Hermitian conjugate of the related FIO1, i.e., 
\begin{align}\label{FIO2=FIO1-dag}
  H^{{\rm FIO2} }_{\varrho \varrho', \eta} = \left( H^{{\rm FIO1} }_{\varrho \varrho', \eta} \right)^\dag.
\end{align}

 By definition, the FIO1 describes the most direct map from 
 the space of $\E_{f\varrho} \otimes \E_{f' \varrho'}$ to the space $\E_B$, 
 subject to three requirements: momentum conservation, layer isomorphism,
 and being through the EvdW symbols
 which are usually indicated as $ \sigma^{\mu}_{ AB'}$.
 Here, \emph{layer isomorphism} means that 
 the direct product of the spinor spaces for the first layers of the two fermions $f$ and $f'$ 
 should be isomorphic to the spinor space for the first layer of the vector boson $B$, 
 and the same is true for their second layers.

 As a map, generically the FIO1 is written as
 \begin{equation}\label{HFIO1-model}
  H^{{\rm FIO1} }_{\varrho \varrho', \eta}
  :=  I_B \, \G^{\rm FIO1}_{\eta} \, I_{f\varrho, f' \varrho'},
 \end{equation}
 where $I_B$ and $ I_{f\varrho, f' \varrho'}$ (multiplied by $\varrho \varrho'$)
 are identity operators for  the space $\E_B$ and the space $\E_{f\varrho} \otimes \E_{f' \varrho'}$
 [see Eqs.(\ref{IB-main}) and (\ref{I-ff'})], respectively,
 and $\G^{\rm FIO1}_{\eta}$ is a symbol that connects kets in $I_{f\varrho, f' \varrho'}$ to bras in $I_B$.
 More exactly, $\G^{\rm FIO1}_{\eta}$ is defined by the elements
 $\la B_{\bk}^{ \lambda}| \G^{\rm FIO1}_{ \eta} | f_{\bp \varrho}^{r}\ra | {f'}_{\bp' \varrho'}^{r'}\ra $,
 which consist of a momentum part and a spinor part, written as
 \begin{align}\label{G1-definition-<|>-mod}
 &  \la B_{\bk}^{ \lambda}| \G^{\rm FIO1}_{ \eta} | f_{\bp \varrho}^{r}\ra 
 | {f'}_{\bp' \varrho'}^{r'}\ra 
 \equiv G_\eta^{\rm mom} G_{\eta}^{\rm spin}.
 \end{align}
 By the above definition, $G_\eta^{\rm mom}$ should represent momentum conservation, 
 meanwhile, $G_{\eta}^{\rm spin}$ should be proportional to the EvdW symbols
 and be subject to  the requirement of layer isomorphism.

 It proves convenient to introduce an 
 operator form of the EvdW symbols $ \sigma^{\mu}_{ AB'}$, 
 which maps the space of $\WW \otimes \ov\WW$
 to the space of $\VV$, denoted by $\sigma$,
 \footnote{See Appendix \ref{sect-vector-abstract} for detailed properties of $\sigma$.}
\begin{equation}\label{sigma}
  \sigma :=  |T_\mu\ra \sigma^{\mu}_{ AB'} \la\la \ov S^{B'}| \la\la  S^A|.
\end{equation}
 Then, the simplest way of fulfilling the two requirements
 on $G_{\eta}^{\rm spin}$ discussed above is to use the following symbol $P_\eta$,  
 \begin{align}\label{P-eta}
  P_\eta = \left\{ 
    \begin{array}{ll}
      \sigma,  &  \text{if the space of ${|\LL^{f}_i \ra |\LL^{f'}_i \ra }$ is isomorphic}
  \\  &  \text{  to that of $|\LL^B_i\ra$ for both $i=1$ and $2$ };
    \\  0,  &  \text{otherwise}.
    \end{array}
    \right. 
 \end{align}
 Moreover, the layer isomorphism suggests that the two layers of
 $ | f_{\bp \varrho}^{r}\ra$ and $| {f'}_{\bp' \varrho'}^{r'}\ra $
 should be treated separately. 
 The simplest method of doing this is to introduce
 a symbol $\overrightarrow D$, defined by
 \begin{align}
  & \label{D}
  \overrightarrow{D}  |\cs_{f\varrho}^{r }(\bp)\ra |\cs_{f'\varrho'}^{r' }(\bp')\ra  \equiv
  \left( \begin{array}{c}   {|\LL^f_1 \ra |\LL^{f'}_1 \ra }
    \\  {|\LL^{f}_2 \ra |\LL^{f'}_2 \ra}  \end{array} \right).
 \end{align}
 Summarizing the above discussions, one gets that
 \footnote{The convention of   $|\psi \phi\ra \equiv |\psi\ra | \phi\ra $ is obeyed. } 
\begin{subequations}\label{G-mom-spin}
  \begin{align} \label{G-mom}
    & G_{\eta}^{\rm mom} = \delta^3(\bp +\bp' -\bk),
 \\ &        G_{\eta}^{\rm spin} =  \la  \cs_{B }^{\lambda }(\bk)|
      P_\eta \overrightarrow{D} | \cs_{f \varrho }^{r }(\bp)  
    \cs_{f' \varrho' }^{r' }(\bp') \ra. \label{G1-spin-eta}
    \end{align}  
\end{subequations}

 One remark: 
 On the rhs of Eq.(\ref{HFIO1-model}), in which 
 $I_B$ and $I_{f\varrho, f' \varrho'}$ are connected by $\G^{\rm FIO1}_\eta$,
 bras in $I_B$ and kets in $I_{f\varrho, f' \varrho'}$ are used to 
 give the interaction amplitude by means of Eq.(\ref{G1-definition-<|>-mod});
 meanwhile, kets in $I_B$ and bras in $I_{f\varrho, f' \varrho'}$ behave like 
 creation and annihilation operators, respectively.
 [As an example for explicit expression of FIO1, see Eq.(\ref{H-FIO1}) to be given later.]

\subsection{Fundamental vacuum fluctuation and a rule for negative-$p^0$ fermion}
\label{sect-FVF}

 The basic idea behind the concept of \textit{fundamental vacuum fluctuation} (FVF) is that
 there is no physical reason to deny the possibility for a pair of a fermion and its antifermion,
 which possesses net zero four-momentum and net zero angular momentum, 
 to \emph{emerge from the vacuum or vanish into the vacuum}. 
 FVF refers to such processes and the fermionic pair involved is called an \textit{FVF pair}. 
 \footnote{Note that FVF thus defined does not involve any boson.
 Hence, it is different from what is usually referred to by the name of ``vacuum fluctuation''
 in QFT as visualized in Feynman diagrams. 
 In fact, as discussed in Appendix \ref{app-Hi-interpret},
 basic Feynman diagrams may be built from FIOs and FVFs. 
 }

 Clearly, in an FVF pair, one of the two fermions should 
 lie in a state with a negative $p^0$. 
 This concept faces the problem
 that such a fermionic state has never been directly observed experimentally.
 We propose to solve this problem by assuming that \emph{no free fermion may lie in
 a negative-$p^0$ state, while, such a state may exist in FIOs and FVFs}.
 We are to call this assumption a \emph{rule for negative-$p^0$ states}.
\footnote{ 
\label{foot-nonegp0-rule}
This assumption for negative-$p^0$ fermion is clearly different from Dirac's
interpretation to negative-energy electron in his theory.
}

 The above discussed rule imposes a mathematical restriction to possible means of using
 FIO and FVF in the construction of a physical theory.
 For example, for an FIO that contains a negative-$p^0$ fermion,
 the rule requires that this FIO should be combined with some FVF or some other FIO,
 in such a way that 
 the negative-$p^0$ fermionic states in them effectively appear \textit{only in the form of scalar product}. 
 In a similar way, an FVF pair should be combined with some FIO and/or FVF.
 We use an \emph{overbrace} to indicate a result that satisfies the above requirement
 imposed by the rule for negative-$p^0$ states.

 To illustrate the exact meaning of the above discussed requirement, as an example, let us consider
 an operator $\OO =|\psi\ra \la f_1|\la f_2|$, where $|\psi\ra$ represents a bosonic state
 and $\la f_1|$ and $\la f_2|$ are fermionic states with $\la f_2|$ possessing a negative $p^0$.
 The action of the operator on a negative-$p^0$ state $|{f}_{\bq s-} \ra$ 
 subject to the above requirement, which is indicated by an overbrace, 
 is written as
\begin{align}\label{overbrace-O}
\overbrace{ \OO |{f}_{\bq s-} \ra }  = \left\{ 
    \begin{array}{ll}
      \la f_2|{f}_{\bq s-} \ra |\psi\ra \la f_1|,  &  \text{if $p^0 >0$ for $\la f_1|$;}
    \\  0,  &  \text{if $p^0 <0$ for $\la f_1|$}.
    \end{array}
    \right. 
\end{align}
 Generically, in the computation of a sequence of kets and bras under an overbrace, 
 those results in which each negative-$p^0$ fermionic state belongs to certain scalar product are retained,
 while, others are abandoned.

 Let us consider the emergence of an FVF pair (from the vacuum)
 and its combination with an operator $\OO$ like an FIO1.
 The emergence feature requires that
 states of the two fermions in the FVF pair should be written as kets.
 It is not difficult to check that the states 
 are generically written as $|{f}_{\bq s -} \ra$ and $|{\ov f}_{\ov \bq s +} \ra$.
 A simplest form for the above mentioned combination, 
 indicated as $V_f(\OO)$ with $f$ referring to a fermion in the FVF pair, is written as
\begin{align} 
V_{f} \left( \OO \right)
 & \equiv \int d\ww q |{\ov f}_{\ov \bq s +} \ra { \OO  {|}{f}_{\bq s-} \ra}. \label{VK1}
\end{align}
 One easily checks that 
\begin{align}\label{Vff'=Vf'f}
 V_{f} \big[ V_{f'}(\cdots ) \big] = V_{f'} \big[ V_{f}(\cdots) \big].
\end{align}
 For the result of $V_f(\OO)$ to satisfy the rule for negative-$p^0$ states, 
 one writes $\overbrace{V_f(\OO)}$.
 Note that, with $|{f}_{\bq s-} \ra$ put on the rhs of the operator $\OO$ 
 in Eq.(\ref{VK1}), this negative-$p^0$ state may form a scalar product with some
 negative-$p^0$ state in $\OO$.

 In a similar way, one may describe the combination of a vanishing FVF pair and some operator, 
 but using bras for the FVF pair. 
 Specifically, one may use the conjugate of $V_f$, denoted by $V_f^\dag$,
 to describe the combination of the FVF pair and an operator $\OO^\dag$, 
\begin{align}
 V^\dag_{f} \left( \OO^\dag \right)
&  \equiv \int d\ww q  { \la {f}_{\bq s -}{|} \OO^\dag }  \la {\ov f}_{\ov\bq s +}|. \label{VK2}
\end{align}

\subsection{Contents of the main result}\label{sect-main-results-2}

 Now, we are ready to present the main result of the first part of this paper,
 the proof of which will be given in the two following sections.
 Basically, it gives a method of building the interaction Hamiltonian operator $H^{\rm gws}_{\rm int,I}$
 by making use of combinations of FIOs and FVFs.
 In each of the combinations, only one FIO is to be used. 
 Since $H^{\rm gws}_{\rm int,I}$ is a Hermitian operator, one may consider only those combinations 
 that contain FIO1s, with the contribution of FIO2s determined by Hermiticity. 
 \footnote{In fact, combinations that involve FIO2 and FVF by means of $V_f^\dag$ may be treated in a way,
 which is similar to that to be given later for FIO1 with $V_f$.
 Detailed discussions of this point in the case of QED are given in Appendix \ref{app-FIO2}.
 }

 Let us consider combination of an FIO1 and some FVF(s). 
 From Eq.(\ref{HFIO1-model}), one sees that the FIO1
 contains two fermionic bras --- one for $f$ and the other for $f'$,
 which are not used in the computation of the interaction amplitude,
 as discussed at the end of Sec.\ref{sect-FIO}.
 This requires that combination of the FIO1 (as $\OO$)
 and FVF(s) should be described by $V_f$ and/or $V_{f'}$
 in Eq.(\ref{VK1}), but, not their conjugates in Eq.(\ref{VK2}). 
 We use $n^V_f$ to indicate the number of $V_f$ that are used in the combination
 and, similarly, $n^V_{f'}$ for the number of $V_{f'}$ used. 
 The final result of the combination, subject to the rule for negative-$p^0$ states,
 is indicated as $V_\omega (\OO)$, with $\omega := ( n^V_f, n^V_{f'})$ and $\OO = {\rm FIO1}$,
 \begin{align}
 & V_\omega (\OO) := \overbrace{ \underbrace{{V_{f'} \big\{ \cdots }} \big[ \underbrace{ V_f( \cdots  } 
 (\OO) \cdots ) \big] \cdots \big\} }.
 \label{F-lambda}
 \\ \notag & \hspace{1.8cm}  n^V_{f'}  \hspace{0.9cm}  n^V_{f}
\end{align}
 Making use of Eq.(\ref{VK1}) and noting Eq.(\ref{overbrace-O}), 
 together with the above mentioned fact that the FIO1 contains only two fermionic bras,
 it is not difficult to check that $V_\omega({\rm FIO1}) =0$ 
 if $n^V_f > 1 $ or $n^V_{f'} > 1$.
 As a consequence, there are only four possibilities of $\omega$ for nonvanishing result, namely, 
 $\omega = (0,0), (1,0), (0,1), (1,1)$. 
 For example, $V_\omega (\OO) = \overbrace{V_f ( \OO )}$ for $\omega = (1,0)$.

 The main result of the first part of this paper is the following expression of $H^{\rm gws}_{\rm int,I}$, 
\begin{align}\label{HIdF-mod}
  H^{\rm gws}_{\rm int,I} 
  = \frac{g}{\sqrt 2} \sum_{\eta, \omega,\varrho, \varrho'} \frac{1}{ \N_B}  
  V_\omega \left( H^{{\rm FIO1} }_{\varrho \varrho', \eta} \right) + {\rm H.c.},
 \end{align}
 where $g$ is a coupling constant,  $\eta$ was defined in Eq.(\ref{eta-define}), 
 ``H.c.'' stands for Hermitian conjugate,
 and $\N_B$ indicate normalization factors of the bosonic states $|\cs_B^\lambda(\bk) \ra$
 [to be given later in Eq.(\ref{N-A-Z0-W})],
 which contain the Weinberg angle [Eq.(\ref{w123})].
 Equation (\ref{HIdF-mod}) shows that, loosely speaking, $H^{\rm gws}_{\rm int,I}$
 is given by the sum of all possible combinations of one FIO and an arbitrary number of FVFs.

\section{Reformulation of the QED interaction Hamiltonian}\label{sect-reformulate-Hint}
 
 In this section, the QED interaction Hamiltonian $H^{\rm QED}_{\rm int}$ 
 is reformulated in a form, which is formally similar to the rhs of Eq.(\ref{HIdF-mod}).
 Specifically, the ordinary formulation of $H^{\rm QED}_{\rm int}$ is recalled in Sec.\ref{sect-int-H-QED}.
 Then, a $V_\omega$-reformulation of $H^{\rm QED}_{\rm int}$ is given in Sec.\ref{sect-Hint-FIOs}. 
 Finally, in Sec.\ref{sect-geometic-FIO1}, 
 it is shown that $H^{\rm QED}_{\rm int}$ contains an FIO1 
 whose  form is formally similar to the rhs of Eq.(\ref{HFIO1-model}). 

\subsection{Ordinary formulation of QED}\label{sect-int-H-QED}

 In the ordinary formulation, $H^{\rm QED}_{\rm int}$ is written in terms of
 quantized electron and photon fields, denoted by
 $\psi_e(\bx)$ and $A_\mu(\bx)$, respectively.
 This formulation contains only fermions with positive $\varrho$
 and hence $\varrho$ is omitted in this section.
 The fields are expanded with annihilation operators for electron, positron, and photon,
 denoted by $b_r(\bp)$, $d_r(\bp)$, and $a_\lambda(\bk)$, respectively,
 together with the corresponding creation operators as their Hermitian conjugates. 
 These operators satisfy well known (anti-)commutation relations, such as
\begin{subequations}\label{bda-bdadag=0}
\begin{align}\label{b-bdag=0}
 &  \{ b_r^{\dag}(\bp) ,  b_s^{\dag}(\bq ) \} =0,
 \quad  \{ d_r^{\dag}(\bp) , d_s^{\dag}(\bq) \} =0,
 \\ &  \{ b_r^{\dag}(\bp) , d_s^{\dag}(\bq) \} =0,
  \quad \{ b_r(\bp) , d_s^{\dag}(\bq) \} =0,
  \\ &  [ a^\dag_\lambda(\bk), a^\dag_{\lambda'}(\bk') ] =0, 
\end{align}
\end{subequations}
 and
\begin{subequations} \label{b-bdag-etc}
\begin{gather}
%   \{ b_r(\bp) , d_s^{\dag}(\bq) \} =0,  \label{bd=0}
  \label{bb-dga}  \{ b_r(\bp) , b_s^{\dag}(\bq) \} = 
  p^0  \delta^{3}(\bp-\bq)  \delta_{rs},
 \\  \{ d_r(\bp) , d_s^{\dag}(\bq) \} =  
 p^0  \delta^{3}(\bp-\bq)  \delta_{rs},  \label{dd-dga}
% \\  [ a_\lambda(\bk) , a_{\lambda'}^{\dag}(\bk') ] 
%= g_{\lambda \lambda'} k^0  \delta^{3}(\bk-\bk'). \label{aa-dga}
\end{gather}
\end{subequations}
 where $p^0 = \sqrt{\bp^2 + m^2}$ with $m$ the electron mass.
 Explicitly, the electron and photon fields are written as follows, 
\footnote{Two remarks:
 (i)  $d\ww p$ is Lorentz invariant 
 and, consistent with this point, the anti-commutators for creation and annihilation operators
 contain a factor $p^0$ [see Eq.(\ref{b-bdag-etc})].
 In the literature, the factor $({1}/{p^0})$ in $d\ww p$ is sometimes written as $({1}/{\sqrt{p^0}})$.
 (ii) Some usually-used constant prefactors of the fields $\psi$ and $A_\mu$ are not written explicitly.
% In fact, these prefactors may be absorbed in the unwritten electronic change [see Eq.(\ref{HI-QED}) given below].
 }
\begin{subequations}\label{fields-QED}
\begin{gather}\label{psi-QED}
  \psi_e(\bx) = \int d\ww p \ b_r(\bp) U^{r}(\bp) e^{i\bp\cdot\bx} + d^{\dag}_r(\bp) V^{r}(\bp) e^{-i\bp\cdot\bx} ,
\\ \psi_e^\dag(\bx) = \int d\ww p \  b^{\dag}_r(\bp)
 U^{\dag r}(\bp) e^{-i\bp\cdot\bx} +d_r(\bp) V^{\dag r}(\bp)e^{i\bp\cdot\bx}, \label{psi-dag-QED}
\\ A_\mu(\bx) = \int d\ww k  a_{\lambda}(\bk) \varepsilon^{\lambda}_\mu(\bk) e^{i\bk\cdot\bx}
  + a^{\dag}_\lambda(\bk) \varepsilon^{\lambda*}_\mu(\bk) e^{-i\bk\cdot\bx}. \label{Amu-QED}
\end{gather}
\end{subequations}
% where  $U^{r}(\bp)$ and $V^{r}(\bp)$ are the ordinarily used stationary solutions of the free Dirac equation
% [see Eqs.(\ref{Up-uv-app}) and (\ref{Vp-uv-app}) in Appendix \ref{sect-Dirac-spinor}],
% $\varepsilon^{\lambda}_\mu(\bk)$ indicate polarization vectors for photons, and

 In terms of the above fields, the QED interaction Hamiltonian is written as
\begin{eqnarray} \label{HI-QED}
  H^{\rm QED}_{\rm int} =  \int d^3x : \psi_e^\dag(\bx) \gamma^0 \gamma^\mu \psi_e(\bx) A_\mu(\bx) :,
\end{eqnarray}
 where ``$: \ldots :$'' indicates normal product.
 Here, for brevity, a prefactor of $H^{\rm QED}_{\rm int}$ is not written explicitly
 \footnote{The unwritten prefactor contains the electronic charge
 and a term ${2m}$ due to the normalization
 condition employed for Dirac spinors [cf.~Eq.(\ref{|UV-IP>})].
 }.
%  (see footnote \ref{note-creation-operator-as-ket}).
% More exactly, in terms of kets of $|e_{\bp r}\ra$, $|\ov e_{\bp r}\ra$, and $|A_{\bk \lambda}\ra$,
% and bras of $\la e_{\bp r}|$, $\la \ov e_{\bp r}|$, and $\la A_{\bk \lambda}|$.
% Consistent with the anticommutators in Eq.(\ref{b-bdag-etc}),
% inner products are written as [cf.~Eq.(\ref{<f'-varrho|f>=delta})]
%\begin{subequations}\label{<e|e><d|d>}
%\begin{align}\label{<e|e>}
% &  \la  e_{\bq s}|  e_{\bp r}\ra =  p^0 \delta^3(\bp-\bq) \delta_{rs},
% \\  \label{<d|d>}
%  &  \la \ov e_{\bq s}| \ov e_{\bp r}\ra =   p^0 \delta^3(\bp-\bq) \delta_{rs}.
%\end{align}
%\end{subequations}
 Substituting Eq.(\ref{fields-QED}) into Eq.(\ref{HI-QED}),
 one gets eight terms,  denoted by $H_i$, with $i=1,\ldots, 8$
 and with a superscript ``QED'' omitted for brevity.
 They are written as follows, in which creation and annihilation operators
 have been replaced by their equivalent forms of kets and bras,
\begin{align}\label{H-sumHi}
 \ H^{\rm QED}_{\rm int} = \sum_{i=1}^8 H_i,
\end{align}
 where
\footnote{ A common prefactor $(2\pi)^3$ in $H_i$ is not written explicitly, for brevity. }
\begin{subequations}\label{Hi}
\begin{align}\label{}
 & H_1 = \int d\ww p d\ww q d\ww k \ |A_{\bk \lambda}\ra  \la \ov e_{\bq s} | \la e_{\bp r}|  h_1, \label{H1}
\\ & H_2 = \int d\ww p d\ww q d\ww k \ |e_{\bq s}\ra  |\ov e_{\bp r}\ra  \la A_{\bk \lambda}| h_2, \label{H2}
 \\ & H_3 = \int d\ww p d\ww q d\ww k \ |A_{\bk \lambda}\ra  |e_{\bq s}\ra   \la e_{\bp r}| h_3, \label{H3}
 \\ & H_4 = \int d\ww p d\ww q d\ww k \ |e_{\bq s}\ra  \la e_{\bp r}|  \la A_{\bk \lambda}|  h_4, \label{H4}
\\ & H_5 = -\int d\ww p d\ww q d\ww k \ |A_{\bk \lambda}\ra  |\ov e_{\bp r}\ra   \la \ov e_{\bq s}| h_5, \label{H5}
 \\ & H_6 = - \int d\ww p d\ww q d\ww k \ |\ov e_{\bp r}\ra  \la \ov e_{\bq s}|  \la A_{\bk \lambda}|   h_6, \label{H6}
\\ & H_7=\int d\ww p d\ww q d\ww k \ |A_{\bk \lambda}\ra  |e_{\bq s}\ra | \ov e_{\bp r}\ra h_7,\label{H7}
 \\ & H_8 = \int d\ww p d\ww q d\ww k \ \la \ov e_{\bq s}| \la e_{\bp r}|  \la A_{\bk \lambda}|   h_8.\label{H8}
\end{align}
\end{subequations}
 Here, $h_i$ of $i=1,\ldots 8$ indicate interaction amplitudes.
 Explicitly, $h_i$ of  odd $i$ are written as
\begin{subequations}\label{hi-odd}
\begin{gather}
  h_1 = V^{\dag s}(\bq) \gamma^0 \gamma^\mu  U^{r}(\bp) \varepsilon^{\lambda*}_\mu(\bk)
  \delta^3(\bp +\bq - \bk), \label{h1-qed}
 \\   h_3 = U^{\dag s}(\bq) \gamma^0 \gamma^\mu  U^{r}(\bp) \varepsilon^{\lambda*}_\mu(\bk)
 \delta(\bp -\bq - \bk),  \label{h3-qed}
 \\   h_5 = V^{\dag s}(\bq) \gamma^0 \gamma^\mu V^{r}(\bp) \varepsilon^{\lambda*}_\mu(\bk)
 \delta^3(\bq -\bp-\bk),  \label{h5-qed}
 \\   h_7 = U^{\dag s}(\bq) \gamma^0 \gamma^\mu V^{r}(\bp) \varepsilon^{\lambda*}_\mu(\bk)
 \delta^3(\bp +\bq+\bk);  \label{h7-qed}
\end{gather}
\end{subequations}
 and those of even $i$ are
\begin{subequations}\label{hi-even}
\begin{gather}
  h_2 =  U^{\dag s}(\bq) \gamma^0 \gamma^\mu  V^{r}(\bp) \varepsilon^{\lambda}_\mu(\bk)
  \delta(\bp +\bq - \bk), \label{h2-qed}
 \\  h_4 = U^{\dag s}(\bq) \gamma^0 \gamma^\mu  U^{r}(\bp) \varepsilon^{\lambda}_\mu(\bk)
 \delta^3(\bq -\bp - \bk),  \label{h4-qed}
 \\ h_6 = V^{\dag s}(\bq) \gamma^0 \gamma^\mu  V^{r}(\bp) \varepsilon^{\lambda}_\mu(\bk)
 \delta^3(\bp -\bq-\bk),  \label{h6-qed}
 \\  h_8  = V^{\dag s}(\bq) \gamma^0 \gamma^\mu U^{r}(\bp) \varepsilon^{\lambda}_\mu(\bk)
 \delta^3(\bp+\bq+\bk). \label{h8-qed}
\end{gather}
\end{subequations}
 Clearly, each $H_i$ corresponds to one basic Feynman diagram
 and $H_{i+1} = H_{i}^\dag$ for $i=1,3,5,7$.

% From Eq.(\ref{Hi}), it is seen clearly that each operator $H_i$ may be geometrically interpreted as 
% mapping the state space of some particle(s) (including the vacuum) to that of some other(s).
% In principle, one may construct $H_i$ based on this geometric property,
% of course, with the help of certain additional rules.
% Unfortunately, such rules turn out quite clumsy,  due to the differences among $H_i$. 
% In this next section, we are to show that the rules may be simplified due to 
% certain relationship among $H_i$. 

\subsection{Reformulation of the QED interaction Hamiltonian}\label{sect-Hint-FIOs}

 To write $H^{\rm QED}_{\rm int}$ in a form similar to the rhs of Eq.(\ref{HIdF-mod}),
 a crucial point is to employ 
 negative-$p^0$ solutions of the free Dirac equation,
 whose spinor parts are written as $U^r_-(\bp)$ and $V^r_-(\bp)$.
 (See Appendix \ref{app-neg-E-UV} for their explicit expressions.)
 In particular, we are to make use of the following relations 
 between positive-$p^0$ and negative-$p^0$ solutions 
 [see Eqs.(\ref{U-(-p)-V-app})-(\ref{V-(-p)-U-app}) of Appendix \ref{app-neg-E-UV}], i.e., 
\begin{subequations}\label{UV-(-p)}
\begin{align}
 & \label{U-(-p)-V}   V^r(\bp) = U^r_-(\ov\bp),
 \\ & \label{V-(-p)-U}  U^r(\bp) =  V^r_-(\ov\bp).
\end{align}
\end{subequations}
 Indicating the sign $\varrho $ of $p^0$ explicitly, 
 the Dirac spinors are written as $U_\varrho^r(\bp)$ and $V_\varrho^r(\bp)$,
 with $U_+^r(\bp) \equiv U^r(\bp)$ and $V_+^r(\bp) \equiv V^r(\bp)$.

 Making use of Eq.(\ref{UV-(-p)}), it is straightforward to write $h_{3,5,7}$ [Eq.(\ref{hi-odd})]
 in a form like that of $h_1$, i.e., 
\begin{subequations}\label{hi-odd-neg}
\begin{gather}
  h_3 = V_-^{\dag s}(\ov\bq) \gamma^0 \gamma^\mu  U^{r}(\bp) \varepsilon^{\lambda*}_\mu(\bk)
 \delta(\bp + \ov\bq - \bk),  \label{h3-qed-neg}
 \\   h_5 = V^{\dag s}(\bq) \gamma^0 \gamma^\mu U_-^{r}(\ov\bp) \varepsilon^{\lambda*}_\mu(\bk)
 \delta^3(\bq + \ov\bp-\bk),  \label{h5-qed-neg}
 \\   h_7 = V_-^{\dag s}(\ov\bq) \gamma^0 \gamma^\mu U_-^{r}(\ov\bp) \varepsilon^{\lambda*}_\mu(\bk)
 \delta^3(\ov\bp +\ov\bq-\bk).  \label{h7-qed-neg}
\end{gather}
\end{subequations}
 Similarly, $h_{4,6,8}$ are written as
\begin{subequations}\label{hi-even-neg}
\begin{gather}
  h_4 = U^{\dag s}(\bq) \gamma^0 \gamma^\mu  V_-^{r}(\ov\bp) \varepsilon^{\lambda}_\mu(\bk)
 \delta^3(\bq + \ov\bp - \bk),  \label{h4-qed-neg}
 \\ h_6 = U_-^{\dag s}(\ov\bq) \gamma^0 \gamma^\mu  V^{r}(\bp) \varepsilon^{\lambda}_\mu(\bk)
 \delta^3(\bp + \ov\bq-\bk),  \label{h6-qed-neg}
 \\  h_8  = U_-^{\dag s}(\ov\bq) \gamma^0 \gamma^\mu V_-^{r}(\ov\bp) \varepsilon^{\lambda}_\mu(\bk)
 \delta^3(\ov\bp+\ov\bq-\bk), \label{h8-qed-neg}
\end{gather}
\end{subequations}
 sharing the same formal form as $h_2$.
 Due to the above formal similarities, we introduce the following amplitudes,
\begin{subequations}\label{h12-FIO}
\begin{align}
  & h^{\rm FIO1}_{\varrho' \varrho} := V^{\dag r'}_{\varrho'}(\bp') \gamma^0 \gamma^\mu
  U^{r}_{\varrho}(\bp) \varepsilon^{\lambda*}_\mu(\bk)  \delta^3(\bp +\bp' - \bk), \label{h1-FIO}
  \\ &  h^{\rm FIO2}_{\varrho' \varrho} :=  U^{\dag r'}_{\varrho'}(\bp') \gamma^0 \gamma^\mu
  V^{r}_{\varrho}(\bp) \varepsilon^{\lambda}_\mu(\bk)  \delta^3(\bp +\bp' - \bk). \label{h2-FIO}
\end{align}
\end{subequations}
 (The reason of writing ``FIO'' in superscripts of the amplitudes will become clear later.)
 It is straightforward to check that
\begin{subequations}\label{h-1-8-hFIO}
\begin{align}\label{}
 &  h_1 = h^{\rm FIO1}_{++} \quad \& \quad  h_2 = h^{\rm FIO2}_{++},  \label{h-1-2-hFIO}
 \\ & h_3 = h^{\rm FIO1}_{-+} \quad \& \quad  h_4 = h^{\rm FIO2}_{+-}, 
 \\ &  h_5 = h^{\rm FIO1}_{+-} \quad \& \quad  h_6 = h^{\rm FIO2}_{-+}, 
 \\ &  h_7 = h^{\rm FIO1}_{--} \quad \& \quad  h_8 = h^{\rm FIO2}_{--}.
\end{align}
\end{subequations}
% under appropriate changes of labels and variables. 

 Then, we introduce two operators, which as to be shown in the next section turn out to satisfy
 the requirements discussed in Sec.\ref{sect-FIO} for FIO.
 We call them FIOs of QED, indicated as FIO1 of
 $H^{\rm FIO1}_{\varrho' \varrho}$ and FIO2 of $H^{\rm FIO2}_{\varrho' \varrho}$,
 being Hermitian conjugate of each other, 
\begin{subequations}\label{H-FIO}
\begin{align}\label{}
 & H^{\rm FIO1}_{\varrho' \varrho} := \int d\ww p d\ww p' d\ww k \ |A_{\bk \lambda}\ra  \la \ov e_{\bp' r' \varrho'} |
 \la e_{\bp r \varrho }|  h^{\rm FIO1}_{\varrho' \varrho}, \label{H-FIO1}
\\ & H^{\rm FIO2}_{\varrho' \varrho} := \int d\ww p d\ww p' d\ww k \ |e_{\bp' r' \varrho' }\ra
|\ov e_{\bp r \varrho }\ra  \la A_{\bk \lambda}| h^{\rm FIO2}_{\varrho' \varrho}. \label{H-FIO2}
\end{align}
\end{subequations}
 After some derivations (see Appendix \ref{app-proof-Vf-Hi}),
 one finds that $H_i$ are expressible in terms of the two FIOs of QED
 and the two superoperators of $V_f$ and $V_f^\dag$ for FVFs in Eqs.(\ref{VK1}) and (\ref{VK2}).
 The results are 
\begin{subequations}\label{H1-8-Vf}
\begin{align}
 & H_1 = H^{\rm FIO1}_{++}, \label{HFIP-H1}
 \\ & H_2 = H^{\rm FIO2}_{++},  \label{HFIP-H2}
 \\ &  H_3 =  \overbrace{ V_{\ov e } \left( H^{\rm FIO1}_{-+} \right) } , \label{HFIP-H3}
 \\ & H_4 =  \overbrace{  V_{\ov e }^\dag \left( H^{\rm FIO2}_{+-} \right)},  \label{HFIP-H4}
 \\ & H_5 = \overbrace{  V_{e } \left( H^{\rm FIO1}_{+-} \right) } , \label{HFIP-H5}
 \\ & H_6 =  \overbrace{ V_{e }^\dag \left( H^{\rm FIO2}_{-+} \right)}  , \label{HFIP-H6}
 \\ & H_7 =  \overbrace{ V_{\ov e } \left( V_{e } (H^{\rm FIO1}_{--}) \right) } , \label{HFIP-H7}
 \\ & H_8 =  \overbrace{ V_{\ov e }^\dag \left( V_{e }^\dag (H^{\rm FIO2}_{--}) \right)}.  \label{HFIP-H8}
\end{align}
\end{subequations}
 Based on the physical meaning of FVF discussed in Sec.\ref{sect-FVF},
 further interpretations to the above expressions of the operators $H_{3,\ldots,8}$ are obtainable, 
 which are discussed in Appendix \ref{app-Hi-interpret} in detail. 
 \footnote{With these interpretations, 
 topological equivalence of the eight simplest Feynman diagrams in QED turns out  to be
 a consequence of the expressions of $H_i$ in Eq.(\ref{H1-8-Vf}).}

 Below, we derive the following expression of $H^{\rm QED}_{\rm int}$, 
 \begin{align}\label{Hint-QED-FF}
  H^{\rm QED}_{\rm int} = \sum_{\omega, \varrho, \varrho'} V_\omega \left( 
    H^{{\rm FIO}1 }_{\varrho' \varrho} \right) + {\rm H.c.},
\end{align}
 where $V_\omega$ is defined in Eq.(\ref{F-lambda}) with $\omega = ( n^V_e, n^V_{\ov e})$. 
 Note that $H^{{\rm FIO}1 }_{\varrho' \varrho}$ 
 in Eq.(\ref{H-FIO1}) contains two fermionic bras;
 meanwhile, $V_f$ [Eq.(\ref{VK1})] contains two fermionic kets, 
 one with a negative $p^0$ and the other with a positive $p^0$.
 As a consequence, as discussed in Sec.\ref{sect-main-results-2},
 under the rule for negative-$p^0$ states, there are only four cases of $\omega$, 
 for which $V_\omega \left( H^{{\rm FIO}1 }_{\varrho' \varrho} \right)$
 does not definitely vanish.
 That is, (i) $n^V_e= n^V_{\ov e}=0$; (ii) $n^V_e=1 \ \& \ n^V_{\ov e}=0$; (iii) $n^V_e=0  \ \& \  n^V_{\ov e}=1$; and,
 (iv) $n^V_e= n^V_{\ov e}=1$.

 In the first case of $n^V_e= n^V_{\ov e}=0$, 
 $V_\omega (H^{{\rm FIO}1 }_{\varrho' \varrho}) = \overbrace{H^{{\rm FIO}1 }_{\varrho' \varrho}}$.
 According the rule for negative-$p^0$ states,
 there is only one nonvanishing $V_\omega(H^{{\rm FIO}1 }_{\varrho' \varrho})$, 
 namely $H^{{\rm FIO}1 }_{++}$, 
 which is equal to $H_1$ [Eq.(\ref{HFIP-H1})].

 In the second case of $n^V_e=1 \ \& \ n^V_{\ov e}=0$,
 $V_\omega (H^{{\rm FIO}1 }_{\varrho' \varrho}) = \overbrace{V_e(H^{{\rm FIO}1 }_{\varrho' \varrho})}$.
 Substituting Eq.(\ref{H-FIO1}) into Eq.(\ref{VK1}), one gets that
\begin{align}\notag
  \overbrace{V_{e} \left( H^{{\rm FIO}1 }_{\varrho' \varrho} \right)}
 &  = \int d\ww q d\ww p d\ww p' d\ww k |{\ov e}_{\ov \bq s +} \ra  |A_{\bk \lambda}\ra 
   h^{\rm FIO1}_{\varrho' \varrho}
 \\ & \times \overbrace{ \left( \la \ov e_{\bp' r' \varrho'} | \la e_{\bp r \varrho }| \right) |{e}_{\bq s-} \ra}. \label{Ve-FIO1}
\end{align}
 For the rhs of Eq.(\ref{Ve-FIO1}) to be nonvanishing, according to Eq.(\ref{<f'-varrho|f>=delta}),
 $\la e_{\bp r \varrho } |$ must form a nonvanishing scalar product with $|{e}_{\bq s-} \ra$,
 which requires that $\varrho =-$.
 Then, the rule for negative-$p^0$ states requires that $\varrho'=+$.
 As a consequence, there is only one nonvanishing result, 
 which corresponds to $(\varrho', \varrho)=(+,-)$, i.e., 
 $\overbrace{V_e (H^{\rm FIO1}_{+-})}$.
 One is ready to see that this gives $H_5$, as shown in Eq.(\ref{HFIP-H5}).

 The third case with $n^V_e=0  \ \& \  n^V_{\ov e}=1$ is 
 similar to the second case.
 Here, the only nonvanishing result is
 $\overbrace{ V_{\ov e} (H^{\rm FIO1}_{-+})}$, 
 which is equal to $H_3$ [Eq.(\ref{HFIP-H3})].

 In the fourth case of $n^V_e=  n^V_{\ov e}=1$,
 one considers the action of $V_{\ov e}[V_{e}(\cdots)]$.
 Applying  $V_{\ov e}$ to the left-hand side of Eq.(\ref{Ve-FIO1}) within the overbrace, 
 one finds that the only nonvanishing result corresponds to $(\varrho', \varrho)=(-,-)$,
 namely $\overbrace{V_{\ov e}[V_{e}(H^{\rm FIO1}_{--})]}$, 
 which is just $H_7$ [Eq.(\ref{HFIP-H7})].

 Summarizing the above discussions and with the Hermitian conjugate terms included, 
 one gets Eq.(\ref{Hint-QED-FF}) from Eq.(\ref{H-sumHi}).

\subsection{Concise expression of the FIO1 of QED}\label{sect-geometic-FIO1}

 In this section, it is shown that the operator $H^{\rm FIO1}_{\varrho' \varrho}$ 
 may be written in a form similar to the rhs of Eq.(\ref{HFIO1-model})
 \footnote{ As shown in Appendix \ref{app-FIO2},
 $H^{\rm FIO2}_{\varrho' \varrho}$ may be treated in a similar way.
 }.
 For this purpose, we are to first write the rhs of Eq.(\ref{h1-FIO})
 for $h^{\rm FIO1}_{\varrho' \varrho}$  in a ket-bra form.

 Making use of the expressions of the Dirac spinors $U^{r}_+(\bp)$ and $V^{r}_+(\bp)$ 
 [see Eqs.(\ref{Up-uv-app}) and (\ref{Vp-uv-app})]
 and noting the relations in Eq.(\ref{UV-(-p)}), it is ready to find the following unified expressions of 
 $U^r_\varrho (\bp)$ and $V^r_\varrho (\bp)$, 
\begin{subequations}\label{UVp-uv-main}
\begin{gather} 
\label{Up-uv-main}
 U^r_\varrho (\bp) = \frac{1}{\sqrt 2} \left( \begin{array}{c} u^{rA}(\varrho \, \bp) 
  \\ \varrho \, \ov v_{B'}^r(\varrho \, \bp) \end{array} \right),
 \\ \label{Vp-uv-main}
 V^r_\varrho (\bp) = \frac{1}{\sqrt 2} \left( \begin{array}{c} u^{rA}(\varrho \, \bp) 
  \\ -\varrho \, \ov v_{B'}^r(\varrho \, \bp) \end{array} \right),
\end{gather} 
\end{subequations}
 where $u^{r,A} \in \WW$ and $\ov v_{B'}^r \in \ov \WW$.
 In the chiral representation of  the $\gamma^\mu$-matrices,
 the product $\gamma^0 \gamma^\mu$ is written as 
 [cf.~Eq.(\ref{gamma0-gamma-mu-app})]
 \footnote{For discussions about the roles played by $\gamma^0$, see the paragraphs including 
 Eqs.(\ref{sig-0AB'-temp})-(\ref{gamma0-gamma-mu-app}) in Appendix \ref{sect-2-spinor}.}
\begin{gather}
 \gamma^0 \gamma^\mu =
 \left( \begin{array}{cc} \ov\sigma^{\mu}_{B'A} & 0 \\ 0 & \sigma^{\mu BA'} \end{array} \right),
  \label{gamma0-gamma-mu}
\end{gather}
 with $\sigma^{\mu AB'}$ the EvdW symbols 
 \cite{Penrose-book,Kim-group,CM-book,Corson,pra16-commu}.
 Making use of  Eqs.(\ref{UVp-uv-main})-(\ref{gamma0-gamma-mu}) 
 and the relation of $ \ov\sigma^{\mu}_{B'A}= \sigma^{\mu}_{AB'}$ [Eq.(\ref{sig-c})],
 it is straightforward to write the Dirac-spinor part of $h^{\rm FIO1}_{\varrho' \varrho}$ [Eq.(\ref{h1-FIO})] 
 in terms of Weyl spinors, getting that
\begin{align}\label{}
 \notag & V^{\dag r'}_{\varrho'}(\bp') \gamma^0 \gamma^\mu U^{r}_{\varrho}(\bp)
 \\ \notag  & = \ov u^{r' B' }(\varrho' \bp') \ov\sigma^{\mu}_{B'A} u^{rA}(\varrho \bp)
 - v^{r' }_{B}(\varrho' \bp') \sigma^{\mu BA'} \ov v^{r}_{A'}(\varrho \bp)
 \\ &  = \sigma^{\mu}_{AB'} \left[  \ov u^{r' B' }(\varrho' \bp')  u^{rA}(\varrho \bp) 
 -  v^{r' A}(\varrho' \bp') \ov v^{r B'}(\varrho \bp)  \right], \label{hFIO1-spinor-part}
\end{align}
 where Eq.(\ref{f-AB}) has been used and some spinor labels have been rearranged
 in the derivation of the second equality.
 Making use of Eq.(\ref{kappa-A-main}), one further writes
 \footnote{
 The rhs of Eq.(\ref{VGU-abs-hFIO1}) is computed by following the explicit order.
 For example, when computing $\big( \la\la \ov S^{B'}| \la\la  S^A| \big)  
 \big[ | u^{r}(\bp)\ra  |\ov u^{s}(\bq)\ra \big]$, 
 one first computes a scalar product in the space $\WW$, 
 namely $\la\la  S^A | u^{r}(\bp)\ra $, then, computes a product in the space $\ov \WW$,
 namely $ \la\la \ov S^{B'} |\ov u^{s}(\bq)\ra$.
 }
\begin{align}\label{}
\notag
  V^{\dag r'}_{\varrho'} & (\bp')  \gamma^0  \gamma^\mu U^{r}_{\varrho}(\bp)  
  =\sigma^{\mu}_{AB'} \la\la \ov S^{B'}| \la\la  S^A|
 \\ & \label{VGU-abs-hFIO1}
  \Big[ | u^{r}(\varrho\bp)\ra  |\ov u^{r'}(\varrho' \bp')\ra   - | v^{r'}(\varrho' \bp')\ra |\ov v^{r}(\varrho\bp) \ra  \Big].
\end{align}

 According to Eq.(\ref{<K|=<<ov-K|}), bras of polarization vectors are written as
\begin{align}\label{<varepsilon|}
 \la \varepsilon_\lambda(\bk)| := \la\la\ov\varepsilon_\lambda(\bk)|,
\end{align}
 where $\la\la \varepsilon_\lambda(\bk)|$ is
 the scalar-product-based bra of $|\varepsilon_\lambda(\bk)\ra$ [cf.~Eq.(\ref{<<K|})], i.e., 
\begin{align}\label{<<ep|=Tmu-epmu}
 \la\la \varepsilon^{\lambda}(\bk)| = \la\la T^\mu| \varepsilon^{\lambda}_{ \mu}(\bk) .
\end{align}
 From Eqs.(\ref{K-mu-main}) and (\ref{<varepsilon|}), one gets that
\begin{align}\label{vepsi-<>}
 \varepsilon^{\lambda*}_\mu(\bk) = \la \varepsilon^{\lambda}(\bk)|T_\mu\ra.
\end{align}
 The vectors $|\varepsilon^{\lambda}(\bk)\ra$ of $\lambda =0,1,2,3$ constitute
 a basis in the space $\VV$, too, like the basis $|T_\mu\ra$.
 Making use of Eqs.(\ref{Tmu-Tnu-main}) and (\ref{<varepsilon|}) and
 noting Eq.(\ref{ovT=T}), it is ready to find that such a basis should satisfy
\begin{equation}\label{<ov-ve|ve>}
 \la  \varepsilon_{\lambda}(\bk)| \varepsilon_{\lambda'}(\bk)\ra = g_{\lambda\lambda'}.
\end{equation}
% where 
%\begin{align}\label{whg=-g}
%  \wh g_{\lambda\lambda'} :=-g_{\lambda\lambda'}.
%\end{align}
% with $[g_{\lambda\lambda'}]$  the Minkovski metric.

 It proves convenient to
 introduce a symbol $\overrightarrow{\cs}$, defined by the following action
 on Dirac spinors, that is, 
\begin{gather}\label{G-dot-right-1}
 \overrightarrow{\cs}  |X   \ov W \ra \equiv |\chi \ra |\ov w\ra  +  |\ov\kappa \ra  |z \ra ,
\end{gather}
 where 
 $|X\ra$ and $|W\ra$ are two arbitrary Dirac spinors written as follows in terms of Weyl spinors, 
\begin{gather}\label{WX}
 |X\ra = \left( \begin{array}{c} |\chi \ra \\ |\ov \kappa \ra \end{array} \right) ,\quad
 |W\ra = \left( \begin{array}{c} |w\ra  \\ |\ov z \ra \end{array} \right).
\end{gather}
% with $|\chi \ra \ \& \ |w\ra \in \WW$ and $|\ov\kappa \ra \ \& \ |\ov z \ra  \in \ov\WW$.
 Note that $|\ov W\ra = \left( \begin{array}{c} |\ov w\ra  \\ |z \ra \end{array} \right)$.
 It is ready to see that the rhs of Eq.(\ref{G-dot-right-1}) is equal to the sum of the two layers 
 of $\overrightarrow{D}|X   \ov W \ra$  [cf.~Eq.(\ref{D})],
 i.e., $ \overrightarrow{\cs}  |X   \ov W \ra  = |\LL^X_1 \ra |\LL^{\ov W}_1 \ra 
  + |\LL^{X}_2 \ra |\LL^{\ov W}_2 \ra$.

% From Eq.(\ref{G-dot-right-1}), one sees that $\overrightarrow{\cs}$
% is in fact  a \emph{spin-mixing operator}, which
% mixes Weyl spinors in $\WW$ and those in $\ov\WW$ by the operation of direct product.
 Then, making use of Eqs.(\ref{VGU-abs-hFIO1})-(\ref{vepsi-<>}) and (\ref{SAB-commu-main}), 
 it is straightforward to check that  $h^{\rm FIO1}_{\varrho' \varrho}$ in Eq.(\ref{h1-FIO}) 
 has the following expression,
\begin{align}\label{}
  h^{\rm FIO1}_{\varrho' \varrho} =  \la  \varepsilon^{\lambda}(\bk)| \sigma 
  \overrightarrow{\cs} | U^{r}_{\varrho}(\bp) \ov U^{r'}_{\varrho'}(\bp') \ra
 \delta^3(\bp +\bp' - \bk). \label{hFIO1-ketbra}
\end{align}
 Here, the Dirac spinors $U^r_\varrho(\bp)$ in Eq.(\ref{UVp-uv-main}) are written in their ket form,
%\begin{subequations}\label{|UV>}
\begin{align}\label{Ur-bp}
 & |U^r_\varrho(\bp)\ra = \frac{1}{\sqrt 2} \left( \begin{array}{c} |u^r(\varrho \, \bp) \ra 
 \\ \varrho \, | \ov v^r(\varrho \, \bp)\ra \end{array} \right).
% \\ & |V^r_\varrho(\bp)\ra = \frac{1}{\sqrt 2} \left( \begin{array}{c} |u^r(\varrho \, \bp) \ra 
% \\ - \varrho \, | \ov v^r(\varrho \, \bp)\ra \end{array} \right).
\end{align}
%\end{subequations}

 It is of interest to note that, in the expression of the amplitude $h^{\rm FIO1}_{\varrho' \varrho}$
 in Eq.(\ref{hFIO1-ketbra}), 
 the positron-spin state is represented by the spinor $|\ov U^{r'}_{\varrho'}(\bp') \ra$.
 This implies that, when the interaction amplitude is expressed in Eq.(\ref{hFIO1-ketbra}),
 one may use $|\ov U^{r}_{\varrho}(\bp) \ra$ \emph{to represent positron-spin states}.
 [This is the reason of writing kets for single-particle states of positron 
 in the form of Eq.(\ref{|d>}).] 
 Moreover, employing this spinor description for positron-spin states has an additional merit:
 It possesses the following inner product, 
\begin{align}\label{ovU-delta-rs}
 \ov{U}_{r\varrho}^{\dag}(\bp)\gamma^0  \ov U_{s\varrho}(\bp) 
 = \varrho \delta_{rs},
\end{align}
 which is in consistency with the inner product of single-positron states as required in 
 Eq.(\ref{<f'-varrho|f>=delta}).
 In contrast, the ordinary spinor description has the property of 
 ${V}_{r\varrho}^{\dag}(\bp)\gamma^0 V_{s\varrho}(\bp) = - \varrho \delta_{rs}$,
 showing a difference in the sign.

 Substituting Eq.(\ref{hFIO1-ketbra}) into Eq.(\ref{H-FIO1})  one finds that
\begin{align}\label{}\notag
  H^{\rm FIO1}_{\varrho' \varrho}
  & = \int d\ww p d\ww p' d\ww k \   \delta^3(\bp +\bp' - \bk)
 |A_{\bk \lambda}\ra   \la  \varepsilon^{\lambda}(\bk)|
  \\ & \times  \sigma \overrightarrow{\cs} | U^{r}_{\varrho}(\bp)  \ov U^{r'}_{\varrho'}(\bp') \ra
    \la \ov e_{\bp' r' \varrho'} |  \la e_{\bp r \varrho }|.
 \label{H-FIO1-geom-1}
\end{align}
 Then, making use of the identity operators in Eqs.(\ref{I-ff'}) and (\ref{IB-main}), 
 one gets a concise expression of $H^{\rm FIO1}_{\varrho' \varrho}$, 
\begin{equation}\label{H1-G}
 H^{\rm FIO1}_{\varrho' \varrho} =  I_A \,  \G^{\rm FIO1} \, I_{e\varrho, \ov e \varrho'},
\end{equation}
 showing formal similarity to that of Eq.(\ref{HFIO1-model}).
 Here, $\G^{\rm FIO1}$ is an operator that maps the space 
 $\E_{e\varrho} \otimes \E_{\ov e \varrho'}$ to the space $I_A$.
 More exactly, it is defined by the following relation,
 which connects kets in $I_{e\varrho, \ov e \varrho'}$
 to bras in $I_A$, that is, 
\begin{align}\label{G1-definition-<|>}
 \la A_{\bk}^{ \lambda}| \G^{\rm FIO1} | e_{\bp \varrho}^r\ra | \ov e_{\bp' \varrho'}^{r'}\ra 
 \equiv G^{\rm mom} G_1^{\rm spin},
\end{align}
 where $  \label{<a|}  \la A_{\bk \lambda}| =  \la\varepsilon_\lambda(\bk)| \la \bk|$,
 $G^{\rm mom}$ is defined in Eq.(\ref{G-mom}), and 
\begin{align}\label{}
 &  G_1^{\rm spin} =  \la  \varepsilon^{\lambda}(\bk)|
   \sigma \overrightarrow{\cs} | U^{r}_{\varrho}(\bp)  \ov U^{r'}_{\varrho'}(\bp') \ra.\label{G1-spin}
\end{align}

 \section{Reformulation of $H^{\rm gws}_{\rm int,I} $}
 \label{sect-EvdW-electroweak}
 
 This section generalizes the above discussions on $H^{\rm QED}_{\rm int}$
 to the GWS lepton-vector-boson interaction Hamiltonian $H^{\rm gws}_{\rm int,I}$.
 Specifically, the ordinary formulation of $H^{\rm gws}_{\rm int,I}$ is recalled in Sec.\ref{sect-recall-GWS}.
 Its reformulation in terms of FIOs and FVFs is given in Sec.\ref{sect-GWS-EvdW-formu}.
 Finally, validity of Eq.(\ref{HIdF-mod}) is shown in Sec.\ref{sect-GWS-two-layer}.
 
\subsection{Ordinary formulation of $H^{\rm gws}_{\rm int,I} $}
\label{sect-recall-GWS}

 In the ordinary formulation, $H^{\rm gws}_{\rm int,I}$ contains six terms
 which are to be labelled by $k=1,2,3,4, 5L, 5R$, with $k=1$ assigned to QED; that is, 
\begin{align} 
  H^{\rm gws}_{\rm int,I} & = \sum_{k}  H^{\text{gws}}_{{\rm int,I},k},
 \label{Hint-gws-I}
\end{align}
 with $H^{\text{gws}}_{{\rm int,I},1} = H^{\rm QED}_{\rm int}$. 
 Densities of $H^{\text{gws}}_{{\rm int,I},k}$, denoted by $\HH^{\text{gws}}_{{\rm int,I},k}$ with
 $H^{\text{gws}}_{{\rm int,I},k} = \int d^3x \HH^{\text{gws}}_{{\rm int,I},k} $,
 are written as \cite{Peskin-p705}
\begin{subequations}\label{HH-gws-k}
\begin{align}
 & \HH^{\text{gws}}_{\rm int,I,1} =   : \xi_1  A_\mu \psi^\dag_e 
 \gamma^0 \gamma^\mu \psi_e :,
 \\ & \HH^{\text{gws}}_{\rm int,I,2} = :  \xi_2 W^-_\mu \psi^\dag_{eL} \gamma^0 \gamma^\mu \psi_{\nu L}: ,
 \\ & \HH^{\text{gws}}_{\rm int,I,3} = :  \xi_3 W^+_\mu \psi^\dag_{\nu L} \gamma^0 \gamma^\mu \psi_{eL}:,
 \\  & \HH^{\text{gws}}_{\rm int,I,4} =  : \xi_4 Z^0_\mu \psi^\dag_{\nu L} \gamma^0 \gamma^\mu  \psi_{\nu L}: ,
 \\  & \HH^{\text{gws}}_{\rm int,I,5L} = :  \xi_{5L} Z^0_\mu \psi^\dag_{eL} \gamma^0 \gamma^\mu  \psi_{eL}:, 
 \\  &  \HH^{\text{gws}}_{\rm int,I,5R} = :  \xi_{5R} Z^0_\mu \psi^\dag_{eR} \gamma^0 \gamma^\mu  \psi_{eR}:.
\end{align}
\end{subequations}
 Here, $\psi_{\nu}$ represents the neutrino field,
 which is a Dirac field like the 
 electron field $\psi_e$ in Eq.(\ref{psi-QED});
 the subscripts ``$L$'' and ``$R$'' indicate the LH and RH parts of a field, respectively;
 $W^\pm_\mu$ and $Z^0_\mu$ represent fields for the $W^\pm$ and $Z^0$ bosons, respectively, 
 all of which are four-component vector fields like
 the photon field $A_\mu$ in Eq.(\ref{Amu-QED}).
 (As is known, $W^\pm_\mu$ contain annihilation operators for the $W^\pm$ bosons
 and creation operators for the $W^\mp$ bosons.)
 The prefactors $\xi_k$ include two coupling constants of $e_0$ and $g$, 
 which are connected by the Weinberg angle $\theta_{W}$;
% , $e_0 = g \sin \theta_{W}$. 
 \begin{subequations}\label{xi-k}
  \begin{align}
   & \xi_1 = e_0 = g \sin \theta_{W},
   \\ & \xi_2 = \xi_3 =  - \frac{g}{\sqrt 2},
   \\ & \xi_4 =  - \frac{g}{2\cos \theta_{W}},
   \\ & \xi_{5L} =  - \frac{g}{\cos \theta_{W}} (-\frac 12 + \sin^2 \theta_{W}),
   \\ & \xi_{5R} =  - \frac{g}{\cos \theta_{W}}  \sin^2 \theta_{W}.
  \end{align}
  \end{subequations}

 All of the operators $\HH^{\text{gws}}_{{\rm int,I},k}$ share a 
 common formal feature,
 each being built from two fermionic fields and one bosonic field, connected by 
 the $\gamma^\mu$-matrices; i.e.,
\begin{align}\label{HHgws-k-concise}
  \HH^{\text{gws}}_{{\rm int,I},k} = : \xi_k  B_\mu \psi_{f' c}^\dag 
  \gamma^0 \gamma^\mu \psi_{f c} : ,
 \end{align}
 where 
\begin{align}\label{B-AWZ}
 B_\mu = \left\{ 
 \begin{array}{cl}
   A_\mu,  &  \text{for $k=1$},
 \\  W^-_\mu,  &  \text{for $k=2$},
 \\  W^+_\mu,  &  \text{for $k=3$},
 \\  Z^0_\mu,  &  \text{for $k=4,5L,5R$},
 \end{array} \right. 
\end{align}
 and $c$ indicates a chiral property of the field $\psi$ defined by
\begin{align}\label{eta}
 c := \left\{ 
 \begin{array}{ll}
 {T},  &  \text{for the total field},
\\  {L},  &  \text{for the LH part},
 \\ {R},  &  \text{for the RH part}.
 \end{array}
 \right. 
\end{align}
 Explicitly, $(f' c, f c)$ has the following dependence on $k$,
\begin{align}\label{F12-enu}
  (f' c, f c)= \left\{ 
\begin{array}{cl}
   (eT, eT)  &  \text{for $k=1$},
 \\   (eL, \nu L)  &  \text{for $k=2$},
 \\   (\nu L, eL)  &  \text{for $k=3$},
 \\   (\nu L, \nu L)  &  \text{for $k=4$},
 \\   (eL(R), e L(R))  &  \text{for $k=5L(R)$}.
\end{array}
  \right. 
\end{align}

 \subsection{Reformulation of $H^{\rm gws}_{{\rm int,I}}$
 with FIO1 and FVF}\label{sect-GWS-EvdW-formu}

 In this section, by generalizing discussions given in Sec.\ref{sect-reformulate-Hint} for QED,
 the interaction Hamiltonian $H^{\rm gws}_{{\rm int,I}}$ is written 
 in a form, which is similar to the rhs of Eq.(\ref{HIdF-mod}).

 For each fixed $k$, $H^{\text{gws}}_{{\rm int,I},k}$ may be divided into eight parts, 
 denoted by $H_{k,i}$, in a way similar to the rhs of Eq.(\ref{H-sumHi}) for $H^{\rm QED}_{\rm int}$
 (as $k=1$), that is,
 \begin{align}\label{Hki}
  H^{\text{gws}}_{{\rm int,I},k} = \sum_{i=1}^8 H_{k,i},
\end{align}
 where $H_{k,i}$ have forms similar to those of $H_i$ in Eq.(\ref{Hi}), respectively. 
 Following a procedure, which is almost the same as 
 that used previously leading to Eq.(\ref{H1-8-Vf}) for QED, 
 one finds that $H_{k,i}$ of $k\ne 1$ may be written in forms similar to the rhs of Eq.(\ref{H1-8-Vf}),
 but, with corresponding changes of particle species.

 In the above generalization, an FIO1 needs to be introduced
 for each $H^{\text{gws}}_{{\rm int,I},k}$, 
 which is to be called the $k$-th electroweak FIO1 and indicated as 
 $H^{{\rm FIO}1 }_{\varrho' \varrho,k}$.
 Note that $H^{{\rm FIO}1 }_{\varrho' \varrho,1}
 = H^{{\rm FIO}1 }_{\varrho' \varrho}$ for QED in Eq.(\ref{H-FIO1}).
 More exactly, the FIO1s are written as
 \footnote{
 Clearly, $H^{{\rm FIO}1 }_{\varrho' \varrho,k}$ of 
 $k=2$ describes $\ov e \nu \to W^+$ 
 (the change of a positron and a neutrino to a $W^+$ boson),
 that of $k=3$ for $e \ov\nu \to W^-$,
% the change of an electron and an antineutrino to a $W^-$ boson,
 that of $k=4$ for $\nu \ov\nu \to Z^0$,
% the change of a neutrino and an antineutrino to a $Z^0$ boson,
 and that of $k=5L$ and $5R$ for $e \ov e \to Z^0$.
 %the change of an electron and a positron to a $Z^0$ boson.
}
 \begin{align}\label{H-FIO1-k}
  & H^{{\rm FIO}1}_{\varrho' \varrho,k} := \int d\ww p d\ww p' d\ww k \ 
  |B_{\bk \lambda}\ra  \la \ov f'_{\bp' r' \varrho'} |
  \la f_{\bp r \varrho }|  h^{{\rm FIO}1}_{\varrho' \varrho,k}, 
 \end{align}
 where 
\begin{align}\label{}
  & h^{\rm FIO1}_{\varrho' \varrho,k} = \xi_k V^{\dag r'}_{\varrho' c'}(\bp') 
  \gamma^0 \gamma^\mu
  U^{r}_{\varrho c}(\bp) \varepsilon^{\lambda*}_\mu(\bk)  \delta^3(\bp +\bp' - \bk). \label{h1-FIO-k}
\end{align}
 Here, dependence of $U^{r}_{\varrho c}$ on $c$ [as a function of $k$
 as shown in Eq.(\ref{F12-enu})] is given by
\begin{align}\label{hrr-k-F}
  U^{r}_{\varrho c}= \left\{ 
  \begin{array}{ll}
    U^{r}_{\varrho},  &  \text{for $c=T$},
  \\  \frac{1-\gamma^5}2 U^{r}_{\varrho},  &  \text{for $c=L$},
  \\  \frac{1+\gamma^5}2 U^{r}_{\varrho},  &  \text{for $c=R$},
  \end{array}
  \right. 
 \end{align}
 and similar for $V^{ r'}_{\varrho' c'}$. 
 Note that the expressions of $H_{k,i}$ thus obtained, which take the same formal
 forms as the rhs of Eq.(\ref{H1-8-Vf}), respectively, use the superoperator $V_f$ 
 of $f=\nu, \ov \nu$ for FVF neutrino-antineutrino pairs,
 which is also defined by Eq.(\ref{VK1}).

 Then, following a procedure similar to that from Eq.(\ref{H1-8-Vf}) to the expression of
 $H^{\rm QED}_{\rm int}$ in Eq.(\ref{Hint-QED-FF}),
 it is not difficult to check that $H^{\text{gws}}_{{\rm int,I},k}$,
 which describes the particle change of $(f,f') \to B$, has the following expression, 
 \begin{align}\label{Hint-FIO-k}
  H^{\text{gws}}_{{\rm int,I},k} = \sum_{\varrho', \varrho,\omega} V_\omega \left(  
    H^{{\rm FIO}1 }_{\varrho' \varrho,k}  \right) + {\rm H.c.},
\end{align}
 with $\omega = ( n^V_f, n^V_{f'})$.

 Like the expression of the QED FIO1 in Eq.(\ref{H1-G}),
 the $k$-th FIO1 of $H^{{\rm FIO}1 }_{\varrho' \varrho,k}$ ($k\ne 1$) may be written 
 in a concise form. 
 In fact, following a procedure similar to that of deriving the expression 
 of $h^{\rm FIO1}_{\varrho' \varrho}$ in Eq.(\ref{hFIO1-ketbra}) from Eq.(\ref{h1-FIO}), 
 one gets the following expression for the $k$-th amplitude 
 $h^{\rm FIO1}_{\varrho' \varrho,k}$, i.e., 
\begin{align}\label{}
  &  h^{\rm FIO1}_{\varrho' \varrho,k} = 
  \xi_k \la  \varepsilon^{\lambda}(\bk)|  \sigma \overrightarrow{\cs_{c}}
   | U^{r}_{\varrho}(\bp) \ov U^{r'}_{\varrho'}(\bp') \ra
  \delta^3(\bp +\bp' - \bk), \label{h1-kb-varrho-k}
 \end{align}
 where $\overrightarrow{\cs_{ c}}$ is a generalization of 
 $\overrightarrow{\cs}$ in Eq.(\ref{G-dot-right-1}), 
\begin{gather}\label{G-dot-right-1-eta}
  \overrightarrow{\cs_{ c}}  |X   \ov W \ra  \equiv  \left\{ 
\begin{array}{ll}
  |\chi \ra |\ov w\ra  + |\ov\kappa \ra  |z \ra  ,  &  \text{for $c = T $},
 \\ |\chi \ra |\ov w\ra  ,  &  \text{for $c = L $},
 \\     |\ov\kappa \ra |z \ra ,  &  \text{for $c = R $}.
\end{array}
  \right. 
\end{gather}
 Substituting Eq.(\ref{h1-kb-varrho-k}) into Eq.(\ref{H-FIO1-k}),
 it is straightforward to find that the $k$-th FIO1 is written in a form
 similar to Eq.(\ref{H1-G}) for QED, but, with appropriate change of 
 the identity operators and of the space-connection operator ($\G$).

 More exactly, the above discussed expression of  the $k$-th FIO1 $H^{{\rm FIO}1 }_{\varrho' \varrho,k} $ is 
 \begin{equation}\label{H1-G-gws}
  H^{{\rm FIO}1 }_{\varrho' \varrho,k} 
  =  I_B \, \G^{\rm FIO1}_{{\rm ew},k} \, I_{f\varrho, f' \varrho'} \quad \forall k, 
 \end{equation}
 where $\G^{\rm FIO1}_{{\rm ew},k}$ is defined  as follows,
 in a way similar to $\G^{\rm FIO1}$ in Eq.(\ref{G1-definition-<|>}) for $k=1$, i.e.,  
\begin{align}\label{G1-definition-<|>-k}
 &  \la B_{\bk}^{ \lambda}| \G^{\rm FIO1}_{{\rm ew},k} 
 | f_{\bp \varrho}^r\ra | {f'}_{\bp' \varrho'}^{r'}\ra 
  \equiv G^{\rm mom} G_{{\rm ew},k}^{\rm spin},
 \\ &  G_{{\rm ew},k}^{\rm spin} := \xi_k  \la  \varepsilon^{\lambda}(\bk)|
 \sigma \overrightarrow{\cs_{ c}} | U^{r}_{\varrho}(\bp)  
 \ov U^{r'}_{\varrho'}(\bp') \ra. \label{G1-spin-k}
 \end{align}
 One remark:
 In the above expression of $H^{{\rm FIO}1 }_{\varrho' \varrho,k}$, 
 single-particle states of the antineutrino 
 are described in the same way as positron in Eq.(\ref{|d>}).
 Making use of Eqs.(\ref{Ur-bp}) and (\ref{G-dot-right-1-eta}), 
 with the dependence of $c$ on $k$ given in Eq.(\ref{F12-enu}),
 it is straightforward to get the following expressions of $G_{{\rm ew},k}^{\rm spin}$
 from Eq.(\ref{G1-spin-k}), i.e., 
\begin{subequations}\label{G-ew-k}
 \begin{align}\label{}
 & G_{{\rm ew},1}^{\rm spin} = \frac{\xi_1}{2} (G_L + G_R),
 \\ & G_{{\rm ew},k}^{\rm spin} = \frac{\xi_k}{2} G_L \quad  \text{for $k=2,3,4,5L$},
 \\ & G_{{\rm ew},5R}^{\rm spin} =  \frac{\xi_{5R}}{2} G_R,
 \end{align}
\end{subequations}
 where $\xi_k$ are given in Eq.(\ref{xi-k}) and 
\begin{subequations}\label{GL-GR}
 \begin{align}\label{}
 &  G_L = \la \varepsilon^\lambda(\bk)| \sigma|u^r(\varrho \, \bp)\ra |\ov u^{r'}(\varrho' \, \bp')\ra,
 \\ &  G_R =  \varrho \varrho' \, \la \varepsilon^\lambda(\bk)| \sigma 
 |\ov v^{r}(\varrho \, \bp)\ra |v^{r'}(\varrho' \, \bp')\ra.
 \end{align}
\end{subequations}

\subsection{$H^{\rm gws}_{\rm int,I} $ expressed with two-layer spinors}
\label{sect-GWS-two-layer}

 This section shows validity of the concise expression of 
 $H^{\rm gws}_{{\rm int,I}}$ in Eq.(\ref{HIdF-mod}).
 This is to be done with appropriate two-layer descriptions for spin states of the particles,
 the spinor-space structures of which have been given in Eqs.(\ref{config-WovW})-(\ref{S-fB-gen}).
 In fact, spin states of electron and positron are already written
 in two-layer forms in previous sections, i.e., [see Eqs.(\ref{single-p-s}) and (\ref{Ur-bp})]
 \begin{subequations}\label{S-eove}
  \begin{align}\label{S-e}
   |\cs_{e \varrho}^{r }(\bp)\ra = |U^r_{ \varrho}(\bp)\ra = \frac{1}{\sqrt 2}
   \left( \begin{array}{c} |u^r(\varrho \, \bp)\ra \\ \varrho \, |\ov v^r(\varrho \, \bp)\ra  \end{array} \right),
   \\ |\cs_{\ov e \varrho }^{r }(\bp)\ra = |{\ov U}^r_{ \varrho}(\bp)\ra = \frac{1}{\sqrt 2}
   \left( \begin{array}{c} |\ov u^r(\varrho \, \bp)\ra \\ \varrho \, |v^r(\varrho \, \bp)\ra \end{array} \right).
   \label{S-ove}
   \end{align}
\end{subequations}

 Regarding spin states of neutrino, since only the LH part of the Dirac spinor $|U^r_{ \varrho}\ra$
 is used in the computation of the GWS interaction amplitudes, 
 there is in fact no need of using the RH part of the Dirac spinor. 
 This is one major reason of assuming the configuration in Eq.(\ref{config-WovW-nu}).
 As a consequence, both layers of the neutrino spin state should be given by the LH part of the Dirac spinor,
 namely, by $|u^r(\varrho \, \bp)\ra$ with certain layer prefactors.
 We found the following expressions useful for reproducing the GWS interaction amplitudes, 
 \begin{subequations}\label{S-nu-ovnu}
  \begin{align}
  \label{S-nu} |\cs_{\nu \varrho}^{r }(\bp)\ra = \frac{i}{\sqrt 2}
    \left( \begin{array}{c} |u^r (\varrho \, \bp)\ra \\ i |u^r(\varrho \, \bp)\ra  \end{array} \right),
    \\ |\cs_{\ov \nu \varrho}^{r }(\bp)\ra = \frac{i}{\sqrt 2}
    \left( \begin{array}{c} |\ov u^r(\varrho \, \bp)\ra 
    \\ i |\ov u^r(\varrho \, \bp)\ra \end{array} \right). \label{S-ovnu}
   \end{align}
\end{subequations}
 Explicit expressions of the bras $\la \cs_{f \varrho}^{r }(\bp)|$ ($f= \nu, \ov \nu$)
 are given in Appendix \ref{app-hatbra-spinors} [Eq.(\ref{S-nu-ovnu-hbra})] and,
 making use of them, it is easy to check validity of Eq.(\ref{<f'-varrho|f>=delta}) 
 for the leptonic states,
 with the terms of $\delta_{ff'}$ and $ \delta_{\varrho \varrho'}$ imposed as assumptions.

 In the computation of the amplitudes $G_{\eta}^{\rm spin}$ in Eq.(\ref{G1-spin-eta}),
 only bras of spin states of vector bosons are used. 
 In consistency with the configurations in Eq.(\ref{config-W-otimes-ovW}),
 we consider the following bras for $W^\pm$ bosons, 
\begin{subequations}\label{S-W}
\begin{align}\label{S-W+}
  &  \la \cs_{W^+}^\lambda(\bk) | = i \N_{W^+}\left( \la \varepsilon^\lambda(\bk) |,
    \la s | \right),
   \\ & \la \cs_{W^-}^\lambda(\bk) | = i \N_{W^-} \left(  \la \varepsilon^\lambda(\bk) |,
   \la \ov s |  \right), \label{S-W-}
 \end{align}
 \end{subequations}
 where $\la s|$ and $\la \ov s|$ indicate spinors in the spaces dual to $\C$ and $\ov \C$, respectively.
 Meanwhile, we consider the following bras for photon and $Z^0$ boson [cf.~Eq.(\ref{photon-Z0-VV})], 
\begin{subequations}\label{S-ZA}
\begin{align}
    \label{S-A} &  \la \cs_{A}^\lambda(\bk) | = \frac{\N_A}{\sqrt 2} 
     \left( w_1 \la \varepsilon^\lambda(\bk) | ,
    w_2 \la \varepsilon^\lambda(\bk) |  \right),
  \\  \label{S-Z}
    &    \la \cs_{Z^0}^\lambda(\bk) | =  \frac{\N_{Z^0}}{\sqrt 2} 
    \left( w_3 \la \varepsilon^\lambda(\bk)|  ,
  - w_4 \la \varepsilon^\lambda(\bk)| \ \right).
 \end{align}
 \end{subequations}
 Here, $w_{1,2,3,4}$ are parameters to be determined
 and $\N_B$ ($B=A,Z^0,W^\pm$) are normalization factors
 which have been used in Eq.(\ref{HIdF-mod}).

 One may note the following properties related to $\sigma$, 
 which are obvious according to its definition in Eq.(\ref{sigma}), i.e., 
 \begin{subequations}\label{sigma-act-0}
  \begin{align}
   & \sigma {|\LL^{f}_i \ra |\LL^{f'}_i \ra } =0,  \, \text{if ${|\LL^{f}_i \ra |\LL^{f'}_i \ra } 
   \in \WW \otimes \WW$
    or $\ov\WW \otimes \ov\WW$},
  \\ &  \la \LL^B_i| \sigma =0, \qquad  \text{if $\la \LL^B_i| \in \C$
    or $\ov \C$}.
  \end{align}
 \end{subequations}
 Due to the above properties, together with the fact that $P_\eta$ contributes $\sigma$ when nonvanishing
 [see Eq.(\ref{P-eta})], the spinors $\la s|$ and $\la \ov s|$ in fact 
 give no contribution to the amplitudes $G_{\eta}^{\rm spin}$.
% in other words, they give no substantial contribution to the theory. 
 For this reason, we neglect them when computing 
 normalization factors of the spin states of $W^\pm$ bosons. 
 It is, then, straightforward to compute the renormalization factors $\N_B$, getting
\begin{subequations}\label{N-A-Z0-W}
 \begin{align} \label{NA}
  & \N_A = \frac{\sqrt 2 }{\sqrt{|w_1|^2 + |w_2|^2}}, 
  \\ &  \N_{Z^0} = \frac{\sqrt 2}{\sqrt{|w_3|^2 + |w_4|^2}}, \label{NZ0}
  \\ & \N_{W^+} = \N_{W^-} =1. \label{NW}
 \end{align}
 \end{subequations}
 With the above explicit expressions, 
 one easily checks Eq.(\ref{<B|B>}) for scalar products of the bosonic states.

 There are totally $4 \times 4 \times 4 = 64$ combinations for $\eta = (f,f',B)$.
 But, due to Eqs.(\ref{P-eta}) and (\ref{sigma-act-0}),
 the amplitude $G_{\eta}^{\rm spin}$ does not definitely vanish only for six of them,
 which are to be indicated by $\eta_l$ of $l =1,\ldots, 6$.
 Explicitly, the six $\eta_l$ are as follows,
  \begin{align}\label{eta-1-6}
 &  \eta_l = \left\{ 
  \begin{array}{ll}
     (e, \ov e, A)  &  \text{for $l=1$},
  \\   (e, \ov \nu, W^-)  &  \text{for $l=2$},
  \\   (\nu, \ov e, W^+)  &  \text{for $l=3$},
  \\   (\nu, \ov \nu, Z^0)  &  \text{for $l=4$},
  \\   (e, \ov e, Z^0)  &  \text{for $l=5$},
  \\   (\nu, \ov \nu, A)  &  \text{for $l=6$}.
  \end{array}
  \right. 
   \end{align}
 Substituting Eqs.(\ref{S-eove})-(\ref{S-ZA}) into Eq.(\ref{G1-spin-eta}),
 one finds the following expressions of the amplitudes $G_{\eta_l}^{\rm spin}$
 (see Appendix \ref{app-derive-G-mod-eta}),
\begin{subequations}\label{G-mod-eta}
 \begin{align}\label{G-mode-1}
 & \frac{1}{\N_{A}} G_{\eta_1}^{\rm spin} = \frac{1}{2\sqrt 2} (w_1 G_L + w_2 G_R),
 \\ & \frac{1}{\N_{W^\pm}} G_{\eta_{2,3}}^{\rm spin} 
 = - \frac{1}{2} G_L ,  \label{G-mode-23}
 \\ & \frac{1}{\N_{Z^0}} G_{\eta_4}^{\rm spin} = - \frac{w_3 + w_4}{2\sqrt 2} G_L   , \label{G-mode-4}
 \\ & \frac{1}{\N_{Z^0}}  G_{\eta_5}^{\rm spin} = \frac{1}{2\sqrt 2} (w_3 G_L - w_4 G_R),
 \label{G-mode-5}
 \\ & \frac{1}{\N_{A}} G_{\eta_6}^{\rm spin} =  -\frac{w_1 - w_2}{2\sqrt 2} G_L. \label{G-mode-6}
 \end{align}
\end{subequations}
 Comparing the rhs of Eq.(\ref{G-mod-eta}) and those of Eq.(\ref{G-ew-k})
 for the GWS amplitudes $G_{{\rm ew},k}^{\rm spin}$ 
 [see also Eq.(\ref{G-ew-k-explicit}) in Appendix \ref{app-derive-G-mod-eta}],
 one finds that, if  $w_{1,2,3,4}$ takes the following values, 
\begin{subequations}\label{w123}
\begin{align}\label{w1-ew}
 & w_1 = w_2 = 2 \sin \theta_{W}, 
 \\ & w_3 = \frac{1 - 2\sin^2 \theta_{W} }{ \cos \theta_{W}},
 \quad  w_4 = \frac{2\sin^2 \theta_{W}}{ \cos \theta_{W}}, \label{w23-ew}
\end{align}
\end{subequations}
 then, these amplitudes have the following relations, 
\begin{subequations}\label{G-l-k}
 \begin{align}\label{G-lk=1-4}
 & \frac{1}{\N_{B}} G_{\eta_l}^{\rm spin} = \frac{\sqrt 2}{g} G_{{\rm ew},k}^{\rm spin},
 \quad \text{for $l=k=1,2,3,4$,}
 \\ & \frac{1}{\N_B} G_{\eta_5}^{\rm spin} 
 = \frac{\sqrt 2}{g} \left( G_{{\rm ew},5L}^{\rm spin} + G_{{\rm ew},5R}^{\rm spin} \right);
 \label{G-lk=5}
 \end{align}
\end{subequations}
 meanwhile, $G_{\eta_6}^{\rm spin} =0$,
 which is in consistency with the fact that 
 neutrinos do not participate in electromagnetic interaction in the GWS theory.
 Then, one is ready to see that the GWS interaction Hamiltonian $H^{\rm gws}_{\rm int,I}$, 
 which is given by Eq.(\ref{Hint-gws-I}) with Eqs.(\ref{Hint-FIO-k}) and (\ref{H1-G-gws}), 
 indeed has the concise expression in Eq.(\ref{HIdF-mod}).

 \section{A unified framework for elementary particles}
 \label{sect-framework}

 The above-obtained concise expression of $\HH^{\rm gws}_{\rm int,I}$ [Eq.(\ref{HIdF-mod})] 
 suggests an alternative way of building
 the lepton-vector-boson interaction Hamiltonian, 
 without resorting to the gauge-symmetry assumption. 
 In this section, based on this, a generic framework is proposed
 which may be useful for the purpose of constructing models for elementary particles and their interactions.
 The framework consists of two parts:
 one containing five basic assumptions (Sec.\ref{sect-framework-A})
 and the other containing certain generic requirements for four model-dependent assumptions
 (Sec.\ref{sect-gen-model}).

 \subsection{Basic assumptions of the framework}
 \label{sect-framework-A}

 The five basic assumptions
 are to be indicated as F1, F2, F3, F4, and F5, with F standing for ``framework's basic assumption''. 
 Most of them are direct generalizations of results of previous sections.

 The first basic assumption is about a multi-layer structure of spin spaces for elementary particles,
 as a generalization of the two-layer case discussed previously. 
\begin{itemize}
 \item \textbf{F1}. 
 The spin space of each particle species has a multi-layer structure,
 as a direct sum of $\WW$/$\ov\WW $ for fermion
 and a direct sum of $\VV$/$\C$/$\ov\C$ for vector boson.
\end{itemize}
 As done before, the first layer of a vector boson is set as $\VV$. 
 In an ideal case, particles species are characterizable by their spinor-space configurations,
 as examplified in the previously discussed case of (first generation of) leptons and vector bosons 
 [see Eqs.(\ref{config-WovW})-(\ref{config-VCC})].

 We use $M$ to indicate an arbitrary particle species  
 and use $n_M$ for its number of layers.
 Not introducing any further degree of freedom, its single-particle states 
 are written as [cf.~Eq.(\ref{single-p-s-gen-2L})]
      \begin{align}\label{}
       \label{|M>} & |M_{\bp \alpha }\ra =  |\bp\ra |\cs_{M\alpha}(\bp)\ra,
      \end{align}
 where $\alpha = (r, \varrho)$ for fermion with $M=f$ and $\alpha = \lambda$ for vector boson with $M=B$.
 The scalar-product expressions of  single-particle states 
 in Eqs.(\ref{<f'-varrho|f>=delta}) and (\ref{<B|B>}) are required to be valid in the generic case;
 in other words, they need to be satisfied when explicit expressions of spin states 
 are assigned to particles. 
 Thus, the label $\varrho$ is, now, introduced as a property of 
 the scalar product in Eq.(\ref{<f'-varrho|f>=delta}).
 The labels of $r$ and $\lambda$ are introduced 
 as indices within the spinor spaces of fermion and boson, respectively,
 which may be done in an ordinary way based on the symmetry of 
 Lorentz group (see, e.g., Ref.\cite{Weinberg-book}).

 According to F1, the spin state $|\cs_{M\alpha}(\bp)\ra$ is written as
  \begin{align}
  \label{S-M} |\cs_{M\alpha }(\bp)\ra = 
    \left( \begin{array}{c} |\LL^{M}_1 \ra \\ \vdots \\ |\LL^{M}_{n_M} \ra  \end{array} \right),
   \end{align}
 with layers $|\LL^{M}_j \ra$ of $j=1,\ldots, n_M$.
 Specifically,  $|\LL^{f}_j \ra \in \WW$ or $\ov\WW$ for a fermion;
 meanwhile, $|\LL^{B}_j \ra \in \VV$, or $\C$, or $\ov\C$  for a vector boson,
 with $|\LL^{B}_1\ra \in \VV$. 
 Generically, one may write
\begin{align}\label{layer-spinor}
 |\LL^{M}_j \ra = c^M_j |\varphi^M_j \ra,
\end{align}
 where $|\varphi^M_j \ra$ are normalized spinors and $c^M_j$ are prefactors.

 For fermions, physically, all the layers within one spin state $|\cs_{f\alpha}(\bp)\ra$ should 
 share a common angular momentum.
 There are two methods of mathematical description for the layer spinors $|\varphi^f_j\ra$. That is, 
\begin{align}\label{}
 &  |\varphi^f_j\ra = \left\{ 
  \begin{array}{l}
 |u^r(\bp)\ra \ \text{for $\WW$-type layers},
 \\   |\ov v^r(\bp)\ra \ \text{ for $\ov\WW$-type layers};
  \end{array}
  \right. 
\end{align}
 or,
\begin{align}\label{}
 &  |\varphi^f_j\ra = \left\{ 
  \begin{array}{l}
 |v^r(\bp)\ra \ \text{for $\WW$-type layers},
 \\   |\ov u^r(\bp)\ra \ \text{ for $\ov\WW$-type layers}.
  \end{array}
  \right. 
\end{align}
 In principle, the above two methods are physically equivalent and 
 we  are to adopt the first one.

 Similarly, for vector bosons, all the $\VV$-type layers within one state 
 $|\cs_{B\lambda }(\bp)\ra$ should share a common angular momentum, implying that
\begin{align}\label{}
 |\varphi^B_j\ra = |\varepsilon_\lambda(\bk) \ra  \quad
 \text{for $\VV$-type layers. }
\end{align}
 The situation with the $\C / \ov\C$-type layers is not so simple,
 but, for a reason to be discussed at the end of this section,
 there is no need of discussing their detailed properties in this paper.

 The second basic assumption is about the free Hamiltonian of a particle $M$, denoted by $H_0^M$.
 As discussed previously, when expanded by means of creation and annihilation operators,
 the interaction Hamiltonian in the GWS theory represents maps between state spaces.
 In fact, this is also true for free Hamiltonians in the theory, 
 except that each maps a state space to itself.
 Therefore, mapping of state spaces supplies a unified viewpoint for the descriptions 
 of both interaction Hamiltonian and free Hamiltonian.

 Based on the above understanding, we propose the following contents for F2.
\begin{itemize}
  \item \textbf{F2}. The free Hamiltonian of a particle $M$ is 
\begin{align}\label{H0f}
 & H_0^M =  \int d\ww p \sqrt{|\bp|^2 + m_M^2} | M_{\bp \alpha}\ra  \la  M_{\bp \alpha}|,
% & H^0_{f\varrho} =  \int d\ww p \sqrt{|\bp|^2 + m_f^2} | f_{\bp r \varrho}\ra  \la  f_{\bp \varrho}^r|.
% \\ &  H^0_{B} =  \int d\ww k |\bk| |B_{\bk \lambda}\ra  \la B_{\bk}^{ \lambda}|.
\end{align}
 where $m_M$ is the mass of $M$.
\footnote{ Since the gauge symmetry is not employed in the proposed approach, 
 there is no reason of forbidding usage of the mass $m_M$ of a particle $M$ 
 at the fundamental level of the theory. }
\end{itemize}
 By definition, $p^0$ is the eigenvalue of $H_0^M$. 
 Making use of Eq.(\ref{<f'-varrho|f>=delta}), one easily checks that 
 $ H_0^{f} | f_{\bp r \varrho}\ra = p^0 | f_{\bp r \varrho}\ra $,
 where $p^0 = \varrho \sqrt{|\bp|^2 + m_f^2}$.
 Thus, $\varrho$ is \emph{the sign of $p^0$},
 guaranteeing consistency with the previous definition of $\varrho$ 
 in Eq.(\ref{p0-varrho}).

 The third basic assumption is about FVF for generic fermions. 
 Here, a generic definition of antiparticle is needed.
 We note that F1 allows a quite simple definition for this concept. 
 That is, two particles $M$ and $\ov M$ are the \emph{antiparticle} of each other, 
 if the spinor space of each layer of $M$ is the complex conjugate of
 that of the corresponding layer of $\ov M$.
 With this generic definition of antiparticle, the concept of FVF may be introduced in the same way 
 as discussed previously in Sec.\ref{sect-FVF}, which has the properties stated below as F3. 
\begin{itemize}
  \item \textbf{F3}. 
  A fermion-antifermion pair, which possesses net zero four-momentum 
  and net zero angular momentum, may emerge from the vacuum as an FVF pair,
 as described by $V_f$ in Eq.(\ref{VK1}); or, it may vanish into the vacuum as an FVF pair,
 as described by $V_f^\dag$ in Eq.(\ref{VK2}). 
\end{itemize}

 The fourth basic assumption is for FIOs concerning generic fundamental interactions.
 It is particularly for FIO1 as a generalization of that given in Eq.(\ref{HFIO1-model}), with FIO2 determined by
 Eq.\eqref{FIO2=FIO1-dag} as a requirement of Hermiticity. 
 To achieve this goal, the main task is a generalization of $\G^{\rm FIO1}_{ \eta}$
 in Eq.(\ref{G1-definition-<|>-mod}), together with the symbols of $P_\eta$ and $\overrightarrow{D}$. 
 Note that a generalization of $\overrightarrow D$ in Eq.(\ref{D}) requires that $n_f = n_{f'}$.
 Obviously, the most direct generalization is written as follows, 
 which is also indicated as $\overrightarrow D$ for brevity, 
 \begin{align}
  & \label{D-gen}
  \overrightarrow{D}  |\cs_{fr\varrho}(\bp)\ra |\cs_{f' r' \varrho'}(\bp')\ra  :=
  \left( \begin{array}{c}   {|\LL^f_1 \ra |\LL^{f'}_1 \ra }
  \\ \vdots
    \\  {|\LL^{f}_{n_f} \ra |\LL^{f'}_{n_f} \ra}  \end{array} \right).
 \end{align}
 In a generic case, it is unnecessary for $n_B$ to be equal to $n_f$.
 In this case, in order to compute an interaction amplitude 
 like $G_{\eta}^{\rm spin}$ in Eq.(\ref{G1-spin-eta}), 
 a $n_B \times n_f$ parameter matrix is needed, which is to be indicated as $T$.
 More exactly, at the place of $P_\eta$ on the rhs of Eq.(\ref{G1-spin-eta}), one should use $P_\eta T$.
 Clearly, the most direct generalization of  $P_\eta$ in Eq.(\ref{P-eta}) is written as follows, 
 \begin{align}\label{P-eta-gen}
  P_\eta = \left\{ 
    \begin{array}{ll}
      \sigma,  &  \text{if the space of ${|\LL^{f}_j \ra |\LL^{f'}_j \ra }$ is isomorphic to}
  \\  &  \text{ that of $|\LL^B_i\ra$ for all $(i,j)$ with $T_{ij} \ne 0$};
    \\  0,  &  \text{otherwise}.
    \end{array}
    \right. 
 \end{align}

 Summarizing the above discussions, one gets the following contents of F4.
\begin{itemize}
 \item \textbf{F4}. 
 For each set $\eta=(f,f',B)$ with $n_f = n_{f'}$, 
 the FIO1 of $H^{{\rm FIO1} }_{\varrho \varrho', \eta}$ is given by 
 Eqs.(\ref{HFIO1-model})-(\ref{G1-definition-<|>-mod}), 
 with the following spinor amplitude $G_{\eta}^{\rm spin}$,
  \begin{align}
  &        G_{\eta}^{\rm spin} =  \la  \cs_{B }^{\lambda }(\bk)|
      P_\eta T \overrightarrow{D} | \cs_{f \varrho }^{r }(\bp)  
    \cs_{f' \varrho' }^{r' }(\bp') \ra, \label{G1-spin-eta-T}
    \end{align}  
 where $\overrightarrow D$ and $P_\eta$ are given in Eqs.(\ref{D-gen})-(\ref{P-eta-gen}).
% And, FIO2 is the Hermitian conjugate of FIO1 [Eq.\eqref{FIO2=FIO1-dag}].
\end{itemize}
 Clearly, like the FIO1 for the GWS theory discussed previously, 
 the generic FIO1 given by F4 describes a direct map  from the state space of $(f,f')$ to that of $B$,
 subject to momentum conservation, layer isomorphism, and $\sigma$.
 Note that the matrix $T$ is model-dependent.

 The fifth assumption is about a generic fermion-vector-boson interaction Hamiltonian, denoted by
 $H^{\rm fvb}_{\rm int}$.
 This interaction should be for a given set of particle species, which is
 model-dependent and is to be denoted by $C_{\text{ps}}$.
 As a direct generalization of the rhs of Eq.(\ref{HIdF-mod}),
 F5 has the following contents. 
\begin{itemize}
  \item \textbf{F5}. 
 \begin{align}\label{HIdF-mod-noN}
  H^{\rm fvb}_{\rm int} =\frac{g}{\sqrt 2}
   \sum_{\eta \subset C_{\text{ps}}} \sum_{\omega, \varrho, \varrho'} \kappa_\eta V_\omega \left(  
    H^{{\rm FIO1} }_{\varrho \varrho', \eta} \right) + {\rm H.c.},
 \end{align}
  where $g$ is a coupling constant, $V_\omega$ with $\omega = ( n^V_f, n^V_{f'})$ 
  is defined in Eq.(\ref{F-lambda}),
  and $\kappa_\eta$ are model-dependent parameters. 
\end{itemize}

 Noting the similarity between the rhs of Eq.(\ref{HIdF-mod-noN}) and that of Eq.(\ref{HIdF-mod})
 and for the same reason as that discussed previously in the two-layer case below Eq.(\ref{sigma-act-0}), 
 one sees that 
 the $\C$-type and $\ov\C$-type layers of vector bosons in fact give no contribution to the 
 generic-FIO1 amplitudes, either. 
 In other words, the interaction amplitudes of $H^{\rm fvb}_{\rm int}$ are independent of 
 spinors in these two types of layers. 
 For this reason, here, there is no need to discuss detailed descriptions for
 spinors in $\C$-type and $\ov\C$-type layers.

\subsection{Generic aspects of model-dependent assumptions}\label{sect-gen-model}

 In order to get a concrete model for elementary particles, 
 further model-dependent assumptions are needed 
 in addition to the five basic assumptions discussed above. 
 At least, four such assumptions are needed,
 which are to be indicated as $A_{\rm MN}^{(i)}$ of $i=1,2,3,4$
 with ``MN'' referring to the model name. 
 In this section, we discuss some generic aspects of $A_{\rm MN}^{(i)}$.

 The first assumption $A_{\rm MN}^{(1)}$ is about
 determination of the exact spinor-space configurations to be employed in a model,
 which determines the set $C_{\text{ps}}$ of the particle species in the model. 
 In an ideal case, each particle species is related to one specific spinor-space
 configuration, such that the former is characterizable by the latter. 

 The second assumption $A_{\rm MN}^{(2)}$ is for determination of 
 the layer factors $c^M_j$ in Eq.(\ref{layer-spinor}).

 The third assumption $A_{\rm MN}^{(3)}$ is for determination of 
 the matrix $T$. 
 In the case of $n_f=n_{f'} = n_B$,
 $T$ may take the simplest form, namely, the $n_f$-dimensional unit matrix with $T_{ij} = \delta_{ij}$.

 The fourth assumption $A_{\rm MN}^{(4)}$ is about the value of $\kappa_\eta$
 in Eq.(\ref{HIdF-mod-noN}). 
 The main reason of introducing $\kappa_\eta$ in F5 is for the possibility of 
 reproducing the GWS lepton-vector-boson interaction Hamiltonian [Eq.(\ref{HIdF-mod})], 
 which is to be discussed in the first model of the next section. 
 Anyway, we note that the most natural choice seems that $\kappa_\eta = 1$ for all $\eta$.

 \section{Three models}
 \label{sect-models}

 In this section, we discuss three models
 as applications of the framework proposed above.
 The first model (Model I) addresses two-layer configurations of spinor space
 and it gives exactly the GWS interaction Hamiltonian $H^{\rm gws}_{\rm int,I}$.
 The second model (Model II) is introduced for removing some flaw observed in Model I. 
 The third model (Model III) discusses a set of three-layer configurations of spinor space,
 which shows a possibility of supplying a description for strong interactions. 
  
\subsection{Model I and its relation to the GWS theory}
\label{sect-model-I}

 Model I contains the following model-dependent assumptions 
 indicated as $A_{\rm MI}^{(i)}$.
 
 \noindent $A_{\rm MI}^{(1)}$. 
 All two-layer independent configurations of spinor space are included.
 They are just those discussed in Sec.\ref{sect-two-layer}.
 In particular, there are totally eight such configurations,
 as given in Eqs.(\ref{config-WovW})-(\ref{config-VCC}).
 They are associated with eight particles: four leptons in the first generation and four vector bosons.
 This fixes the set $C_{\text{ps}}$.
 
 \noindent   $A_{\rm MI}^{(2)}$. 
 Values $c_j^M$ are those given in Eqs.(\ref{S-eove})-(\ref{S-ZA}).

 \noindent   $A_{\rm MI}^{(3)}$. 
 The parameter matrix $T$ in Eq.(\ref{G1-spin-eta-T}) is the $2 \times 2$ unit matrix,
 with $T_{ij} = \delta_{ij}$.

 \noindent $A_{\rm MI}^{(4)}$. 
 The parameters $\kappa_\eta$ are given by $\kappa_\eta = \frac 1{\N_B}$,
 with $B$ representing the boson in the set $\eta$.

 Making use of the above assumptions, 
 it is straightforward to check that this model is equivalent to 
 the part of the GWS electroweak theory discussed in Sec.\ref{sect-GWS-two-layer}.
 In particular, Model I predicts the same lepton-vector-boson interaction 
 Hamiltonian, i.e., $H^{\rm fvb}_{\rm int} = H^{\rm gws}_{\rm int,I}$.
 Moreover, making use of F2 of the framework, 
 one gets the free Hamiltonian for the eight particle species, denoted by $H_{{\rm MI},0}$,
\begin{align}\label{H0-lepton}
 & H_{{\rm MI},0} = \sum_{M=e,\ov e, \nu, \ov \nu,{\rm photon},Z^0, M^\pm} H_0^M,
\end{align}
 where $H_0^M$ is defined in Eq.(\ref{H0f}).
 One sees that $H_{{\rm MI},0}$ is equal to the free Hamiltonian 
 in the GWS theory (after the Higgs mechanism is used).

 One interesting question is whether the $SU(2)\otimes U(1)$-gauge-invariant Lagrangian
 in the GWS theory, which is to be indicated as $\LL^{\rm gws}_{\rm gauge}$,
 may be gotten formally by some extension of Model I (with additional assumptions introduced). 
 Below, we discuss one method of doing this, without touching the subtle topic of quantization.

 To achieve the goal mentioned above,
 one needs to write the two Hamiltonians of $H^{\rm fvb}_{\rm int}$ 
 and $H_{{\rm MI},0}$ of Model I in terms of fields. 
 As discussed above, 
 these two Hamiltonians are equal to the GWS lepton-vector-boson interaction Hamiltonian
 $H^{\rm gws}_{\rm int,I}$ and the GWS free Hamiltonian, respectively.
 Thus, from them, it is straightforward to get that part of the Lagrangian in the GWS theory,
 which is for ``free'' particles and the lepton-vector-boson interactions
 (in a form after the Higgs mechanism is used).
 We use $\LL_{1}$ to denote this Lagrangian.
 Then, reversing the procedure of Higgs mechanism 
 (with appropriate assumptions introduced), from $\LL_1$ one gets a Lagrangian
 denoted by $\ww\LL_1$.
 Compared with $\LL^{\rm gws}_{\rm gauge}$, $\ww\LL_1$ lacks only 
 interactions among the vector bosons.

 It is not difficult to check that, neglecting its part for free vector bosons, 
 $\ww\LL_1$ possesses the gauge symmetry of  $SU(2) \otimes U(1)$.
 Note that, here, this symmetry is a consequence of Model I of the proposed framework. 
 However, the framework does not guarantee this symmetry for the bosonic fields.
 In order to get interactions among vector bosons, 
 one may assume this gauge symmetry for the vector-boson fields. 
 Then, by the standard procedure, one gets $\LL^{\rm gws}_{\rm gauge}$.
 Thus, at least formally, 
 the GWS gauge-invariant Lagrangian $\LL^{\rm gws}_{\rm gauge}$ may be gotten in a hybrid way,
 particularly by introducing an $SU(2) \otimes U(1)$-gauge-symmetry assumption
 for bosonic fields to Model I.

 Clearly, the above hybrid construction of $\LL^{\rm gws}_{\rm gauge}$ is unsatisfactory
 from the viewpoint of  the proposed framework. 
 In its perspective, a better method of getting interactions among vector bosons
 should be by means of introducing some natural generalization of the proposed framework.
 One possible way of doing this will be briefly discussed in Sec.\ref{sect-illus-inter-boson}.

\subsection{Model II: small modification to Model I}
\label{sect-model-II}

 From the perspective of the proposed framework, a flaw is seen in Model I.
 That is, the parameters $\kappa_\eta$ contain the normalization factors ${ \N_B}$,
 though the spin states used in the FIO1s have already been normalized.
 Logically, there should be no reason for the normalization factors to appear in an independent manner
 in the interaction Hamiltonian. 
 Moreover, with $w_1=w_2$ [Eq.(\ref{w1-ew})], the photon spin state
 in Eq.(\ref{S-A}) should be written as
\begin{align}
      \label{S-A-mod} &  \la \cs_{A}^\lambda(\bk) | = \frac{1}{\sqrt 2} 
       \left( \la \varepsilon^\lambda(\bk) | ,
      \la \varepsilon^\lambda(\bk) |  \right),
\end{align}
 no need of using the normalization factor $\N_A$.

 The second model is proposed for removing the above-discussed flaw of Model I, 
 by introducing a modification as small as possible.
 For this purpose, 
 we observe that, if $w_3$ and $w_4$ in Eq.(\ref{w23-ew}) could be equal to each other, 
 then, the spin state of $Z^0$ boson in Eq.(\ref{S-Z}) would be written as 
\begin{align}
  \label{S-Z-mod}
      &    \la \cs_{Z^0}^\lambda(\bk) | =  \frac{1}{\sqrt 2} 
      \left( \la \varepsilon^\lambda(\bk)|,   -  \la \varepsilon^\lambda(\bk)| \ \right).
\end{align}
 As a result, like the photon state in Eq.(\ref{S-A-mod}), 
 there is no need of introducing the normalization factor $\N_{Z^0}$, either. 
 Furthermore, from  Eq.(\ref{w23-ew}) one sees that the equality of $w_3 = w_4$ 
 requires a fixed value of $\sin^2 \theta_{W} = 1/4$ for the Weinberg angle, 
 which is not far from the value computed in the GWS theory from experimental data.

 For the above discussed reasons, we propose to consider a model II, which contains the following
 assumptions $A_{\rm MII}^{(i)}$.
\begin{itemize}
 \item  $A_{\rm MII}^{(1)} = A_{\rm MI}^{(1)}$. 
 \item  $A_{\rm MII}^{(2)}$. 
 The coefficients $c_j^M$ are equal to those given in
 Eqs.(\ref{S-eove})-(\ref{S-W}) and (\ref{S-A-mod})-(\ref{S-Z-mod}).
 \item  $A_{\rm MII}^{(3)} = A_{\rm MI}^{(3)}$. 
 \item  $A_{\rm MII}^{(4)}$. 
 $\kappa_\eta = 1$ for all $\eta$.
\end{itemize}
 It is then ready to find that the lepton-vector-boson interaction Hamiltonian in Model II is written as 
 \begin{align}\label{HIdF-mod-II}
  H^{\rm fvb}_{\rm int} =\frac{g}{\sqrt 2}
   \sum_{\eta \subset C_{\text{ps}}} \sum_{\omega, \varrho, \varrho'}  V_\omega \left(  
    H^{{\rm FIO1} }_{\varrho \varrho', \eta} \right) + {\rm H.c.}.
 \end{align}
 Clearly, Model II contains merely one coupling constant $g$,
 which looks attracting in the perspective of grand unification.
 Moreover, $\la \cs_{A}^\lambda(\bk) | \cs_{Z^0}^\lambda(\bk) \ra =0$,
 which looks interesting though not compulsory due to the term $\delta_{BB'}$
 on the rhs of Eq.(\ref{<B|B>}).

 One question of relevance is about relationship between Model II and the GWS theory. 
 An interesting observation is that the interaction Hamiltonian density of Model II
 may be written in a form like that of the GWS theory.
 More exactly, it is equal to the sum of the GWS densities in Eq.(\ref{HH-gws-k}),
 but, with some of the prefactors $\xi_k$ appropriately changed. 
 Indeed, following procedures similar to those given in Sec.\ref{sect-EvdW-electroweak}, 
 one may check that
 the interaction Hamiltonian $H^{\rm fvb}_{\rm int}$ of Model II [Eq.(\ref{HIdF-mod-II})] is 
 obtainable from the GWS interaction Hamiltonian densities given in Eq.(\ref{HH-gws-k}), 
 if the following two changes are made to the coefficients $\xi_k$. 
 That is, (i) replace all the terms of ``$\sin^2 \theta_{W}$''  in Eq.(\ref{xi-k}) by $\frac 14$, 
 with a corresponding change for ``$\sin \theta_{W}$'' in $\xi_1$;
 and, (ii) replace all the terms of ``$\cos \theta_W$'' by $\frac 12$.

 Regarding the first change discussed above, 
 one may note that  $\sin^2 \theta_{W} = 0.25$ 
 is not far from $\sin^2 \theta_{W} \simeq 0.23$, 
 the latter of which is obtained from experimental data 
 by the renormalization scheme employed in the GWS theory.
 In fact, since Model II contains one coupling constant only, 
 the renormalization scheme employed for Model II should not be exactly the same
 as that of the GWS theory. 
 This implies that explanations of the two theories to experimental data may show some differences.

 Concerning the second change mentioned above,
 the term $\cos \theta_{W}$ in the GWS theory is related to the mass ratio of $W^\pm$ and $Z^0$ bosons.
 For the same reason as that discussed above,
 explanation of Model II to experimental data related to the mass ratio
 may show some difference from that given by the GWS theory, too.

 Therefore, further investigations in future are needed, 
 in order to make clear whether the above discussed two changes of $\xi_k$ for Model II
 could be compatible with experimental data. 
 In other words, it is worth investigation whether Model II, which contains one coupling constant only,
 could be appropriate for the description of electroweak interactions.

\subsection{Model III: three-layer configurations}
\label{sect-model-III}

 This section discusses Model III, which is for three-layer configurations of spinor space.
 Since it is unclear at the present stage whether such configurations could be 
 related to real particles, 
 we use the name of ``mode'' to refer to ``particle'' in this model.

 Let us first discuss fermions in Model III.
 We recall that the four leptons in the first generation may be divided into two classes, 
 which behave quite differently in interaction,
 one class including electron and positron and the other including electron neutrino and antineutrino. 
 From discussions given in previous sections,
 we note that the two classes of leptons are related to different types of two-layer configurations 
 of spinor space.
% , such that the latter are also divided into two corresponding classes. 
 Indeed, according to Eq.(\ref{config-WovW-e}) for $(e,\ov e)$
 and Eq.(\ref{config-WovW-nu}) for $(\nu, \ov \nu)$,
 each configuration for the former class 
 contains both $\WW$ and $\ov\WW$,
 while, each for the latter contains either two $\WW$ or two $\ov\WW$.

 Similarly, three-layer configurations, the total number of which is eight, 
 may be divided into two classes.
 The first class includes six three-layer configurations, 
 each of which contains both types of Weyl-spinor spaces.
 Three of them are to be referred to as $Q$-type,
 with the corresponding modes indicated as $Q_j$ of $j=1,2,3$;
 and the rest three as $\ov Q$-type, which have a complex-conjugation relationship to the $Q$-type,
 with modes indicated as  $\ov Q_j$.
 That is, 
\begin{subequations}\label{config-WovW-t1-3}
 \begin{align}
 \label{config-WovW-t1}
 & Q_1: \left( \begin{array}{c} \WW \\ \ov\WW  \\ \ov\WW  \end{array} \right), \ 
 Q_2: \left( \begin{array}{c} \WW \\ \WW  \\ \ov\WW  \end{array} \right), \ 
 Q_3: \left( \begin{array}{c} \WW \\ \ov \WW  \\ \WW  \end{array} \right), \ 
 \\ \label{config-WovW-t2}
&  \ov Q_1: \left( \begin{array}{c} \ov \WW \\ \WW  \\ \WW  \end{array} \right), \ 
  \ov Q_2: \left( \begin{array}{c} \ov \WW \\ \ov \WW  \\ \WW  \end{array} \right), \ 
  \ov Q_3: \left( \begin{array}{c} \ov \WW \\ \WW  \\ \ov \WW  \end{array} \right). \ 
 \end{align}
\end{subequations}
 Clearly, $Q_j$ and $\ov Q_j$ are the antimode of each other. 
 The above six fermionic modes are to be referred to as \textit{$Q$-modes}.

 The second class includes the rest two configurations,
 each containing either three $\WW$ or three $\ov\WW$. 
 We use $D_\nu$ and $\ov D_\nu$ to indicate the two modes related to them, 
 \begin{align}
 \label{config-WovW-D1}
 D_\nu: \left( \begin{array}{c} \WW \\ \WW  \\ \WW  \end{array} \right), \ \quad
  \ov D_\nu: \left( \begin{array}{c} \ov \WW \\ \ov \WW  \\ \ov \WW  \end{array} \right), \ 
 \end{align}
 which are the antimode of each other.

 Due to the relevance of spinor-space configuration in the
 computation of the interaction amplitudes [cf.~Eq.(\ref{G1-spin-eta-T})], 
 it seems reasonable to 
 expect that the previously-discussed classification for two-layer configurations
 may be useful in the study of three-layer configurations, too.
 This suggests that interactions involving $Q_j$ and $\ov Q_j$
 may behave quite differently from those for $D_\nu$ and $\ov D_\nu$. 
 In Model III, we consider the first class of configuration only.
 Thus, the fermionic subset of $C_{\text{ps}}$ of Model III contains the $Q$-modes only,
 i.e., $ \{ Q_j, \ov Q_j: j=1,2,3 \}$.

 Next, we discuss bosonic modes in Model III.
 It is reasonable to consider only those three-layer vector-bosonic modes, 
 which may interact with the $Q$-modes.
 We are to call them \textit{$G$-modes}.
 The $G$-modes constitute the bosonic subset of $C_{\text{ps}}$.
 The parameter matrix $T$ is assumed to be the unit matrix,

 Below, we derive exact configurations of $G$-modes.
 To see whether a bosonic mode may interact with $Q$-modes,
 one should consider the spinor part of the FIO1 amplitude given in Eq.(\ref{G1-spin-eta-T}),
 particularly, the symbol $P_\eta$ in it. 
 According to the expression of $P_\eta$ in Eq.(\ref{P-eta-gen}),
 layers of each $G$-mode should be isomorphic to the direct products 
 of the corresponding layers of a pair of $Q$-modes, respectively.
 Since the first layer of  the $G$-mode should be of the $\VV$-type, 
 the two modes in the pair should be written as $Q_j$ and $\ov Q_{j'}$. 
 We use $G_{j\ov{j'}}$ to indicate the related $G$-mode.
 Thus, there are nine $G$-modes.

 Suppressing other labels for brevity, the above discussed isomorphic relationship is written 
 as follows for spinors, indicated by ``$\leftrightarrow$'', 
\begin{align}
 |\cs_{G_{j\ov{j'}} }\ra  {\longleftrightarrow }
 \overrightarrow{D} | \cs_{Q_j }\ra | \cs_{Q_{j'} } \ra.
\end{align}
 Making use of the above relationship, it is straightforward to find the following nine
 spinor-space configurations of the $G$-modes, i.e., 
\begin{subequations}\label{config-B}
 \begin{align}
 \label{config-g1}
 G_{1 \ov 1}: \left( \begin{array}{c} \VV \\ \ov\VV  \\ \ov\VV  \end{array} \right), \ 
 G_{1 \ov 2}: \left( \begin{array}{c} \VV \\ \ov\C  \\ \ov\VV  \end{array} \right), \ 
 G_{1 \ov 3}: \left( \begin{array}{c} \VV \\ \ov\VV  \\ \ov \C  \end{array} \right), \ 
 \\ \label{config-g2}
 G_{2 \ov 1}: \left( \begin{array}{c} \VV \\ \C  \\ \ov\VV  \end{array} \right), \ 
 G_{2 \ov 2}: \left( \begin{array}{c} \VV \\ \VV  \\ \ov\VV  \end{array} \right), \ 
 G_{2 \ov 3}: \left( \begin{array}{c} \VV \\ \C  \\ \ov\C  \end{array} \right), \ 
 \\ \label{config-g3}
 G_{3 \ov 1}: \left( \begin{array}{c} \VV \\ \ov\VV  \\ \C  \end{array} \right), \ 
 G_{3 \ov 2}: \left( \begin{array}{c} \VV \\ \ov\C  \\ \C  \end{array} \right), \ 
 G_{3 \ov 3}: \left( \begin{array}{c} \VV \\ \ov\VV  \\ \VV  \end{array} \right). \ 
 \end{align}
\end{subequations}

 Based on the above discussions, Model III contains the following assumptions. 
\begin{itemize}
 \item  $A_{\rm MIII}^{(1)}$. 
 $C_{\text{ps}} = \{ Q_j, \ov Q_j,  G_{j\ov{j'}}: j,j'=1,2,3 \}$.
 \item  $A_{\rm MIII}^{(2)}$. 
 $c_j^M$ are undetermined.
 \item  $A_{\rm MIII}^{(3)}$. 
 $T$ is the $3 \times 3$ unit matrix. 
 \item  $A_{\rm MIII}^{(4)}$. 
 $\kappa_\eta = 1$ for all $\eta$.
\end{itemize}

 Finally, let us compare Model III and 
 the quantum chromodynamics (QCD) for the description of strong interactions among quarks.
 If relating the label $j$ of $Q_j$ to the quantum number of color,
 one observes certain formal similarities between them, 
 particularly, the four discussed below. 
 
 Firstly, there are three $Q$-modes of $Q_j$ ($j=1,2,3$), together with their antimodes;
 meanwhile, each quark $q$ has three colors, $q_c$ of $c=r,b,g$, with antiquarks $\ov q_c$. 

 Secondly, there are six $G$-modes, which mediate interactions between $Q$-modes with 
 different labels of $j$, i.e., 
\begin{align}\label{Gij-config}
 G_{1\ov{2}}, G_{1\ov{3}}, G_{2\ov{1}}, G_{2\ov{3}}, G_{3\ov{1}} ,G_{3\ov{2}}.
\end{align}
 Meanwhile, in QCD, there are six gluons, which mediate interactions of quarks of different colors, 
 written in the following configurations, 
\begin{align}\label{gluon-config}
 r\ov b, r\ov g, b\ov g, b\ov r, g\ov r, g\ov b.
\end{align}

 Thirdly, there are three $G$-modes, which mediate interactions
 of $Q$-modes with a same label of $j$; i.e., 
 $G_{1\ov{1}}$, $G_{2\ov{2}}$, and $G_{3\ov{3}}$.
 Although there are only two independent gluons for quark interactions with a same color, 
 as seen from their descriptions of 
 $\frac{r\ov r - b\ov b}{\sqrt 2}$ and $\frac{r\ov r + b\ov b - 2 g\ov g}{\sqrt 6}$,
 they need to involve three configurations.
 Moreover, if imposing certain restriction to the three $G_{j\ov{j}}$ modes,
 it may happen that only two of them are independent regarding interactions with the $Q$-modes.

 Fourthly, reversing discussions given in previous sections,
 one sees that densities of the interaction Hamiltonian $H^{\rm fvb}_{\rm int}$ 
 [Eq.(\ref{HIdF-mod-noN})] of Model III are written in forms similar to the rhs of Eq.(\ref{HHgws-k-concise}).
 In QCD, the fermion-vector-boson interaction Hamiltonian densities
 are also written in forms like  the rhs of Eq.(\ref{HHgws-k-concise}).

 In view of the above discussed formal similarities, it should be worth future investigation
 to study whether Model III could be further developed for supplying a useful description for
 strong interactions. 
% In such a development, the $Q$-type configurations in Eq.(\ref{config-WovW-t1-3}) 
% may be used as blocks in the construction of spinor-space configurations that may be  related to quarks. 

\section{Summary and discussions}\label{sect-conclusion}

\subsection{Summary}
 
 This paper consists of two parts. 
 In the first part, the GWS lepton-vector-boson interaction Hamiltonian $H^{\rm gws}_{\rm int,I}$,
 which describes interactions between the four leptons in the first generation
 and the four vector bosons,  is reformulated in a concise form. 
 This reformulation reveals a method, by which $H^{\rm gws}_{\rm int,I}$ may be built
 without making use of gauge symmetry. 
 In the second part, based on the above discussed reformulation of $H^{\rm gws}_{\rm int,I}$, 
 a framework is proposed, which may be useful for constructing models
 of elementary particles and their interactions without making use of gauge symmetry
 at the fundamental level.

 The reformulation given in the first part is based on two concepts:
 FIO (fundamental interaction operator) and FVF (fundamental vacuum fluctuation). 
 Loosely speaking, each FIO describes either the most direct map from the state space of two fermions 
 to that of one boson, or the reverse (with the relationship of Hermitian conjugation). 
 An FVF refers to either emergence from the vacuum or vanishing into the vacuum,
 of a fermion-antifermion pair that possesses
 net zero four-momentum and net zero angular momentum. 
 The reformulation shows that $H^{\rm gws}_{\rm int,I}$ is basically given by the sum of all 
 those FIO-FVF combinations, each of which contains one FIO and an arbitrary number of FVF,
 subject to a rule of no free leptonic state with  negative $p^0$.

 In the above-mentioned reformulation, a two-layer structure is adopted for spin spaces 
 of all the particles involved.
 Loosely speaking, the spin space of each lepton is a direct sum of two layers, each as a Weyl-spinor space;
 while, that of each vector boson is a direct sum of two layers,
 each as a direct product of two Weyl-spinor spaces.
 There exist eight independent such configurations of spinor space, 
 which are assigned to the eight particle species, respectively.

 The framework proposed in the second part of the paper contains five basic assumptions (F1, F2, 
 F3, F4, and F5).
 Most of them are direct generalizations of results of the first part,
 particularly, of the above discussed reformulation of $H^{\rm gws}_{\rm int,I}$.
 Loosely speaking, F1 is about a multi-layer structure for spin spaces of particles, 
 F2  about free Hamiltonians of particles, 
 F3  about FVF, F4 about FIO, and F5 gives a method of building 
 a fermion-vector-boson interaction Hamiltonian. 
 The framework also includes some generic aspects of further assumptions, which are needed 
 for the purpose of constructing concrete models for particles.
 (Moreover, as to be discussed briefly in the next section, in principle,
 the framework is generalizable to include interactions merely among vector bosons.)

 As applications of the proposed framework, three models have been constructed and studied.
 The first model predicts the same lepton-vector-boson interaction Hamiltonian as 
 that predicted by  the GWS theory, namely, $H^{\rm gws}_{\rm int,I}$.
 However, the first model suffers a serious flaw;
 that is, normalization factors of spin states of vector bosons are used in
 the interaction Hamiltonian in a way, which is independent of the states themselves. 
 To get rid of this flaw, the second model is proposed by
 introducing a small modification to the first model.
 Interestingly, as a byproduct of the modification, the second model contains only one coupling constant, 
 with the Weinberg angle fixed at $\sin^2 \theta_{W} = 0.25$.

 The third model considers spinor spaces that possess a three-layer structure. 
 Interactions predicted by this model show certain formal similarities to strong interactions,
 with the quantum number of color in QCD interpreted as  certain 
 type of spinor-space configuration. 

\subsection{Illustration of a method of predicting 
interaction among vector bosons}\label{sect-illus-inter-boson}

 As discussed in Sec.\ref{sect-model-I}, with some clumsy assumptions external to the framework 
 introduced, an extended version of model I may predict interactions merely among vector bosons;
 and, this hybrid construction of model is unsatisfactory in view of the proposed framework. 
 In this section, we give a brief discussion for a method of 
 naturally generalizing the proposed framework, which may describe 
 interactions among vector bosons.

 The point lies in that,
 in the interaction Hamiltonian $H^{\rm fvb}_{\rm int}$ given by the framework 
 [Eq.(\ref{HIdF-mod-noN})],
 each superoperator $V_{\omega}$ contains only one FIO1.
 Note that each FIO1 contains only one bosonic state and that FVF does not involve any boson.
 Hence, each term in $H^{\rm fvb}_{\rm int}$ describes an interaction process
 that involves only one vector boson. 
 In other words, in order to describe interactions among vector bosons,
 one needs to make use of more than one FIOs in each term.

 As an illustration, let us discuss a method of describing
 the change of a pair of $W^+$ and $W^-$ bosons into one $Z^0$ boson,
 $(W^+,W^-) \to Z^0$, by making use of several FIOs and FVF(s). 
 In such a description, two $W^\pm$ bosons need to be annihilated, which should 
 correspond to two bras; meanwhile, one $Z^0$ boson need to be generated, 
 which should correspond to one ket. 
 According to the explicit expression of $H^{{\rm FIO}1}_{\varrho' \varrho,k}$ in Eq.(\ref{H-FIO1-k}) 
 for FIO1 and its Hermitian conjugate for FIO2, 
 the above description should contain two FIO2 with $W^\pm$ bosons and one FIO1 with $Z^0$ boson.

 We use FIO2$^+$ to indicate the FIO2 that involves the $W^+$ boson.
 It is described by $H^{{\rm FIO}1 \dag}_{\varrho' \varrho,2}$,
 the Hermitian conjugate of the related FIO1.
 And, it corresponds to the change of $W^+ \to (\nu, \ov e)$ and contains
 the ket-bra sequence of $|\nu\ra |\ov e\ra \la W^+|$, with other labels suppressed for brevity.
 Similarly, we use FIO2$^-$ to indicate the other FIO2, 
 described by $H^{{\rm FIO}1 \dag}_{\varrho' \varrho,3}$, 
 which corresponds to $W^- \to (\ov\nu, e)$ and contains $|\ov \nu\ra | e\ra \la W^-|$.
 Suppose that the neutrino and antineutrino states, which belong to FIO2$^+$ and FIO2$^-$, respectively,
 may form an FVF pair.
 Then, these two leptons may vanish into the vacuum, described a superoperator $V^\dag_\nu$.
 Furthermore, it is possible for the electron and positron states left in the two FIO2s
 to change to a $Z^0$ boson;
 this may be described by both FIO1s of 
 $H^{{\rm FIO}1}_{\varrho' \varrho,5L(5R)}$,
 corresponding to $(e,\ov e) \to Z^0$ and containing $|Z^0\ra \la e | \la \ov e|$.
 Combining the above changes, one gets $(W^+,W^-) \to Z^0$ as the net effect.

 The above example shows that a Hamiltonian for interactions merely among vector bosons may be built
 by making use of more than one FIOs.
 But, to build a Hamiltonian of this type and give a
 meaningful comparison with $H^{\rm gws}_{\rm int,II}$, 
 both in the formal form and in compatibility with experimental data, 
 lots of investigations in future are needed.

\subsection{Some topics for future study}

 In this section, in perspective of the proposed framework we briefly discuss several topics,
 which are worth attention of investigation in future for the purpose of 
 developing a complete theory for elementary particles. 
 Specifically, they are the following: (i) spinor-space characterization of particle species, 
 (ii) spinor-space configurations that may be related to quarks, 
 and (iii) the $D_\nu$ and $\ov D_\nu$ modes.

 Firstly, in the above discussed reformulation of $H^{\rm gws}_{\rm int,I}$,  
 the first-generation leptons and the four vector bosons  (photon, $Z^0$, and $W^\pm$)
 are characterized by configurations of their spinor spaces. 
 This suggests that at least some of the quantum numbers in the ordinary formulation
 of the particle theory, which are employed 
 for the purpose of distinguishing between particle species, may be associated with 
 spinor-space configurations.
 For example, the difference between electron and electron neutrino
 may be related to their spinor-space configurations [see Eq.(\ref{config-WovW})]. 
 It should be of interest in future to study whether such a 
 spinor-space characterization of particle species may be generalizable 
 to other elementary particles
 \footnote{The basic assumption F1 in the proposed framework allows such a 
 characterization, though does not request it.  
 }.

 Clearly, the above-mentioned task is not an easy one. 
 For example, in such a generalization (if possible), 
 one needs to consider leptons in the second and third generations.
 Two strategies may be employed when dealing with them.
 One is to associate them with spinor-space configurations different from those of the first generation,
 while, the other is to take them as excited states of the first generation.

 Secondly, we give a very brief discussion on properties of
 spinor-space configurations that might be used for quarks in the first generation, namely, 
 for up quark $u$ and down quark $d$. 
 As discussed in Model III (Sec.\ref{sect-model-III}),
 the spinor-space configurations of $Q$-modes may be potentially related to the quantum number of color.
 However, $Q$-modes may not be directly interpreted as $u$ and $d$ quarks,
 because there are not enough three-layer configurations
 for accommodating two quark species. 
 One method of solving this problem seems to use the $Q$-type configurations 
 as building blocks in the construction of configurations for quarks. 
 For example, one may replace the Weyl-spinor spaces in the leptonic configurations
 [see Eq.(\ref{config-WovW})] by $Q$-type configurations.

 Thirdly, we discuss the two three-layer modes of $D_\nu$ and $\ov D_\nu$ in 
 Eq.(\ref{config-WovW-D1}).
 Both of them have identical layers and, in this sense, are 
 similar to electron neutrino and antineutrino, respectively. 
 By analogy to the difference between $(e,\ov e)$ and $(\nu, \ov \nu)$, 
 one may suspect that interactions participated by $D_\nu$ and $\ov D_\nu$
 might be much weaker than those of the $Q$-modes.
 Supposing that spinor-space configurations for quarks might be built from $Q$-modes, 
 then, in a similar way, the $D$-modes may be used to construct some other configurations. 
 An interesting question, which deserves future investigation, is 
 whether the latter type of configuration might be related to some candidate for dark matter.

 \acknowledgements

 The author is grateful to Yan Gu and Guijun Ding for valuable discussions and suggestions.
 This work was partially supported by the Natural Science Foundation of China under Grant
 Nos.~12175222, 11535011, and 11775210.

\appendix

\section{Spinors  in an abstract notation}\label{sect-2-spinor}

 In this appendix, we recall basic properties of spinors 
 \cite{Penrose-book,CM-book,Corson,Kim-group,pra16-commu}
 and write them in the abstract notation as discussed in Ref.\cite{pra16-commu}.
 Specifically, we discuss basic properties of Weyl spinors 
 in Sec.\ref{sect-recall-Weyl-spinor}
 and discuss stationary solutions of the Dirac equation in Sec.\ref{sect-Dirac-spinor}, both in the abstract notation.
 We recall basic properties of four-component vectors
 in Sec.\ref{sect-recall-vector} and discuss their abstract ket-bra expressions
 in Sec.\ref{sect-vector-abstract}.

\subsection{Basic properties of Weyl spinors}\label{sect-recall-Weyl-spinor}

 In the spinor theory, there are two smallest nontrivial representation spaces of the $SL(2,C)$ group,
 which are spanned by two types of two-component Weyl spinors, respectively,
 with the relationship of complex conjugation.
 In this section, we give a brief discussion for Weyl spinors written in the ket-bra notation.
 \footnote{See Ref.\cite{pra16-commu} for more detailed discussions, except for the last paragraph
 of this section.}

 We use $\WW$ to denote one of the two spaces mentioned above.
 In terms of components, a Weyl spinor in $\WW$ is written as, say,
 $\kappa_A$ with an index $A=0,1$.
 In the abstract notation of ket, a basis in the space $\WW$ is written as $|S^{A}\ra $
 and the ket form of the above spinor is written as $|\kappa\ra$, with the expansion
\begin{equation}\label{|kappa>}
  |\kappa\ra = \kappa_A|S^{A}\ra,
\end{equation}
 with a summation over $A$ implied.
 \footnote{In Ref.\cite{pra16-commu}, different from Eq.(\ref{|kappa>}),
 $|\kappa\ra$ is expanded as $|\kappa\ra = \kappa^A|S_{A}\ra$,
 which is equal to $-\kappa_A|S^{A}\ra$.
 A disadvantage of using the expansion of $|\kappa\ra = \kappa^A|S_{A}\ra$ is that it brings
 a minus sign to the rhs of the two equalities in Eq.(\ref{kappa-A}).
  }

 One may introduce a space that is dual to $\WW$, composed of bras with a basis written as $\la\la  S^A|$.
 This type of bra is called a \emph{scalar-product-based bra} in the main text, 
 because it may form a scalar product with a ket. 
 More exactly, in order to construct a product that is a scalar under $SL(2,C)$ transformations,
 the bra dual to the ket $|\kappa\ra$ should be written as
\begin{equation}\label{<kappa|}
  \la\la  \kappa | =  \la\la  S^{A}|  \kappa_A,
\end{equation}
 which has the same components as $|\kappa\ra$ in Eq.(\ref{|kappa>}),
 but not their complex conjugates.
 (See Appendix \ref{sect-SL2C-transf} for basic properties of $SL(2,C)$ transformations,
 particularly  Eq.(\ref{wwXK=XK}).)

 Scalar products of the basis spinors satisfy
\begin{equation}\label{SA-SB}
  \la\la  S^{A}|S^{B}\ra = \epsilon^{A B},
\end{equation}
 where
\begin{equation}\label{epsilon}
 \epsilon^{AB} = \left( \begin{array}{cc} 0 & 1 \\ -1 & 0 \end{array} \right).
\end{equation}
 It proves convenient to introduce another matrix $\epsilon_{AB}$, which has the same
 elements as $\epsilon^{AB}$.
 These two matrices can be used to raise and lower indexes of components,  say,
\begin{equation}\label{A-raise}
  \kappa^A = \epsilon^{AB} \kappa_B, \quad \kappa_A = \kappa^B \epsilon_{BA},
\end{equation}
 as well as for the basis spinors, namely,
\begin{gather}\label{}
 |S^A\ra = \epsilon^{AB}|S_B\ra, \quad |S_A\ra = |S^B\ra \epsilon_{BA} .
\end{gather}
 It is not difficult to verify that (i) $\la\la  S_{A}|S_{B}\ra = \epsilon_{A B}$; (ii)
\begin{equation}\label{f-AB}
  {f_{\ldots}^{\ \ \ A}\ (g)^{  \cdots }}_{ A } = - {f_{\ldots A}\ (g)^{\cdots A}};
\end{equation}
 and (iii) the symbols $\epsilon_C^{\ \ A} =\epsilon^{BA} \epsilon_{BC}$ and
 $\epsilon^A_{\ \ C} =\epsilon^{AB} \epsilon_{BC}$ satisfy the relation
\begin{gather}\label{eps-delta}
 \epsilon_C^{\ \ A} = -\epsilon^A_{\ \ C} = \delta^A_C,
\end{gather}
 where $\delta^A_B=1$ for $A=B$ and $\delta^A_B=0$ for $A \ne B$.
 (The $\delta$-symbols for other types of labels to be discussed below are defined in the same way.)

 The scalar product of two generic spinors $|\chi\ra$ and $|\kappa\ra$, written as
 $\la\la \chi|\kappa\ra$, has the expression of
\begin{equation}\label{<chi|kappa>}
 \la\la \chi|\kappa\ra = \chi_A \kappa^A.
\end{equation}
 The anti-symmetry of $\epsilon_{AB}$ implies that
\begin{equation}\label{ck=-kc}
  \la\la  \chi |\kappa\ra = -\la\la \kappa |\chi\ra
\end{equation}
 and, as a consequence, $\la\la  \kappa |\kappa\ra =0$ for all $|\kappa\ra$.
 Moreover, we note the following two properties:
 (a) The identity operator in the space $\WW$, denoted by  $I_{\WW}$, is written as
\begin{eqnarray}\label{I}
 I_{\WW} = |S^{A}\ra \la\la  S_{A}|,
\end{eqnarray}
 satisfying $I_{\WW}|\kappa\ra =|\kappa\ra $ for all $|\kappa\ra \in \WW$;
 and (b) the components of $|\kappa\ra$
 have the following expressions,
\begin{equation}\label{kappa-A}
  \kappa^A = \la\la  S^{A}|\kappa\ra, \quad \kappa_A = \la\la  S_{A}|\kappa\ra.
\end{equation}

 An operation of complex conjugation may be introduced,
 which converts $\WW$ to a space denoted by $\ov\WW$.
 $\ov\WW$ is the second representation space of the $SL(2,C)$ group mentioned
 in the beginning of this section.
 This  operation changes spinors $|\kappa\ra$ in $ \WW$
 to spinors in $\ov\WW$, denoted by $|\ov\kappa\ra$.
 Corresponding to the basis $|S_{A}\ra \in \WW$, the space $\ov \WW$ has a basis
 denoted by $|\ov S_{A'}\ra$ with a primed index $A' = 0', 1'$.
 On the basis of $|\ov S_{A'}\ra$, $|\ov\kappa\ra$ is written as
\begin{equation}\label{ov-kappa}
  |\ov\kappa \ra = {\ov\kappa}_{A'}|\ov S^{A'}\ra,
\end{equation}
 where
\begin{equation}\label{}
  \ov\kappa^{A'} := (\kappa^A)^* .
\end{equation}
 Similar to the $\epsilon$ metrices discussed above,
 one introduces metrices $\epsilon^{A'B'}$ and $\epsilon_{A'B'}$,
 which have the same elements as $\epsilon^{AB}$ and are used to raise and lower primed labels.
 The identity operator in the space $\ov\WW$, denoted by $I_{\ov\WW}$,
 has the form of $I_{\ov\WW} = |\ov S^{A'}\ra \la\la  \ov S_{A'}|$.
 When a spinor $\kappa^A$ is transformed by an $SL(2,C)$ matrix,
 the spinor $\ov\kappa^{A'}$ is transformed by the complex-conjugate matrix
 (see Appendix \ref{sect-SL2C-transf}).

 We also consider the direct-product space $\WW \otimes\ov\WW$.
 Basis spinors in this space are written as $|S_{AB'}\ra \equiv |S_A\ra | \ov S_{B'}\ra $.
 As mentioned in the main text, sometimes we write $ |\kappa\ra | \ov \chi\ra $ as $ |\kappa \ov\chi \ra $.
 In this space, in consistency with the well-known anticommutation relation of 
 fermionic states, it is natural to assume that
\begin{align}\label{SAB-commu}
 |S_A\ra | \ov S_{B'}\ra = - |\ov S_{B'}\ra |S_A\ra | . 
\end{align}
 For the space dual to $\WW \otimes\ov\WW$,
 we write the basis spinors as $\la\la  S_{B'A}| \equiv \la\la  \ov S_{B'}| \la\la  S_A|$.

\subsection{Dirac spinors in the abstract notation}\label{sect-Dirac-spinor}

 In this section, we briefly discuss the abstract ket-bra notation for Dirac spinors as solutions to the Dirac equation.
 We write them as combinations of Weyl spinors.
 \footnote{See Ref.\cite{pra16-commu} for more detailed discussions, except the part
 between Eq.(\ref{Vp-uv-app}) and Eq.(\ref{gamma0-gamma-mu-app}).}
% Although the relationship between Dirac spinors and Weyl spinors is
% well known, detailed properties of Weyl spinors are not widely discussed in the literature of Physics.

 As is well known, the Dirac equation for a free electron with a mass $m$
 has two plane-wave solutions labelled by a Lorentz invariant index $r=0,1$, i.e.,
\begin{gather}\label{phi-e}
 \varphi^r_{\rm elec} (x)  = U^r(\bp) e^{-ipx},
\end{gather}
 where $\bp$ indicates a three-momentum and $p$ a four-momentum,
 $p \equiv p^\mu = (p^0,\bp)$  with $\mu=0,1,2,3$,
 satisfying $p^\mu p_\mu =m^2$ with $p^0>0$.
 Here, $U^r(\bp)$ are four-component spinors  satisfying
\begin{equation}\label{stat-DE}
  (\gamma^\mu p_\mu -m) U^r(\bp)=0.
\end{equation}
 In the chiral representation of the $\gamma^\mu$-matrices,
 a four-component Dirac spinor $U^r(\bp)$ is decomposed into two Weyl spinors,
 as its left-handed (LH) part and right-handed (RH) part, respectively
 \cite{Penrose-book,Corson,CM-book,Peskin,Itzy}.
 Specifically, the spinor $U^r(\bp)$ is written as
\begin{gather} \label{Up-uv-app}
 U^r(\bp) = \frac{1}{\sqrt 2} \left( \begin{array}{c} u^{r,A}(\bp) \\ \ov v_{B'}^r(\bp) \end{array} \right).
\end{gather}

 With labels for two-component spinors, the $\gamma^\mu$-matrices are written as
\begin{gather} \label{gamma-mu}
 \gamma^\mu = \left( \begin{array}{cc} 0 & \sigma^{\mu AB'}
 \\ \ov\sigma^{\mu}_{A'B} & 0\end{array}\right),
\end{gather}
 where $\sigma^{\mu AB'}$ are the so-called Enfeld-van der Waerden symbols, in short, \emph{EvdW symbols}
 \cite{Penrose-book,Kim-group,CM-book,Corson,pra16-commu}.
 Note that $\ov\sigma^{\mu}_{A'B}$ indicates the complex conjugate of $\sigma^{\mu}_{AB'}$, namely
 $\ov\sigma^{\mu}_{A'B} =(\sigma^{\mu}_{AB'})^*$.
 A set of  explicit expressions often used for these symbols is written as
\begin{eqnarray}\notag
 \sigma^{0AB'} = \left(\begin{array}{cc} 1 & 0 \\ 0 & 1 \\ \end{array} \right),
 \sigma^{1AB'}  =  \left(\begin{array}{cc} 0 & 1 \\ 1 & 0 \\ \end{array} \right),
 \\ \sigma^{2AB'} = \left(\begin{array}{cc} 0 & -i \\ i & 0 \\ \end{array} \right),
 \sigma^{3AB'}  =  \left(\begin{array}{cc} 1 & 0 \\ 0 & -1 \\ \end{array} \right).
 \label{sigma^AB}
\end{eqnarray}
 The stationary Dirac equation (\ref{stat-DE}) is, then, split into
 two equivalent subequations, namely,
\begin{subequations}\label{Deq-2p}
\begin{eqnarray}
 \label{sv-u-1m} u^{r,A}(\bp) = \frac 1m p_\mu \sigma^{\mu AB'} \ov v_{B'}^r(\bp),
 \\ \label{sv-u-2m} \ov v_{B'}^r(\bp) = \frac 1{m} p_\mu  \ov\sigma^{\mu}_{B'A} u^{r,A}(\bp).
\end{eqnarray}
\end{subequations}

 Similarly, a solution for a free positron with a four-momentum $p$
 ($p^0>0$) is usually written as
\begin{gather}\label{phi-p}
 \varphi^r_{\rm posi} (x)  = V^r(\bp) e^{ipx},
\end{gather}
 where $V^r(\bp)$ satisfies
\begin{equation}\label{stat-DE-V}
  (\gamma^\mu p_\mu +m) V^r(\bp)=0
\end{equation}
 and is written as
\begin{gather} \label{Vp-uv-app}
 V^r(\bp) = \frac{1}{\sqrt 2} \left( \begin{array}{c} u^{r,A}(\bp) \\ -\ov v_{B'}^r(\bp) \end{array} \right).
\end{gather}

 When explicitly writing two-component-spinor labels, special attention should be paid to the symbol $\gamma^0$.
 In fact, this symbol functions in two different ways:
 one as a component of $\gamma^\mu$,
 and the other as an ingredient that appears in some Lorentz-covariant quantities,
 such as $U^\dag \gamma^0 U'$ and $\psi^\dag \gamma^0 \gamma^\mu \psi K_\mu$.
 When playing the second function, this symbol can not take the expression of
 $\gamma^\mu$ in Eq.(\ref{gamma-mu}) with $\mu=0$.
 Indeed, e.g., doing this would lead to the following expression for $\gamma^0 \gamma^\mu $,
\begin{gather}\label{sig-0AB'-temp}
 \left( \begin{array}{cc} \sigma^{0 AB'}  \ov\sigma^{\mu}_{B'C} & 0 \\
  0& \ov\sigma^{0}_{A'B} \sigma^{\mu BC'}  \end{array}\right),
\end{gather}
 which implies that the spinor part of the product $\psi^\dag \gamma^0 \gamma^\mu \psi K_\mu$
 would contain a term like
%\begin{gather}\label{}\notag
 $\ov u^{r,A'}(\bp) \sigma^{0 AB'}  \ov\sigma^{\mu}_{B'C}  u^{s,C}(\bq) K_\mu$;
%\end{gather}
 the point lies in that this term is not Lorentz invariant due to the two labels $A'$ and $A$.

 In fact, in the second function discussed above, the sole role of $\gamma^0$ is to
 exchange positions of the LH and RH parts of Dirac spinors.
 For the sake of clearness in presentation,
 we write $\gamma_c^0$ for $\gamma^0$ in this case.
 It has the following matrix expression, without involving any two-component-spinor label,
\begin{gather}  \label{gamma-c}
 \gamma_c^0 = \left( \begin{array}{cc} 0 & 1 \\ 1 & 0 \end{array} \right).
\end{gather}
 Then, the matrix product $\gamma^0 \gamma^\mu$ used in the interaction Hamiltonian has the following form,
\begin{gather}
 \gamma_c^0 \gamma^\mu =
 \left( \begin{array}{cc} \ov\sigma^{\mu}_{B'A} & 0 \\ 0 & \sigma^{\mu BA'} \end{array} \right).
  \label{gamma0-gamma-mu-app}
\end{gather}

 Below, we discuss the abstract notation.
 In this notation, the Weyl spinors $u^{r,A}(\bp)$ and $ \ov v_{B'}^r(\bp)$ are written as
\begin{subequations} \label{|uv>}
\begin{align}\label{|u>-uA}
 & |u^r(\bp)\ra = u^{r}_{A}(\bp)|S^A\ra = -u^{r,A}(\bp)|S_A\ra,
 \\ & |\ov v^r(\bp)\ra = \ov v^{r}_{B'}(\bp)|S^{B'}\ra.
\end{align}
\end{subequations}
They satisfy the following relations,
\begin{subequations} \label{|u-vp>}
\begin{align}
\label{uu-eps}
 & \la\la  u^r(\bp) |u^{s}(\bp)\ra =  \la\la  \ov v^{r}(\bp)| \ov v^{s}(\bp)\ra = \epsilon^{rs}.
 \\ & \la\la  v^r(\bp) |u^{s}(\bp)\ra = \delta^{rs}. \label{vu-eps}
\end{align}
\end{subequations}
 The Dirac spinors $U^r(\bp)$ and $V^r(\bp)$ in ket are written as
\begin{subequations} \label{|U-Vp>}
\begin{align}\label{}
 |U^r(\bp)\ra = \frac{1}{\sqrt 2} \left( \begin{array}{c} |u^r(\bp) \ra \\ |\ov v^r(\bp)\ra \end{array} \right),
 \\ |V^r(\bp)\ra = \frac{1}{\sqrt 2} \left( \begin{array}{c} |u^r(\bp) \ra \\ -|\ov v^r(\bp)\ra \end{array} \right).
\end{align}
\end{subequations}
 To be consistent with the scalar-product-based bra in Eq.(\ref{<kappa|})
 for Weyl spinors,
 scalar-product-based bras corresponding to the above two Dirac kets should be written as
\begin{subequations}\label{<U-Vp|}
\begin{gather}
 \la\la  U^r(\bp)| = \frac{1}{\sqrt 2} \left(  \la\la  u^r(\bp)| , \la\la \ov v^r(\bp)| \right),
 \\ \ \la\la  V^r(\bp)| =  \frac{1}{\sqrt 2} \left( \la\la  u^r(\bp)| , -\la\la  \ov v^r(\bp)| \right),
\end{gather}
\end{subequations}
 without taking complex conjugation for the two-component spinors.
 Direct derivation shows that
\begin{gather}\label{<Ur|Us>}
 \la\la  U^r(\bp)|U^s(\bp)\ra = \la\la  V^r(\bp)|V^s(\bp)\ra = \epsilon^{rs}.
\end{gather}
 The complex conjugates of $|U^r(\bp)\ra$ and $\la\la  U^r(\bp)|$ are written as
\begin{subequations}\label{|wh-U>}
\begin{gather}
 |\ov U^r(\bp)\ra =\frac{1}{\sqrt 2} \left( \begin{array}{c} |\ov u^r(\bp) \ra \\ |v^r(\bp)\ra \end{array} \right),
 \\  \la\la  \ov U^r(\bp)| =\frac{1}{\sqrt 2} \left( \la\la \ov u^r(\bp)| , \la\la  v^r(\bp)| \right),
\end{gather}
\end{subequations}
 and similar for $|V^r(\bp)\ra$ and $\la\la  V^r(\bp)|$.

 Although a product $\la\la U(\bp)|U'(\bp)\ra$ is a Lorentz scalar, it is not an inner product,
 because the procedure of taking a scalar-product-based bra does not involve complex conjugation. 
 In fact, Eq.(\ref{<Ur|Us>}) implies that $\la\la U(\bp)|U(\bp)\ra =0$.
 In the ordinary notation, the inner product of two Dirac spinors is written as, say, $U^\dag \gamma^0 U'$.
% in this paper, we write it as $\wh U U'$, where
%\begin{gather} \label{whU-whV}
% \wh U := U^{\dag } \gamma^0 .
%\end{gather} 
 As shown in Ref.\cite{pra16-commu},
 to write the inner product in the abstract notation, one may make use of the following matrix $\gamma_c$,
\begin{equation} \label{gamma-c-app}
 \gamma_c =  \left( \begin{array}{cc} 0 & -1 \\ 1 & 0 \end{array} \right),
\end{equation} 
 and introduce \emph{hat-bras} as defined below,
 \footnote{In fact, the hat-bra $\la \wh U(\bp)|$ corresponds to $U^\dag(\bp) \gamma^0$
 in the ordinarily used notation.}
\begin{subequations} \label{<whUV|-app}
\begin{gather}\label{wh-U-app}
  \la\wh U(\bp)| := \la\la \ov U(\bp)|  \gamma_c =  (\la\la v(\bp)|,-\la\la \ov u(\bp)|),
 \\ \label{wh-V-app}   \la\wh V(\bp)| := \la\la \ov V(\bp)|  \gamma_c =  (-\la\la v(\bp)|,-\la\la \ov u(\bp)|).
\end{gather}
\end{subequations}
 It is straightforward to check the following scalar products,
\begin{subequations} \label{|UV-IP>}
\begin{gather}
 \label{UrUs}   \la \wh U^{r}(\bp)|U^{s}(\bp)\ra = \delta^{rs},
 \\ \la {\wh{V}^{r}}(\bp)|V^{s}(\bp)\ra = -\delta^{rs},\label{VrVs}
 \\ \la {\wh{U}^{r}}(\bp)|V^{s}(\bp)\ra = 0,
 \\ \la {\wh{V}^{r}}(\bp)|U^{s}(\bp)\ra = 0.
\end{gather}
\end{subequations}
 It is seen that $\la \wh U^{r}(\bp)|U^{s}(\bp)\ra$ for electron is an inner product.
 While, $\la {\wh{V}^{r}}(\bp)|V^{s}(\bp)\ra$ for positron is a negative inner product,
 due to the minus sign on the rhs of Eq.(\ref{VrVs}).

 One remark: In fact, functions of $\gamma^0_c$ and $\gamma_c$ are essentially equivalent.
 Their difference lies in that $\gamma^0_c$ is used when Dirac spinors 
 are written in the component notation,
 while, $\gamma_c$ is used in the abstract notation.

 To introduce Dirac spinors with indices written in the lower position, due to the relation in Eq.(\ref{UrUs}),
 one may do in the ordinary way.
 That is,  a label $r$ in the upper position is lowered  by $\delta_{rs}$;
 reversely, $r$ in the lower position is raised by $\delta^{rs}$.
 Explicitly, one writes
\begin{gather}\label{r-position-change}
 |U_{s}(\bp)\ra = |U^{r}(\bp)\ra \delta_{rs}, \quad |U^{s}(\bp)\ra = \delta^{sr} |U_{r}(\bp)\ra.
\end{gather}
 For the sake of consistency, lower labels of the spinors $|V\ra$ should be defined in the same way
 as in Eq.(\ref{r-position-change}).
 Thus, one gets that
\begin{gather}\label{U^rU_s}
 \la \wh U^{r}(\bp)|U_{s}(\bp)\ra = \delta^{r}_{s},
 \quad \la \wh V^{r}(\bp)|V_{s}(\bp)\ra = -\delta^{r}_{s}.
\end{gather}
 Making use of  Eq.(\ref{U^rU_s}),
 it is easy to verify that the identity operator on the four-dimensional
 space of Dirac spinors, denoted by $I_\D$, has the following expression,
\begin{gather}\label{ID-UV}
 I_\D = |U^{r}(\bp)\ra \la \wh U_{r}(\bp)| - |V^{r}(\bp)\ra \la {\wh{V}_{r}}(\bp)|.
\end{gather}

 \subsection{Basic properties of four-component vectors}\label{sect-recall-vector}

 In this section, we recall basic properties of four-component vectors given in the theory of spinors
 \cite{Penrose-book,Corson,CM-book}.
 We use the ordinary notation in this section and
 will discuss the abstract notation in the next section.

 A basic point is a one-to-one mapping, given by the EvdW symbols discussed above,
 between the direct-product
 space $\WW \otimes\ov\WW$ and a four-dimensional space denote by $\VV$.
 For example,  a spinor $\phi_{AB'}$ in the space
 $\WW \otimes\ov\WW$ is mapped to a vector $K^\mu$ in the space $\VV$ by
\begin{equation}\label{map-WW-V}
  K^\mu = \sigma^{\mu AB'} \phi_{AB'}.
\end{equation}
 In the space $\VV$, of particular importance is a symbol denoted by $g^{\mu\nu}$, which is defined
 by the following relation to the $\epsilon$-symbols discussed previously,
\begin{equation}\label{g-sig}
 g^{\mu\nu} =  \sigma^{\mu AB'} \sigma^{\nu CD'} \epsilon_{AC} \epsilon_{B'D'}.
\end{equation}
 One may introduce a lower-indexed symbol $g_{\mu\nu}$,
 which has the same matrix elements as $g^{\mu\nu}$, namely, $[g^{\mu\nu}] = [g_{\mu\nu}]$.
 These two symbols $g$, like the symbols $\epsilon$ for the space $\WW$,
 may be used to raise and lower indexes, e.g.,
\begin{equation}\label{mu-raise}
  K_\mu =  K^\nu g_{\nu \mu}, \quad K^\mu = g^{\mu\nu} K_\nu.
\end{equation}
 Making use of the antisymmetry of the symbol $\epsilon$,
 it is easy to verify that $g^{\mu\nu}$ is symmetric, i.e.,
\begin{equation}\label{g-sym}
 g^{\mu\nu} = g^{\nu\mu}.
\end{equation}
 Due to this symmetry, the upper/lower positions of repeated
 indexes ($\mu$) are exchangeable, namely
\begin{equation}\label{f-munu}
  {F_{\ldots}^{\ \ \ \mu}(f)^{  \cdots }}_{\mu } = {F_{\ldots \mu}(f)^{\cdots \mu}}.
\end{equation}

 The EvdW symbols have the following properties,
\begin{equation}\label{st-delta}
 \sigma^{AB'}_\mu \sigma_{CD'}^\mu = \delta^{AB'}_{CD'}, \quad
 \sigma_{AB'}^\mu \sigma^{AB'}_\nu  = \delta_\nu^\mu,
\end{equation}
 where  $ \delta^{AB'}_{CD'} :=  \delta^{A}_{C} \delta^{B'}_{D'}$.
 Making use of the relations in Eq.(\ref{st-delta}), it is not difficult to check that the map
 from $\WW\otimes\ov\WW$ to $\VV$ given in Eq.(\ref{map-WW-V}) is reversible.
 Moreover, using Eq.(\ref{eps-delta}), one finds that
\begin{gather}\label{ss-ee-2}
 \sigma_{\mu AB'} \sigma_{CD'}^\mu = \epsilon_{AC} \epsilon_{B'D'}.
\end{gather}
 Then, substituting the definition of $g^{\mu\nu}$ in Eq.(\ref{g-sig})
 into the product $g^{\mu\nu} g_{\nu \lambda}$, after simple algebra, one gets that
\begin{gather} \label{ggd}
 g^{\mu\nu} g_{\nu \lambda} = g^\mu_{\ \ \lambda} = g^{\ \ \mu}_{\lambda} = \delta^\mu_\lambda .
\end{gather}

 When an $SL(2,C)$ transformation is carried out on the space $\WW$,
 a related transformation should be applied to the space $\VV$.
 Requiring invariance of the EvdW symbols,
 transformations on the space $\VV$ can be fixed, which turn out to constitute a (restricted) Lorentz group
 and the space $\VV$ is a four-component vector space
 (see Appendix \ref{sect-SL2C-transf}).
 In fact, substituting the explicit expressions of the EvdW symbols in Eq.(\ref{sigma^AB})
 into Eq.(\ref{g-sig}), one gets
\begin{eqnarray}\label{g-munu}
 g^{\mu\nu} = \sigma^{\mu}_{AB'} \sigma^{\nu AB'}
 =\left(\begin{array}{cccc} 1 & 0 & 0 & 0 \\ 0 & -1 & 0 & 0 \\ 0&0 & -1 &0
 \\ 0 &0 &0 & -1 \end{array} \right),
\end{eqnarray}
 which is just the Minkovski's metric.

 As shown in Appendix \ref{sect-SL2C-transf}  [Eq.(\ref{<K|J>})], the following product, 
\begin{gather}\label{Kmu-Jmu}
  J_\nu K^\nu = J^\mu g_{\mu\nu} K^\nu,
\end{gather}
 is a scalar under Lorentz transformations.
 Physically, of more interest is a product, in which one of the two vectors takes a complex-conjugate form, say, 
\begin{gather}\label{Kmu-Jmu*}
 J^*_\nu K^\nu = J^{\mu *} g_{\mu\nu} K^\nu.
\end{gather}
 Similarly,  one finds that this product is also a scalar.

 \subsection{Abstract notation for four-component vectors}\label{sect-vector-abstract}

 In the abstract notation of ket,
 a basis in the space $\VV$ is written as $|T_\mu\ra $.
 The index of the basis may be raised by $g^{\mu\nu}$,
 i.e., $|T^\mu\ra = g^{\mu\nu} |T_\nu\ra $,  and similarly $ |T_{\mu}\ra = g_{\mu \nu}|T^\nu\ra $.
 A generic four-component vector $|K\ra$ in the space $\VV$ is expanded as
\begin{equation}\label{|K>-expan}
 |K\ra = K_\mu |T^\mu\ra = K^\mu |T_\mu\ra.
\end{equation}
 In consistency with the scalar-product-based bra in Eq.(\ref{<kappa|}),
 the scalar-product-based bra corresponding to $|K\ra$ is written as
\begin{gather}\label{<K|}
 \la\la  K| = \la\la  T_\mu| K^\mu .
\end{gather}
 Similar to the case of Weyl spinors in Eq.(\ref{SA-SB}), one requires that
\begin{eqnarray}\label{Tmu-Tnu}
  \la\la  T_\mu|T_\nu\ra =  g_{\mu\nu}.
\end{eqnarray}
 Then, it is easy to check that the scalar product $J_\nu K^\nu$ is written as
 $\la\la  J|K\ra$, namely,
\begin{gather}\label{}
 \la\la  J|K\ra = J_\nu K^\nu.
\end{gather}

 It is not difficult to verify the following properties.
 (i) Making use of Eq.(\ref{ggd}), one finds that the identity operator
 in the space $\VV$, denoted by $I_{\VV}$, is written as
\begin{equation}\label{I-vector}
 I_{\VV} = |T_\mu\ra \la\la  T^\mu | = |T^\mu\ra \la\la  T_\mu|.
\end{equation}
 (ii) The components $K^\mu$ and $K_\mu$ have the following expressions,
\begin{equation}\label{K-mu}
  K^\mu = \la\la  T^\mu |K\ra , \quad K_\mu = \la\la  T_\mu |K\ra .
\end{equation}
 And (iii) the symmetry of $g^{\mu\nu}$ implies that $\la\la  T_\mu|T_\nu\ra = \la\la  T_\nu|T_\mu\ra$,
 as a result,
\begin{eqnarray}\label{<K|J>=<J|K>}
  \la\la  K|J\ra = \la\la  J|K\ra.
\end{eqnarray}

 Since $\ov{|S_{AB'}\ra} =|S_{A'B}\ra= -|S_{BA'}\ra$, with $|S_{AB'}\ra \equiv |S_A\ra | \ov S_{B'}\ra $,
 the operation of complex conjugation
 maps the space $\VV$ into itself.
 We use $\ov{|T_\mu\ra }$ to denote  the complex conjugate of $|T_\mu\ra $.
 Since $\ov{|T_\mu\ra }$ and $|T_\mu\ra $ lie in the same space,
 it is unnecessary to introduce any change to the label $\mu$.
 Hence, $\ov{|T_\mu\ra }$ can be written as $|\ov T_\mu\ra $ with the
 label $\mu$ unchanged.

 It proves convenient to introduce an operator related to the EvdW symbols,
 denoted by $\sigma$, namely,
\begin{equation}\label{sigma-app}
 \sigma :=  |T_\mu\ra \sigma^{\mu AB'} \la\la  S_{B'A}|.
\end{equation}
 This operator $\sigma$ has a simple geometric meaning; that is,
 it maps a product space $\WW \otimes\ov\WW$ to a vector space $\VV$.
 Using $\ov\sigma$ to indicate the complex conjugate of $\sigma$, one has
\begin{equation}\label{ov-sigma}
 \ov\sigma =  |\ov T_\mu\ra \ov\sigma^{\mu A'B} \la\la  S_{BA'}|,
\end{equation}
 where $\ov\sigma^{\mu A'B} \equiv (\sigma^{\mu AB'})^*$.
 Making use of the explicit expressions of the EvdW symbols in Eq.(\ref{sigma^AB}),
 it is easy to verify that
\begin{equation}\label{sig-c}
 \ov\sigma_{\mu}^{B'A}= \sigma_{\mu}^{AB'}.
\end{equation}

 There is some freedom in the determination of the relationship between $\sigma$ and $\ov \sigma$
 and, relatedly, between $|T_\mu\ra$ and $|\ov T_\mu\ra$.
 The simplest assumption is that $|T_\mu\ra $ is ``real'', i.e.,
\begin{equation}\label{ovT=T}
 {|\ov T_\mu\ra }= |T_\mu\ra .
\end{equation}
 Making use of Eqs.(\ref{sigma}), (\ref{SAB-commu}), and (\ref{ov-sigma})-(\ref{sig-c}),
 it is not difficult to verify that
\begin{equation}\label{ov-s=s}
 \ov \sigma = -\sigma.
\end{equation}
 Thus, the complex conjugates of $|K\ra$ and $\la\la  K|$ are written as
\begin{gather}\label{|ovK>}
 |\ov K\ra = K^{\mu *} |T_\mu\ra \quad \&
 \quad \la\la  \ov K | = \la\la  T_\mu| K^{\mu *}.
\end{gather}
%\end{equation}
 Then, the scalar product in Eq.(\ref{Kmu-Jmu*}) is written as
\begin{gather}\label{<Kmu-Jmu*>}
 \la\la  \ov J |K\ra = J^*_{\mu} K^\mu.
\end{gather}
 It is easy to verify that
\begin{gather}\label{JK-KJ*}
 \la\la \ov K|J\ra ^* = \la\la \ov J|K\ra.
\end{gather}

% that makes transformation between kets and bras,
% which we call \emph{transposition} and denote by a superscript $T$.
% Specifically, the operation is defined by
%\begin{subequations}\label{transpose}
%\begin{gather}
% |\phi \psi\ra^T := \la\la  \psi \phi|, \ \la\la  \psi \phi|^T :=  |\phi \psi\ra, \
%\\ (|\phi\ra \la\la  \psi|)^T := |\psi\ra \la\la  \phi |.
%\end{gather}
%\end{subequations}
% Note that, e.g., the transposition of $|S_{AB'}\ra$ is $\la\la  S_{B'A}|$, but not $\la\la  S_{AB'}|$.
 We use $\sigma^T$ to denote the transposition of  $\sigma$,
\begin{equation}\label{sigma-T}
 \sigma^{T} := |S_{AB'}\ra \sigma^{\mu AB'} \la\la  T_\mu|.
\end{equation}
 Computing the product of $\sigma$ and $\sigma^T$,
 with the help of Eqs.(\ref{Tmu-Tnu}), (\ref{st-delta}), and (\ref{I}),  it is easy to verify that
\begin{equation}\label{s-T-s}
  \sigma^T\sigma = \sigma\sigma^T =I,
\end{equation}
 which implies that $\sigma^T$ is the reverse of $\sigma$.
% We use $\sigma^\dag$ to denote the complex conjugate of the transposition of $\sigma$, namely,
%\begin{gather}\label{sigma-dag}
% \sigma^\dag := \ov\sigma^T.
%\end{gather}

\section{SL(2,C) transformations and Lorentz transformations}\label{sect-SL2C-transf}

 In this appendix, we recall the relation between SL(2,C) transformations and Lorentz transformations.
 Particularly, when SL(2,C) transformations are carried out on a
 space $\WW$, the corresponding transformations on the space $\VV$ are Lorentz transformations.

 We recall that
 the group SL(2,C) is composed of $2\times 2$ complex matrices with unit determinant
 \cite{Penrose-book,CM-book,Corson,pra16-commu,Kim-group}, written as
\begin{equation}\label{h-AB}
  h^{A}_{\ \ B} = \left( \begin{array}{cc} a & b \\ c & d \end{array} \right)
  \quad \text{with} \  ad-bc=1.
\end{equation}
 Under a transformation given by $h^{A}_{\ \ B}$,
 a two-component spinor $\kappa^A$ is transformed to
\begin{equation}\label{}
  \ww \kappa^A = h^{A}_{\ \ B} \kappa^B,
\end{equation}
 where we use tilde to indicate the result of a SL(2,C) transformation.

 It is straightforward to verify that $\epsilon^{AB}$ is
 invariant under SL(2,C) transformations, that is,
 $\ww\epsilon^{AB}= h^{A}_{\ \ C} h^{B}_{\ \ D} \epsilon^{CD}$
 has the same matrix form as $\epsilon^{AB}$ in Eq.(\ref{epsilon}).
 Direct computation can verify the following relations,
\begin{eqnarray} \label{h-property-1}
 & h^A_{\ \ B} h_{C}^{\ \ B} = h_{B}^{\ \ A} h^B_{\ \ C} = -\epsilon_C^{\ \ A}.
\\ & h_{AD} h^A_{\ \ C}  = \epsilon_{DC}, \quad h^A_{\ \ B} h^{C B} = \epsilon^{AC}.
\label{h-property-2}
\end{eqnarray}
 It is not difficult to verify that the product $\chi_A \kappa^A$ is a scalar product,
 that is,
\begin{align}\label{wwXK=XK}
 \ww\chi_A \ww\kappa^A = \chi_A \kappa^A.
\end{align}

 When $\kappa^A$ is transformed by a matrix $h^{A}_{\ \ B}$,
 $\ov\kappa^{A'}$ is transformed by its complex-conjugate matrix, namely,
\begin{equation}\label{}
  \ww {\ov\kappa}^{A'} = \ov h^{A'}_{\ \ B'} \ov\kappa^{B'},
\end{equation}
 where
\begin{equation}\label{}
 \ov h^{A'}_{\ \ B'} := (h^{A}_{\ \ B})^*.
%  \quad \text{with} \ A=A', B= B'.
\end{equation}

 Now, we discuss relationship between SL(2,C) transformations and Lorentz transformations.
 Related to a SL(2,C) transformation $h^{A}_{\ \ B}$ performed on a space $\WW$,
 we use $\Lambda^\mu_{\ \ \nu}$ to denote the corresponding transformation on the space $\VV$,
\begin{equation}\label{ww-K}
   \ \ww K^\mu = \Lambda^\mu_{\ \ \nu} K^\nu .
\end{equation}
 It proves convenient to require invariance of the EvdW symbols under SL(2,C) transformations,
 namely,
\begin{equation}\label{wwsig=sig}
 \ww \sigma^{\mu A'B} = \sigma^{\mu A'B},
\end{equation}
 where
\begin{gather}\label{ww-sig}
 \ww \sigma^{\mu A'B} = \Lambda^\mu_{\ \ \nu} \ov h^{A'}_{\ \ C'}
 h^{B}_{\ \ D} \sigma^{\nu C'D}.
\end{gather}
 This requirement can fix the form of $\Lambda^\mu_{\ \ \nu}$.
 In fact, substituting Eq.(\ref{ww-sig})
 into Eq.(\ref{wwsig=sig}) and rearranging the positions of some labels, one gets
\begin{gather}\label{sigma-Lam-int1}
 \sigma^{\mu}_{\ A'B} = \Lambda^\mu_{\ \ \nu} \ov h_{A' C'} h_{B D} \sigma^{\nu C'D}.
\end{gather}
 Multiplying both sides of Eq.(\ref{sigma-Lam-int1})
 by $ \ov h^{A'}_{\ \ E'}h^B_{\ \ F} \sigma_{\nu}^{E'F}$, the rhs gives
\begin{gather}\label{}\notag
 \Lambda^\mu_{\ \ \eta} \ov h_{A' E'} h_{B F} \sigma^{\eta E'F}
  \ov h^{A'}_{\ \ C'} h^B_{\ \ D} \sigma_{\nu}^{C'D}
  \\ = \Lambda^\mu_{\ \ \eta} \epsilon_{E'C'} \epsilon_{FD} \sigma^{\eta E'F} \sigma_{\nu}^{C'D}
  = \Lambda^\mu_{\ \ \nu},
\end{gather}
 where Eq.(\ref{h-property-2}) and Eq.(\ref{st-delta}) have been used.
 Then, one gets the following expression for $\Lambda^\mu_{\ \ \nu}$,
\begin{equation}\label{Lam-s-h}
  \Lambda^\mu_{\ \ \nu} = \sigma^{\mu}_{A'B}  \ov h^{A'}_{\ \ C'} h^B_{\ \ D} \sigma_{\nu}^{C'D}.
\end{equation}

 Substituting  Eq.(\ref{Lam-s-h}) into the product
 $\Lambda^\mu_{\ \ \eta} \Lambda^\nu_{\ \ \xi} g^{\eta\xi}$, one gets
\begin{gather*}
 \sigma^{\mu}_{A'B}  \ov h^{A'}_{\ \ C'} h^B_{\ \ D} \sigma_{\eta}^{C'D}
 \sigma^{\nu}_{E'F}  \ov h^{E'}_{\ \ G'} h^F_{\ \ H} \sigma_{\xi}^{G'H} g^{\eta\xi}.
\end{gather*}
 Using Eq.(\ref{ss-ee-2}), this gives
\begin{gather*}
 \sigma^{\mu}_{A'B}  \ov h^{A'}_{\ \ C'} h^B_{\ \ D}
 \sigma^{\nu}_{E'F}  \ov h^{E'C'} h^{FD}.
\end{gather*}
 Then, noting Eqs.(\ref{h-property-2}) and (\ref{g-sig}), one gets the first equality in
 the following relations,
\begin{gather}\label{LLg=g}
 \Lambda^\mu_{\ \ \eta} \Lambda^\nu_{\ \ \xi} g^{\eta\xi} = g^{\mu\nu},
 \quad \Lambda^\mu_{\ \ \eta} \Lambda^\nu_{\ \ \xi} g_{\mu \nu} = g_{\eta \xi}.
\end{gather}
 The second equality in (\ref{LLg=g}) can be proved in a similar way.
 Therefore, the transformations $\Lambda^\mu_{\ \ \nu}$ constitute the
 (restricted) Lorentz group
 and the space $\VV$ is composed of four-component vectors.

 The transformations $\Lambda$ and the matrix $g$ have the following properties.
 (i) The inverse transformation of $\Lambda^\mu_{\ \ \nu}$, denoted by $\Lambda^{-1}$
 has the simple expression,
\begin{equation}\label{Lambda-1}
 (\Lambda^{-1})^\nu_{\ \ \mu} = \Lambda_\mu^{\ \ \nu} \Longleftrightarrow
 (\Lambda^{-1})_{\nu \mu} = \Lambda_{\mu \nu}.
\end{equation}
 In fact, substituting Eq.(\ref{Lam-s-h}) into the product
 $\Lambda^\mu_{\ \ \nu} \Lambda_\lambda^{\ \ \nu}$ and making use of
 Eqs.(\ref{h-property-1}), (\ref{st-delta}), and (\ref{f-AB}), it is straightforward
 to verify Eq.(\ref{Lambda-1}).

 (ii) Equation (\ref{LLg=g}) implies that the matrix
 $g^{\mu\nu}$ is invariant under the transformation $\Lambda$, that is,
\begin{equation}\label{wwg=g}
  \ww g^{\mu\nu} = g^{\mu\nu}.
\end{equation}

 (iii) The product $K^\mu g_{\mu\nu} J^\nu = K_\mu J^\mu$ is a scalar under the
 transformation $\Lambda$, i.e.,
\begin{equation}\label{<K|J>}
 \ww K_\mu \ww J^\mu = K_\mu J^\mu,
\end{equation}
 which can be readily proved making use of Eq.(\ref{LLg=g}).

 (iv) Making use of Eq.(\ref{sig-c}), it is straightforward to show that
the transformation $\Lambda $ is real, namely,
\begin{equation}\label{real-Lam}
 \Lambda^\mu_{\ \ \nu} = (\Lambda^\mu_{\ \ \nu})^*.
\end{equation}
 Then, it is easy to check that $K^*_\mu J^\mu$ is also a scalar product.

\section{Negative-$p^0$ solutions of Dirac equation and their relationship to positive-$p^0$ solutions}
\label{app-neg-E-UV}

 In this appendix, we discuss properties of negative-$p^0$ stationary solutions of the Dirac equation,
 which are used in the main text.

 Let us first recall properties of stationary solutions of the Dirac equation with positive $p^0$. 
 Two types of such solutions have been 
 discussed in  Appendix \ref{sect-Dirac-spinor}, 
 one taking the form of $\psi(x) = U(\bp) e^{-ipx}$ [Eqs.(\ref{phi-e})-(\ref{stat-DE})]
 and the other of $\psi(x) = V(\bp) e^{ipx}$ [Eqs.(\ref{phi-p})-(\ref{stat-DE-V})].
 In the rest frame of reference, the above spinors $U(\bp)$ and $V(\bp)$ are written as
\begin{gather}\label{U0-V0}
 U_0 = \left( \begin{array}{c} \xi^A \\ \ov\eta_{B'} \end{array} \right)
 \ \ \text{and} \ \ V_0 = \left( \begin{array}{c} \xi^A \\ -\ov\eta_{B'} \end{array} \right),
\end{gather}
 respectively.
 Changing to a reference frame in which the particle moves with a momentum $\bp$,
 its four-momentum $(m,0,0,0)$ is changed to $p=(p^0,\bp)$ ($p^0>0$), meanwhile,
 the spinors $U_0$ and $V_0$ are changed to $U(\bp)$ and $V(\bp)$, respectively,
 by a Lorentz transformation $\Lambda(\bp)$, with
\begin{gather}\label{U-uv-Lamb}
 U(\bp) = \Lambda(\bp) U_0 =
 \frac{1}{\sqrt 2} \left( \begin{array}{c} u^A(\bp) \\ \ov v_{B'}(\bp) \end{array} \right),
 \\ V(\bp) = \Lambda(\bp) V_0 =
 \frac{1}{\sqrt 2} \left( \begin{array}{c} u^A(\bp) \\ -\ov v_{B'}(\bp) \end{array} \right). \label{V-uv-Lamb}
\end{gather}

 A negative-$p^0$ solution of the $U$-type discussed above
 is written as $\psi(x) = U_-(\bp) e^{-ipx}$ with $p_0 <0$.
 Its spinor part $U_-(\bp)$ also satisfies Eq.(\ref{stat-DE}).
 Straightforward derivation shows that,
 in the rest frame in which the particle has a four-momentum $(-m,0,0,0)$,
 the solution takes the form
\begin{gather}\label{}
 U_{0-} = \left( \begin{array}{c} \xi^A \\ -\ov\eta_{B'} \end{array} \right).
\end{gather}
 The Lorentz transformation, which brings the four-momentum $(-m,0,0,0)$ to $p=(-|p^0|,\bp)$,
 should bring $(m,0,0,0)$ to $(|p^0|,-\bp)$.
 Hence,  for the spin degree of freedom, this transformation is
 written as $\Lambda(-\bp)$, bringing $\xi^A$ to
 $u^A(-\bp)$ and $\ov\eta_{B'}$ to $\ov v_{A'}(-\bp)$.
 Then,
\begin{gather}
 U_-(\bp) = \Lambda(-\bp) U_{0-} =
 \frac{1}{\sqrt 2} \left( \begin{array}{c} u^A(-\bp) \\ -\ov v_{B'}(-\bp) \end{array} \right).
\end{gather}
 Comparing with Eq.(\ref{V-uv-Lamb}), it is seen that
\begin{gather}\label{U-(-p)-V-app}
 U_-(-\bp) = V(\bp).
\end{gather}

 Following arguments similar to those given above, one may discuss 
 negative-$p^0$ solutions of the $V$-type, i.e., $\psi(x) = V_-(\bp) e^{ipx}$ with $p_0 <0$.
 One finds that
\begin{gather}\label{V-(-p)-U-app}
 V_-(-\bp) = U(\bp).
\end{gather}

 In the abstract notation and with the label $r$ written explicitly, 
 the above-discussed Dirac spinors are written as
\begin{subequations}\label{|UV-mimus-p>}
\begin{align}\label{}
% |U(\bp)\ra = \left( \begin{array}{c} |u(\bp) \ra \\ |\ov v(\bp)\ra \end{array} \right),
 & |U^r_-(\bp)\ra = \frac{1}{\sqrt 2} \left( \begin{array}{c} |u^r(-\bp) \ra \\ -|\ov v^r(-\bp)\ra \end{array} \right),
% |V(\bp)\ra = \left( \begin{array}{c} |u(\bp) \ra \\ -|\ov v(\bp)\ra \end{array} \right),
 \\ & |V^r_-(\bp)\ra = \frac{1}{\sqrt 2} \left( \begin{array}{c} |u^r(-\bp) \ra \\ |\ov v^r(-\bp)\ra \end{array} \right).
\end{align}
\end{subequations}

\section{Hat-bras for Dirac and Weyl spinors}\label{app-hatbra-spinors}

 In this appendix, we write hat-bras of Dirac spinors in a form 
 which is valid for both signs of $\varrho$.
 We also discuss hat-bras for Weyl spinors.

 Let us first discuss Dirac spinors. 
 The scalar-product-based bra of the Dirac ket $|U^r_\varrho(\bp)\ra$ in Eq.(\ref{Ur-bp}) is written as
\begin{align}\label{<<Ur+|}
 \la\la U^r_\varrho (\bp)| = \frac{1}{\sqrt 2} \Big( \la\la u^r(\varrho \, \bp)| , 
 \varrho \, \la\la \ov v^r(\varrho \, \bp)|\Big).
\end{align}
 It is impossible to get a scalar product of $\la\la \ov U^r_\varrho(\bp)|$
 (the complex conjugate of $\la\la U^r_\varrho(\bp)|$)
 and  $|U^s_\varrho(\bq)\ra$.
 In fact, neither their first layers nor their second layers lie in a same Weyl-spinor space.
 For example, the first layers of $\la\la \ov U^r_\varrho(\bp)|$ and $|U^s_\varrho(\bq)\ra$ 
 lie in $\ov\WW$ and $\WW$, respectively.
 To solve this problem, as done in the ordinary construction of inner product for Dirac spinors,
 one may make use of the matrix $\gamma_c$ in Eq.(\ref{gamma-c-app}),
 whose basic role is to exchanges positions of the two layers in a Dirac spinor. 
 Thus, one gets the hat-bra $\la \wh U^r_\varrho(\bp)|$
 [cf.~Eq.(\ref{wh-U-app})],
\begin{gather}\label{wh-U-rho}
  \la\wh U^r_\varrho (\bp)| =  \la\la \ov U^r_\varrho(\bp)|  \gamma_c , 
\end{gather}
 or, explicitly, 
\begin{gather}\label{wh-U-explicit}
  \la\wh U^r_\varrho (\bp)|  = \frac{1}{\sqrt 2} (\varrho \, \la\la \, v^r( \varrho \bp)|,
  -\la\la  \, \ov u^r( \varrho \bp)|).
\end{gather}
 It is easy to check that 
 \begin{align}
 \label{UrUs-main}   \la \wh U^{r}_\varrho(\bp)|U^{s}_\varrho(\bp)\ra 
 = \varrho \, \delta^{rs}.
\end{align}
 One sees that $\la\wh U^s_+ (\bp)| U^r_+ (\bp)\ra$ is the well-know inner product
 of Dirac spinors, while, $\la\wh U^s_- (\bp)| U^r_- (\bp)\ra$ is a negative inner product.

 Next, we discuss hat-bras of Weyl spinors. 
 Let us consider the ket $|u^r(\bp)\ra$, whose scalar-product-based bra is written as $\la\la u^r(\bp)|$.
 (The case of $|v^r(\bp)\ra$ may be treated in a similar way.)
 Clearly, it is impossible for the complex conjugate bra $\la\la \ov u^r(\bp)| \in \ov\WW$
 to directly form a scalar product with any ket $|u^s(\bp)\ra \in \WW$.
 Moreover, the above discussed treatment to Dirac spinors,
 which makes use of the matrix $\gamma_c$, is inapplicable to the case of Weyl spinors.

 In order to construct a scalar product which contains $\la\la \ov u^r(\bp)|$ and $|u^s(\bp)\ra$, 
 an operation is needed which may map, say, $\la\la \ov u^r(\bp)|$ to the space $\WW$. 
 From the physical point of view, it is reasonable to assume that such a map should not change 
 the angular momentum. 
 Noting that the two layers in one Dirac spinor [see $|\ov U^r(\bp)\ra$ 
 in Eq.(\ref{Ur-bp}), or the component form in Eq.(\ref{Up-uv-app})] possess a same angular momentum,
 one may map $\la\la \ov u^r(\bp)|$ to $\la\la v^r(\bp)|$.
 That is, the hat-bra for $|u^r(\bp)\ra$ may be taken as 
 $\la \wh u^r(\bp) | =  \la\la  v^r(\bp) |$.

 The above discussions are  generalizable to include both signs of $\varrho$. 
 Specifically, one may use $\varrho$-independent Weyl-spinor kets, namely, 
\begin{align}
 |u^r_\varrho ( \bp)\ra = |u^r (\bp)\ra;
 \label{vu-ip-rho}
\end{align}
 while, use $\varrho$-dependent hat-bras defined by
\begin{align}
 & \la \wh u^r_\varrho (\bp) | = \varrho  \la\la  v^r(\bp) |. \label{hat-bra-u-rho}
\end{align}
 Then, making use of Eq.(\ref{vu-eps}), it is ready to check that 
\begin{align}
 & \la \wh u^r_\varrho (\bp) |u^{s}_\varrho (\bp)\ra = \varrho \delta^{rs}. \label{vu-ip}
\end{align}
 One sees that, as expected, $|u^r_\varrho ( \bp)\ra$ of $\varrho=+$ and $\varrho = -$ 
 possess inner product and negative inner product, respectively.

 It is then straightforwardly to get (hat-)bras for the spin states of neutrino and antineutrino in 
 Eq.(\ref{S-nu-ovnu}), i.e., 
 \begin{subequations}\label{S-nu-ovnu-hbra}
  \begin{align}
  \label{S-nu-hbra}
  & \la \cs_{\nu \varrho}^{r }(\bp)| = - \frac{i}{\sqrt 2}
  \left(  \la \wh u^r_\varrho (\varrho \, \bp), - i \la \wh u^r_\varrho(\varrho \, \bp)  \right),
  \\ & \la \cs_{\ov \nu \varrho}^{r }(\bp)| = - \frac{i}{\sqrt 2}
    \left( \la \wh {\ov u}^r_\varrho(\varrho \, \bp),
    - i \wh {\ov u}^r_\varrho(\varrho \, \bp)|  \right). \label{S-ovnu-hbra}
   \end{align}
\end{subequations}
 Furthermore, one easily checks that single-particle states of neutrino and antineutrino
 with the above spin states satisfy Eq.(\ref{<f'-varrho|f>=delta}).

\section{Proof of Eq.(\ref{H1-8-Vf})}\label{app-proof-Vf-Hi}

 In this appendix, we give detailed derivation of Eq.(\ref{H1-8-Vf}). 
 It is easy to see that $H_1 = H^{\rm FIO1}_{++}$ and $H_2 = H^{\rm FIO2}_{++}$.
  For the rest $H_i$,
 let us first compute $ \overbrace{V_{\ov e } \left( H^{\rm FIO1}_{-+} \right)} $. 
 The operator $H^{\rm FIO1}_{-+}$ is given by Eq.(\ref{H-FIO1}), 
 with $(\varrho',\varrho)=(-+)$ and an amplitude given in Eq.(\ref{h1-FIO}).
 Substituting this operator as $\OO$ into Eq.(\ref{VK1}) with $f=\ov e$,
 one gets that
\begin{align}\label{}
\notag
 & \overbrace{ V_{\ov e } \left( H^{\rm FIO1}_{-+} \right)  }
  =  \int d\ww q' |{e}_{\ov \bq' s' +} \ra \Big(
 \int d\ww p d\ww q d\ww k \ V^{\dag s}_{-}(\bq) \gamma^0 \gamma^\mu  U^{r}_{+}(\bp)
  \\ & \times  \varepsilon^{\lambda*}_\mu(\bk)  \delta^3(\bp +\bq - \bk)
\overbrace {
 |A_{\bk \lambda}\ra  \la \ov e_{\bq s -} | \la e_{\bp r + }|
 \Big) |{\ov e}_{\bq' s'-} \ra }. \label{Vf-H3-1}
\end{align}
 Making use of the anticommutability of electron state and positron state
 and Eq.(\ref{<f'-varrho|f>=delta}) for the scalar product of positron states with negative $\varrho$, 
 the terms under the big brace in the above equality is computed as follows,
\begin{align}\label{}\notag
 & \overbrace {
  |A_{\bk \lambda}\ra  \la \ov e_{\bq s -} | \la e_{\bp r + }| \Big) |{\ov e}_{\bq' s'-} \ra }
  \\ & \quad = |q^0|  \delta^3(\bq - \bq') \delta_{ss'} |A_{\bk \lambda}\ra  | \la e_{\bp r + }|. \label{app-D2}
\end{align}
 Substituting Eq.(\ref{app-D2}) into Eq.(\ref{Vf-H3-1}), one gets that
\begin{gather}\notag
 \overbrace{ V_{\ov e } \left( H^{\rm FIO1}_{-+} \right)  }
 =  \int d\ww q d\ww p  d\ww k |{e}_{\ov \bq s +} \ra    \ |A_{\bk \lambda}\ra    \la e_{\bp r + }|
   \\ \times V_-^{\dag s}(\bq) \gamma^0 \gamma^\mu  U^{r}(\bp) \varepsilon^{\lambda*}_\mu(\bk)
  \delta^3(\bp +\bq - \bk).
\end{gather}
 Then,  one gets Eq.(\ref{HFIP-H3}), i.e.,  $\overbrace{V_{\ov e } \left( H^{\rm FIO1}_{-+} \right)}  = H_3$
 for $H_3$ defined in Eq.(\ref{H3}) with $h_3$ given in Eq.(\ref{h3-qed-neg})
 (with the replacement of $\bq \to \ov \bq$).

 Next,  we compute $ \overbrace{V_{e } \left( H^{\rm FIO1}_{+-} \right)}$.
 Similar to Eq.(\ref{Vf-H3-1}), but for $f=e$ in the superoperator $V_f$ and with exchange of $\varrho$ and 
 $\varrho'$, one gets that
\begin{align}\label{} \notag
 & \overbrace{V_{e } \left( H^{\rm FIO1}_{+-} \right)}
  =  \int d\ww q' |{\ov e}_{\ov \bq' s' +} \ra \Big(
\int d\ww p d\ww q d\ww k \ 
  V^{\dag s}_{+}(\bq) \gamma^0 \gamma^\mu U^{r}_{-}(\bp) 
  \\ \notag & \times 
  \varepsilon^{\lambda*}_\mu(\bk)  \delta^3(\bp +\bq - \bk)
 \overbrace{|A_{\bk \lambda}\ra  \la \ov e_{\bq s +} | \la e_{\bp r - }|
 \Big) |{ e}_{\bq' s'-} \ra }
 \\ \notag  & = - \int d\ww p d\ww q d\ww k \  |A_{\bk \lambda}\ra  |{\ov e}_{\ov \bp r +} \ra  \la \ov e_{\bq s +} | 
  \\ & \times V^{\dag s}_{+}(\bq) \gamma^0 \gamma^\mu
  U^{r}_{-}(\bp) \varepsilon^{\lambda*}_\mu(\bk)  \delta^3(\bp +\bq - \bk).
\end{align}
% With the replacement of $\bp \to \ov \bp$, 
 This gives Eq.(\ref{HFIP-H5}), i.e., $\overbrace{V_{e } \left( H^{\rm FIO1}_{+-} \right)}  = H_5$
 for $H_5$ defined in Eq.(\ref{H5}) with $h_5$ given in Eq.(\ref{h5-qed-neg}).

 Then, we discuss $\overbrace{V_{e } \left( V_{\ov e } (H^{\rm FIO1}_{--}) \right)}$.
 Following a procedure similar to that used above, one finds that
\begin{align}\label{} \notag 
 & \overbrace{ V_{e } \left( V_{\ov e } (H^{\rm FIO1}_{--}) \right)}
  =  \int d\ww q'' |{\ov e}_{\ov \bq'' s'' +} \ra \Big[  \int d\ww q' |{e}_{\ov \bq' s' +} \ra 
 \\  \notag  & \times \Big( \int d\ww p d\ww q d\ww k \ 
   V^{\dag s}_{-}(\bq) \gamma^0 \gamma^\mu
  U^{r}_{-}(\bp) \varepsilon^{\lambda*}_\mu(\bk)  \delta^3(\bp +\bq - \bk)
 \\ & \notag \times 
 \overbrace{|A_{\bk \lambda}\ra  \la \ov e_{\bq s -} | \la e_{\bp r - }|
 \Big) |{\ov e}_{\bq' s'-} \ra \Big]|{ e}_{\bq'' s''-} \ra} 
 \\  \notag  & = -\int d\ww p d\ww q d\ww k \ |A_{\bk \lambda}\ra 
 |{\ov e}_{\ov \bp r +} \ra |{e}_{\ov \bq s +} \ra 
   \\ & \times  V^{\dag s}_{-}(\bq) \gamma^0 \gamma^\mu
  U^{r}_{-}(\bp) \varepsilon^{\lambda*}_\mu(\bk)   \delta^3(\bp +\bq - \bk).
\end{align}
 Noting that  $ |{\ov e}_{\ov \bp r +} \ra |{e}_{\ov \bq s +} \ra  =
 - |{e}_{\ov \bq s +} \ra   |{\ov e}_{\ov \bp r +} \ra $ and Eq.(\ref{Vff'=Vf'f}),
 one gets Eq.(\ref{HFIP-H7}), i.e., $\overbrace{V_{e } \left( V_{\ov e } (H^{\rm FIO1}_{--}) \right)} = H_7$
 for $H_7$ defined in Eq.(\ref{H7}) with $h_7$ given in Eq.(\ref{h7-qed-neg})
 (with the replacements of $\bp \to \ov \bp$ and $\bq \to \ov \bq$).

 Following arguments similar to those given above, but for $H^{\rm FIO2}_{\varrho' \varrho}$, 
 it is straightforward to get Eqs.(\ref{HFIP-H4}), (\ref{HFIP-H6}), and (\ref{HFIP-H8}). 
 For example, for $\overbrace{ V_{\ov e } \left( H^{\rm FIO2}_{+-} \right)}$, 
 also making use of Eqs.(\ref{H-FIO}), (\ref{VK2}), and (\ref{<f'-varrho|f>=delta}), one gets that
\begin{align}\label{}\notag
 & \overbrace{ V_{\ov e } \left( H^{\rm FIO2}_{+-} \right)}
  = \int d\ww q'  \overbrace{ \la {\ov e}_{\bq' s' -}| \Big(
 \int d\ww p d\ww q d\ww k \ |e_{\bq s + }\ra
|\ov e_{\bp r - }\ra  \la A_{\bk \lambda}|}
 \\ & \times U^{\dag s}_{+}(\bq) \gamma^0 \gamma^\mu
  V^{r}_{-}(\bp) \varepsilon^{\lambda}_\mu(\bk)  \delta(\bp +\bq - \bk)
  \Big) \la {e}_{\ov\bq' s' +}| \notag
\\ & = \int d\ww q d\ww p  d\ww k  \ |e_{\bq s + }\ra  \la {e}_{\ov\bp r +}|
 \la A_{\bk \lambda}| \notag
 \\ & \times U^{\dag s}_{+}(\bq) \gamma^0 \gamma^\mu
  V^{r}_{-}(\bp) \varepsilon^{\lambda}_\mu(\bk)  \delta(\bp +\bq - \bk).
\end{align}
 This gives Eq.(\ref{HFIP-H4}), i.e., $\overbrace{V_{\ov e } \left( H^{\rm FIO2}_{+-} \right) } = H_4$
 for $H_4$ in Eq.(\ref{H4}) with $h_4$ given in Eq.(\ref{h4-qed-neg}).

\section{Operators $H_{i}$ interpreted with FVF}\label{app-Hi-interpret}

 In this appendix,  we discuss interpretations to the operators $H_i$ as
 expressed in Eq.(\ref{H1-8-Vf}) in terms of FIO and FVF.
 Clearly, $H_1$ and $H_2$ are just FIOs.
 Below, we discuss $H_i$ of $i=3, \ldots, 8$. 

 Firstly, let us discuss $ \overbrace{V_{\ov e} \left( H^{\rm FIO1}_{-+} \right)}$ for $H_3$. 
 As seen from Eq.(\ref{H-FIO1}) with $(\varrho',\varrho) = (-,+)$, 
 the FIO1 of $H^{\rm FIO1}_{-+}$ contains a negative-$p^0$ positron and a positive-$p^0$ electron. 
 To compute  $\overbrace{V_{\ov e} \left( H^{\rm FIO1}_{-+} \right)}$, 
 one may substitute Eq.(\ref{H-FIO1}) into Eq.(\ref{VK1}) with $f=\ov e$.
 This gives an integral whose integrand contains the following kets and bras,  
\begin{align} \notag
  & |{e}_{\ov \bq s +} \ra \overbrace{ \Big( |A_{\bk \lambda}\ra  \la \ov e_{\bp' r' \varrho'} |
 \la e_{\bp r \varrho }| \Big) |{\ov e}_{\bq s-} \ra}
 \\ & = - \Big( \la \ov e_{\bp' r' - }| {\ov e}_{\bq s-} \ra \Big)
 |{e}_{\ov \bq s +} \ra  |A_{\bk \lambda}\ra  \la  e_{\bp r +} |,
 \label{Vf-FIO1-interp}
\end{align}
 where Eqs.(\ref{overbrace-O}) and (\ref{<f'-varrho|f>=delta}) have been used.
 Since the scalar product $ \la \ov e_{\bp' r' - }|{\ov e}_{\bq s-} \ra$ on the rhs of Eq.(\ref{Vf-FIO1-interp})
 is proportional to $\delta^3(\bp'-\bq) \delta_{r's}$,
 the negative-$p^0$ positron in the FIO1 of $H^{\rm FIO1}_{-+}$
 should be just the positron in the FVF described by $V_{\ov e}$.

 Then, we get the following interpretation to $\overbrace{V_{\ov e} \left( H^{\rm FIO1}_{-+} \right)}$.
 That is, an electron-positron pair in the state of $(|{e}_{\ov \bq s +} \ra, |{\ov e}_{\bq s-} \ra)$
 emerges as a vacuum fluctuation (represented by the two kets in $V_{\ov e}$).
 And, the positron  in this  pair and some other electron,
 the latter of which already exists lying in a state described by $\la e_{\bp r +} |$,
 combine and change to a photon in a state $|A_{\bk \lambda}\ra$, as described
 by $H^{\rm FIO1}_{-+}$;
 meanwhile, the electron in the FVF pair leaves as a free particle. 
 Thus, the net effect is that a positive-$p^0$ electron
 in a state $\la e_{\bp r +} |$ changes to a state $|{ e}_{\ov \bq s +} \ra$ 
 and emits a photon in a state $|A_{\bk \lambda}\ra $.
 This is just what is described by $H_3$ in its original form in Eq.(\ref{H3}).

 Secondly, we discuss $\overbrace{V_{\ov e}^\dag \left( H^{\rm FIO2}_{+-} \right)}$  for $H_4$.
 As seen from Eq.(\ref{H-FIO2}) with $(\varrho',\varrho) = (+,-)$, 
 the FIO2 of $H^{\rm FIO2}_{+-}$ also contains a negative-$p^0$ positron and a positive-$p^0$ electron. 
 Substituting Eq.(\ref{H-FIO2}) into Eq.(\ref{VK2}) with $f=\ov e$, 
 one gets an integrand that contains the following ket-bras,
\begin{align}\notag
 & \overbrace{ \la {\ov e}_{\bq s -}{|} \Big( |e_{\bp' r' \varrho' }\ra
|\ov e_{\bp r \varrho }\ra  \la A_{\bk \lambda}| \Big) }  \la {e}_{\ov\bq s +}|
 \\  &  = - \Big( \la  {\ov e}_{\bq s-} | \ov e_{\bp r - } \ra \Big)
   | e_{\bp' r' +} \ra  \la {e}_{\ov \bq s +}|  \la A_{\bk \lambda}|.
 \label{Vf-FIO2-interp}
\end{align}
 The scalar product on the rhs of (\ref{Vf-FIO2-interp}) 
 indicates that the negative-$p^0$ positron, which participates in the FIO of $H^{\rm FIO2}_{+-}$,
 should belong to the electron-positron pair described by $V_{\ov e}^\dag$, which vanishes into the vacuum.

 Thus, one gets the following interpretation to $\overbrace{V_{\ov e}^\dag \left( H^{\rm FIO2}_{+-} \right)}$.
 That is, there exists an electron in a state $ \la {e}_{\ov \bq s +}|$; meanwhile, 
 a photon in a state $\la A_{\bk \lambda}|$ changes to an electron-positron pair 
 in the state of $( | e_{\bp' r' +} \ra, | \ov e_{\bp r - } \ra) $ with $\bp = \bq$ and $r =s$.
 The electron  existing and the positron coming from the photon
 possess opposite four-momentum and opposite angular momentum and, hence,
 they may vanish into the vacuum as a vacuum fluctuation.
 The total net effect is, then, that a positive-$p^0$ electron  in a state $ \la {e}_{\ov \bq s +}|$ 
 absorbs a photon and changes to a state of $| e_{\bp' r' +} \ra$, as described by $H_4$
  in its original form in Eq.(\ref{H4}).

 Thirdly, $ \overbrace{V_{e} \left( H^{\rm FIO1}_{+-} \right)}$ may be interpreted in a way similar to 
 $\overbrace{V_{\ov e} \left( H^{\rm FIO1}_{-+} \right)}$ discussed above. 
 The net effect is that one positive-$p^0$ positron emits a photon as 
 described by $H_5$.
 Fourthly, $\overbrace{V_{e}^\dag \left( H^{\rm FIO2}_{-+} \right)}$ may be interpreted in a way similar to 
 $\overbrace{V_{\ov e}^\dag \left( H^{\rm FIO2}_{+-} \right)}$, 
 with the total net effect  that a positive-$p^0$ positron absorbs a photon as 
 described by $H_6$.

 Fifthly, we discuss $\overbrace{V_{\ov e}[V_{e} \left( H^{\rm FIO1}_{--} \right) ]}$  for $H_7$.
 In this case, instead of the terms in (\ref{Vf-FIO1-interp}), one gets the following ones, 
\begin{align}\label{}
 &  |{e}_{\ov \bq' s' +} \ra |{\ov e}_{\ov \bq s +} \ra \overbrace{ \Big( |A_{\bk \lambda}\ra  \la \ov e_{\bp' r' -} |
 \la e_{\bp r -}| \Big) |{e}_{\bq s-} \ra |{\ov e}_{\bq' s' -} \ra }.
 \label{Vf-FIO1-interp-2}
\end{align}
 From (\ref{Vf-FIO1-interp-2}), one gets the following interpretation to 
 $\overbrace{V_{\ov e}[V_{e} \left( H^{\rm FIO1}_{- -} \right) ]}$. 
 That is,  two electron-positron pairs emerge from the vacuum as vacuum fluctuations, 
 one containing a negative-$p^0$ electron and the other 
 containing a negative-$p^0$ positron;
 and, these two negative-$p^0$ fermions change a photon. 
 The net effect is then emergence of one electron, one positron, and one photon from the vacuum, 
 as described by $H_7$.

 Sixthly and finally, for $\overbrace{V_{\ov e}^\dag [V_{e}^\dag \left( H^{\rm FIO2}_{--} \right) ]}$,
 one finds that it contains the following ket-bras, 
\begin{align}\label{} 
 & \overbrace{  \la {\ov e}_{\bq' s'-}|  \la {e}_{\bq s-}| \Big( |e_{\bp' r' \varrho' }\ra
|\ov e_{\bp r \varrho }\ra  \la A_{\bk \lambda}| \Big) } \la {\ov e}_{\ov \bq s +}| \la {e}_{\ov \bq' s' +}|,
 \label{Vf-FIO2-interp-2}
\end{align}
 and gets the following interpretation
 regarding three particles --- one positive-$p^0$ electron, one positive-$p^0$ 
 positron, and one photon.
 That is, the photon changes to one electron and one positron,
 whose four-momenta and angular momenta are exactly opposite to
 those of the existing  positron and electron, respectively.
 The two resulting electron-positron pairs then vanish into the vacuum. 
 Thus, the net effect of $\overbrace{V_{\ov e}^\dag [V_{e}^\dag \left( H^{\rm FIO2}_{--} \right) ]}$
 is vanishing of one electron, one positron, and one photon into the vacuum, 
 as described by $H_8$.

\section{Properties of the FIO2 of QED}\label{app-FIO2}

 In this appendix, it is shown that the QED FIO2
 [$H^{\rm FIO2}_{\varrho' \varrho}$ in Eq.(\ref{H-FIO2})]
 may be treated in a way similar to that given in Sec.\ref{sect-geometic-FIO1} 
 for the FIO1 of $H^{\rm FIO1}_{\varrho' \varrho}$.

 Let us first discuss the amplitude $h_2$  in Eq.(\ref{h2-qed}). 
 Following a procedure similar to that used in the main text for getting Eq.(\ref{hFIO1-spinor-part}), 
 one finds that the Dirac-spinor part of $h_2$ is written as follows, 
\begin{align}\label{}\notag
 & U^{\dag s}(\bq) \gamma^0 \gamma^\mu  V^{r}(\bp)
 \\ \notag  & = \ov u^{s B' }(\bq) \ov\sigma^{\mu}_{B'A} u^{rA}(\bp)
 - v^{s }_{B}(\bq) \sigma^{\mu BA'} \ov v^{r}_{A'}(\bp)
 \\  &  =  \left[ \ov u^{s B' }(\bq)  u^{rA}(\bp)
 -  v^{s A}(\bq)  \ov v^{rB'}(\bp) \right] \sigma^{\mu}_{AB'},  \label{h2-spinor-part} 
% \\ & =   \cs_{\odot}^- \left( U^{r}(\bp) \ov U^{s}(\bq) \right) \sigma^{\mu}_{AB'}.
\end{align}
 where $\varrho=\varrho'=+$ are not written explicitly,  for brevity.
 One further writes
\begin{align}\label{}\notag
  U^{\dag s}(\bq) \gamma^0 \gamma^\mu  V^{r}(\bp)
 &  =  \Big[ \la\la \ov u^{s  }(\bq) | \la\la  u^{r}(\bp)|  -\la\la \ov v^{r}(\bp)|  \la\la v^{s }(\bq)  \Big]
 \\ & \label{R2-immet1} \times |S^A \ra | \ov S^{B'}\ra \sigma^{\mu}_{AB'}.
\end{align}
 Then, using the relation of
\begin{align}\label{}
 \varepsilon^{\lambda}_\mu(\bk) = \la\la T_\mu | \varepsilon^{\lambda}(\bk)\ra,
\end{align}
 one finds that
\begin{align}
  h_2 =  \la\la \ov U^{s}(\bq) U^{r}(\bp)| \overleftarrow{\cs} \sigma^T | \varepsilon^{\lambda}(\bk) \ra
 \delta^3(\bp +\bq - \bk), \label{h2-ketbra}
\end{align}
 where $\overleftarrow{\cs}$ is defined in a way similar 
 to $\overrightarrow{\cs}$ in Eq.(\ref{G-dot-right-1}), but acting to the left, i.e., 
\begin{gather}\label{G-dot-left}
 \la\la  \ov W X | \overleftarrow{\cs} \equiv \la\la \ov w|  \la\la \chi |  +  \la\la z |   \la\la \ov\kappa |,
\end{gather}
 and $\sigma^T$ is the transposition of $\sigma$, as defined in Eq.(\ref{sigma-T}), namely, 
\begin{equation}\label{sigma-T-FIO2}
 \sigma^{T} := |S^A \ov S^{B'}\ra  \sigma^{\mu}_{ AB'} \la\la T_\mu|.
\end{equation}
 Here,  $\la\la  X | $ and $\la\la  W| $ are scalar-product-based bras of  
 $|X\ra$ and $|W\ra$ in Eq.(\ref{WX}), respectively, 
\begin{gather}\label{WX-bra}
 \la\la X| = \left( \la\la \chi |, \la\la \ov \kappa |  \right) ,\quad
 \la\la W| = \left( \la\la w|, \la\la \ov z |  \right).
\end{gather}

 One may further write $\la\la \ov U^{s}(\bq) U^{r}(\bp)|$
 on the rhs of Eq.(\ref{h2-ketbra}) in terms of hat-bra. 
 In fact, from Eq.(\ref{wh-U-rho}) in Appendix \ref{app-hatbra-spinors}
 (with $\gamma_c$ defined in Eq.(\ref{gamma-c-app})), 
 one finds that 
\begin{align}\label{<<U|-whU}
  \la\la U_{r }(\bp)| = -\la \wh{\ov U}_{r }(\bp)| \gamma_c.
\end{align}
 Making use of Eq.(\ref{<<U|-whU}), 
 it is straightforward to check that
\begin{align}\label{cs-=cs--}
 \la\la \ov U^{s}(\bq)  U^{r}(\bp) | \overleftarrow{\cs}
=  \la \wh U^{s}(\bq) \wh{\ov U}^{r}(\bp)|\overleftarrow{\cs}.
\end{align}
Then, Eq.(\ref{h2-ketbra}) is written as follows, 
\begin{align}\label{}
  & h_2 =  \la \wh U^{s}(\bq) \wh{\ov U}^{r}(\bp)
  | \overleftarrow{\cs} \sigma^T | \varepsilon^{\lambda}(\bk) \ra
  \delta^3(\bp +\bq - \bk). \label{h2-kb-S2}
  \end{align}

 By the formal similarity between $h^{\rm FIO2}_{\varrho' \varrho}$  
 in Eq.(\ref{h2-FIO}) and $h_2$   in Eq.(\ref{h2-qed}), from the expression
 of $h_2$ in Eq.(\ref{h2-kb-S2}), one finds that
\begin{align}\label{}
& h^{\rm FIO2}_{\varrho' \varrho} =  \la \wh U^{r'}_{\varrho'}(\bp') \wh{\ov U}^{r}_{\varrho}(\bp)
| \overleftarrow{\cs} \sigma^T | \varepsilon^{\lambda}(\bk) \ra
\delta^3(\bp +\bp' - \bk). \label{h2-kb-varrho}
\end{align}
 Substituting Eq.(\ref{h2-kb-varrho}) into Eq.(\ref{H-FIO2}), one gets that
\begin{align}\label{} \notag
   H^{\rm FIO2}_{\varrho' \varrho} &  = - \int d\ww p d\ww p' d\ww k \
 \delta^3(\bp +\bp' - \bk) |\ov e_{\bp r \varrho }\ra |e_{\bp' r' \varrho' }\ra
 \\ &   \times  \la \wh U^{r'}_{\varrho'}(\bp') \wh{\ov U}^{r}_{\varrho}(\bp)
 | \overleftarrow{\cs}  \sigma^T 
 | \varepsilon^{\lambda}(\bk) \ra   \la A_{\bk \lambda}|,
\label{H-FIO2-geom2-1}
\end{align}
 where the minus sign is due to a swap of the positions of 
 $|\ov e_{\bp r \varrho }\ra$ and $ |e_{\bp' r' \varrho' }\ra$.

 Finally, one gets a concise expression of $H^{\rm FIO2}_{\varrho' \varrho}$, 
 which gives a map reverse to that of the QED FIO1 in Eq.(\ref{H1-G}), 
\begin{equation}\label{H2-G}
 H^{\rm FIO2}_{\varrho' \varrho} =  I_{e\varrho', \ov e \varrho} \, \G^{\rm FIO2}  \, I_A,
\end{equation}
 where $\G^{\rm FIO2}$ is an operator similar to $\G^{\rm FIO1}$, but, giving the
 reverse mapping of state spaces.
 More exactly, by definition, $\G^{\rm FIO2}$ gives the following connection,
\begin{align}\label{}
 \la e_{\bp' \varrho'}^{r'} | \la \ov e_{\bp \varrho}^{r} | \G^{\rm FIO2} | A_{\bk}^{ \lambda} \ra 
 \equiv G^{\rm mom} G_2^{\rm spin},
\end{align}
 where
\begin{align}\label{G2-spin}
 &  G_2^{\rm spin} = - \la \wh U^{r'}_{\varrho'}(\bp') \wh{\ov U}^{r}_{\varrho}(\bp)
 | \overleftarrow{\cs}  \sigma^T  | \varepsilon^{\lambda}(\bk) \ra .
\end{align}
 Kets and bras in $I_{e\varrho', \ov e \varrho}$ and $I_A$ on the rhs of Eq.(\ref{H2-G})
 function in ways similar to those discussed at the end of Sec.\ref{sect-geometic-FIO1}
 for the rhs of Eq.(\ref{H1-G}), but, with a ket-bra swap. 
 Note that, on the rhs of Eq.(\ref{G2-spin}), the positron spin state is represented by 
 a hat-bra which corresponds to the ket $|{\ov U}^{r}_{\varrho}(\bp)\ra$, in consistency with what has been
 discussed in the main text.

\section{Explicit expressions of $G_{\eta_l}^{\rm spin}$ and $G_{{\rm ew},k}^{\rm spin}$}
\label{app-derive-G-mod-eta}

 In this appendix, we derive the expressions of the interaction amplitudes 
 $G_{\eta_l}^{\rm spin}$ in Eq.(\ref{G-mod-eta}),
 as well as some expressions for the GWS interaction amplitudes $G_{{\rm ew},k}^{\rm spin}$.

 To get Eq.(\ref{G-mod-eta}),
 one may substitute the spin states in Eqs.(\ref{S-eove})-(\ref{S-ZA})
 into the definition of $G_{\eta}^{\rm spin}$ in Eq.(\ref{G1-spin-eta}).
 Below, we discuss the six cases of $\eta_l$ [$l=1,\ldots,6$ as defined in Eq.(\ref{eta-1-6})], separately.

 Let us first discuss the case of $\eta =1$, for which $(f,f',B) = (e, \ov e, A)$. 
 Making use of the  electron and positron spin states given in Eq.(\ref{S-eove}) and 
 the definition of the symbol $\overrightarrow{D}$ in Eq.(\ref{D}), one gets that
\begin{align} \label{D-eove}
  \overrightarrow{D} | \cs_{e \varrho }^{r }(\bp) \cs_{\ov e \varrho' }^{r' }(\bp') \ra
 = \frac{1}{ 2}
   \left( \begin{array}{c} |u^r(\varrho \, \bp)\ra |\ov u^{r'}(\varrho' \, \bp')\ra
   \\ \varrho \varrho' \, |\ov v^r(\varrho \, \bp)\ra |v^{r'}(\varrho' \, \bp')\ra  \end{array} \right).
\end{align}  
 Substituting this result and Eq.(\ref{S-A}) for the spin state of photon  into Eq.(\ref{G1-spin-eta}),
 and noting the definition of the symbol $P_\eta$ in Eq.(\ref{P-eta}), one gets that
\begin{align} \notag
   \notag  G_{\eta_1}^{\rm spin} = & \frac{\N_A}{2\sqrt 2} \Big(
 w_1 \la \varepsilon^\lambda(\bk)| \sigma
 |u^r(\varrho \, \bp)\ra |\ov u^{r'}(\varrho' \, \bp')\ra 
 \\ & + w_2 \varrho \varrho' \la \varepsilon^\lambda(\bk)| \sigma 
  |\ov v^r(\varrho \, \bp)\ra |v^{r'}(\varrho' \, \bp')\ra 
 \Big) \notag
 \\ & = \frac{\N_A}{2\sqrt 2} (w_1G_L + w_2G_R),
   \end{align}
 where Eq.(\ref{GL-GR}) has been used.

 Next, we discuss the case of $\eta =2$, for which $(f,f',B) = (e, \ov \nu, W^-)$.
 With the antineutrino spin state in Eq.(\ref{S-ovnu}), one finds that
\begin{align}
  \overrightarrow{D} | \cs_{e \varrho }^{r }(\bp) \cs_{\ov \nu \varrho' }^{r' }(\bp') \ra
 = \frac{i}{ 2}
   \left( \begin{array}{c} |u^r(\varrho \, \bp)\ra |\ov u^{r'}(\varrho' \, \bp')\ra
   \\ i \varrho  \, |\ov v^r(\varrho \, \bp)\ra |\ov u^{r'}(\varrho' \, \bp')\ra  \end{array} \right).
\end{align}  
 Substituting this result and Eq.(\ref{S-W-}) into Eq.(\ref{G1-spin-eta})
 and noting Eq.(\ref{sigma-act-0}), one finds that
\begin{align} 
  \notag G_{\eta_2}^{\rm spin} = & -\frac{\N_{W^-}}{2} 
\la \varepsilon^\lambda(\bk)| \sigma
 |u^r(\varrho \, \bp)\ra |\ov u^{r'}(\varrho' \, \bp')\ra  
 \\ &= -\frac{\N_{W^-}}{2} G_L .
   \end{align}
 Similarly, in the case of $\eta =3$ with $(f,f',B) = (\nu, \ov e, W^+)$,
 noting the neutrino spin state in Eq.(\ref{S-nu}), one gets
\begin{align}
  \overrightarrow{D} | \cs_{\nu \varrho }^{r }(\bp) \cs_{\ov e \varrho' }^{r' }(\bp') \ra
 = \frac{i}{2}
   \left( \begin{array}{c} |u^r(\varrho \, \bp)\ra |\ov u^{r'}(\varrho' \, \bp')\ra
   \\ i \varrho'  \, |u^r(\varrho \, \bp)\ra |v^{r'}(\varrho' \, \bp')\ra  \end{array} \right).
\end{align}  
 Substituting this result and Eq.(\ref{S-W+}) into Eq.(\ref{G1-spin-eta}), one gets 
\begin{align} \notag
  G_{\eta_3}^{\rm spin} = & -\frac{\N_{W^+}}{2} 
\la \varepsilon^\lambda(\bk)| \sigma
 |u^r(\varrho \bp)\ra |\ov u^{r'}(\varrho' \bp')\ra 
 \\ & = -\frac{\N_{W^+}}{2} G_L.
   \end{align}

 In the case of $\eta =4$ with $(f,f',B) = (\nu, \ov \nu, Z^0)$, one gets
\begin{align}\label{D-nu-ovnu}
  \overrightarrow{D} | \cs_{\nu \varrho }^{r }(\bp) \cs_{\ov \nu \varrho' }^{r' }(\bp') \ra
 = -\frac{1}{ 2}
   \left( \begin{array}{c} |u^r(\varrho \, \bp)\ra |\ov u^{r'}(\varrho' \, \bp')\ra
   \\ - |u^r(\varrho \, \bp)\ra |\ov u^{r'}(\varrho' \, \bp')\ra  \end{array} \right).
\end{align}  
 Substituting this result and Eq.(\ref{S-Z}) into Eq.(\ref{G1-spin-eta}), one gets 
\begin{align} \notag
  G_{\eta_4}^{\rm spin} = & - \frac{\N_{Z^0}}{2\sqrt 2} (w_3 +w_4) 
 \la \varepsilon^\lambda(\bk)| \sigma |u^r(\varrho \, \bp)\ra |\ov u^{r'}(\varrho' \, \bp')\ra 
  \\ & =- \frac{\N_{Z^0}}{2\sqrt 2} (w_3 +w_4) G_L .
   \end{align}
 For $\eta =5$ with $(f,f',B) = (e, \ov e, Z^0)$, making use of Eq.(\ref{D-eove}) one gets that
\begin{align} \notag
   \notag  G_{\eta_5}^{\rm spin} = & \frac{\N_{Z^0}}{2\sqrt 2} 
 \Big( w_3 \la \varepsilon^\lambda(\bk)| \sigma
 |u^r(\varrho \, \bp)\ra |\ov u^{r'}(\varrho' \, \bp')\ra 
 \\ \notag & - w_4 \la \varepsilon^\lambda(\bk)| \sigma    
 \varrho \varrho' \, |\ov v^r(\varrho \, \bp)\ra |v^{r'}(\varrho' \, \bp')\ra \Big)
  \\ & =  \frac{\N_{Z^0}}{2\sqrt 2} (w_3 G_L - w_4 G_R).
   \end{align}
 For $\eta =6$ with $(f,f',B) = (\nu, \ov \nu, A)$, making use of Eq.(\ref{D-nu-ovnu}) one finds that
\begin{align} \notag
  G_{\eta_6}^{\rm spin}  = & -\frac{\N_A}{2\sqrt 2} 
\Big( w_1 \la \varepsilon^\lambda(\bk)| \sigma
 |u^r(\varrho \, \bp)\ra |\ov u^{r'}(\varrho' \, \bp')\ra 
 \\ & - w_2 \la \varepsilon^\lambda(\bk)| \sigma  |u^r(\varrho \, \bp)\ra |\ov u^{r'}(\varrho' \, \bp')\ra 
 \Big)  \notag
 \\ & =  -\frac{\N_A}{2\sqrt 2} (w_1 - w_2)G_L.
   \end{align}
 Summarizing the above results, one is ready to get Eq.(\ref{G-mod-eta}).

 At last, we derive some expressions of
 the GWS electroweak interaction amplitudes $G_{{\rm ew},k}^{\rm spin}$.
 Substituting Eq.(\ref{xi-k}) into Eq.(\ref{G-ew-k}), one gets that
\begin{subequations}\label{G-ew-k-explicit}
  \begin{align}\label{G-ew-1}
  & \frac{\sqrt 2}{g} G_{{\rm ew},1}^{\rm spin} = \frac{\sin \theta_{W}}{\sqrt 2}  (G_L + G_R),
  \\ & \frac{\sqrt 2}{g} G_{{\rm ew},k}^{\rm spin} = - \frac{1}{2} G_L 
  \hspace{1cm} \text{(for $k =2,3$)},   \label{G-ew-23}
  \\ & \frac{\sqrt 2}{g}  G_{{\rm ew},4}^{\rm spin} 
  = - \frac{1}{2 \sqrt{2} \cos \theta_{W}}   G_L, \label{G-ew-4}
  \\ \notag & \frac{\sqrt 2}{g} \left( G_{{\rm ew},5L}^{\rm spin}  + G_{{\rm ew},5R}^{\rm spin} \right)
  \\ & \hspace{1.1cm}    =   \frac{(1 - 2\sin^2 \theta_{W}) G_L 
   -  2\sin^2 \theta_{W} G_R}{2\sqrt{2} \cos \theta_{W}}.  \label{G-ew-56}
  \end{align}
 \end{subequations}

%\end{multicols}{2}


\begin{thebibliography}{99}

 \bibitem{Weinberg-book} S.~Weinberg, {\it The Quantum Theory of
 Fields} (Cambridge University Press, Now York, 1996).
 \bibitem{Peskin} M.E. Peskin and D.V. Schroeder, {\it In Introduction to Quantum Field Theory}
 (Westview Press, 1995).
 \bibitem{Itzy} C.~Itzykson and J.~B.~Zuber,
 {\it Quantum Field Theory} (McGraw-Hill, New York, 1980).


% \bibitem{Pauli40} W.~Pauli, Phys.Rev.~{\bf 58}, 716 (1940).
% \bibitem{SW64} R.F.~Streater and A.S.~Wightman, {\it PCT, Spin and Statistics, and
% All That} (Benjamin/Cummings, Reading, Mass., 1964).


\bibitem{IWW-RMP24} G. Isidori, F. Wilsch,  and D. Wyler, 
\textit{The standard model effective field theory at work},
Rev.~Mod.~Phys. {\bf 96}, 015006 (2024).
\bibitem{FR-RMP21} F. Feruglio and A. Romanino, 
\textit{Lepton flavor symmetries},
Rev.~Mod.~Phys. {\bf 93}, 015007 (2021).
\bibitem{BB-RMP21} D. B\"{o}deker and W. Buchm\"{u}ller, 
\textit{Baryogenesis from the weak scale to the grand unification scale},
Rev.~Mod.~Phys. {\bf 93}, 035004 (2021).
\bibitem{FADOSV-RMP21} C.  A. Arg\"{u}elles, A. Diaz, 
A. Kheirandish, A. Olivares-Del-Campo, I. Safa, and A. C. Vincent,  
\textit{Dark matter annihilation to neutrinos},
Rev.~Mod.~Phys. {\bf 93}, 035007 (2021).
\bibitem{DeGrand-MP16} T.  DeGrand, 
\textit{Lattice tests of beyond standard model dynamics},
Rev.~Mod.~Phys. {\bf 88}, 015001 (2016).

\bibitem{book-fundam-QFT} I.-O. Stamatescu and E. Seiler (Eds.),
{\it Approaches to Fundamental Physics}, Lect. Notes Phys. 721
(Springer, Berlin, Heidelberg 2007).

 \bibitem{SV06} A. Strumia and F. Vissani, 
 \textit{Neutrino masses and mixings and...},
 arXiv:hep-ph/0606054v3.
 \bibitem{GN03} M. C. Gonzalez-Garcia and Yosef Nir, 
 \textit{Neutrino masses and mixing: evidence and implications},
 Rev. Mod. Phys. {\bf 75}, 345 (2003).


 \bibitem{KO01} Y. Kuno and Y. Okada, 
 \textit{Muon decay and physics beyond the standard model},
 Rev.~Mod.~Phys. {\bf 73}, 151 (2001).
 \bibitem{GGS99} M.K. Gaillard, P.D. Grannis, and F.J. Sciulli, 
 \textit{The standard model of particle physics}
 Rev.Mod.Phys. {\bf 71}, S96 (1999).

 \bibitem{Polch-book} J.~Polchinski, {\it String Theroy} (Cambridge University Press 1998).
 \bibitem{Raby93} {\it The Building Blocks of Creation}, S.Raby, ed.,
 (World Scientific, Singapore, 1993).

 \bibitem{Ross84} G.G.~Ross, {\it Grand Unified Theories},
 (Benjamin / Cummings, Menlo Park, California, 1984).


 \bibitem{Penrose-book} R.~Penrose and W.~Rindler, {\it Spinors and
 space-time} (Cambridge University Press, London, 1984).
 \bibitem{CM-book} M.~Carmeli and S.~Malin, {\it Theory of spinors: an introduction}
 (World Scientific Publishing, Singapore, 2000).
 \bibitem{Corson} E.M. Corson,
 {\it Introduction to Tensors, Spinors, and Relativistic Wave-Equations}
 (Blackie $\&$ Son Limited, London and Glasgow, 1953).
 \bibitem{Kim-group} Y.S. Kim and M.E. Noz, {\it Theory and Application of the Poincar\'{e} Group}
 (D.Reidel Publishing Company, Dordrecht, 1986).

 \bibitem{pra16-commu} W.-g. Wang, 
 \textit{Anticommutation relations for fermionic fields derived from basic properties of the inner product},
 Phys.Rev.A {\bf 94}, 012112 (2016).

% \bibitem{arXiv-EWS} Wen-ge Wang, arXiv:2203.09478.

 \bibitem{Peskin-p705} See, e.g., p.705 in the textbook of Ref.\cite{Peskin} for the interaction 
 Lagrangian, which differs by a minus sign from the interaction Hamiltonian discussed here. 

%\bibitem{foot-ovU} As discussed above, the operation of complex conjugation is indicated by
% an overline for 2-component spinors and also for the EvdW symbols (similar for 4-component vectors
% to be discussed in a later section).
% In view of this fact, it seems more convenient to write the complex conjugate of $|U^r(\bp)\ra$ as
% $|\ov U^r(\bp)\ra$.
% However, this might cause confusion with an notation conventionally used in QED,
% that is, $\ov \psi = \psi^\dag \gamma^0$.
% To avoid this type of confusion, we write the complex conjugate of
% a Dirac spinor $|U^r(\bp)\ra$ as $|\ov U^r(\bp)\ra$.

% \bibitem{foot-ov-T} In principle, one may take other assumptions.
% For example, one may assume that ${|\ov T_\mu\ra }=-|T_\mu\ra$;
% this will lead to the relation $\ov\sigma =\sigma$.


%\bibitem{Gupta} S.N. Gupta, Proceedings of Physical Society A {\bf 63}, 681 (1950).

%\bibitem{foot-virtual} A negative-$p^0$ particle
% is different from a virtual particle, the latter of which
% refers to an off-mass-shell particle in an intermediate interaction process.

% \bibitem{foot-commu} Indeed, writing $b_r^{\dag}(\bp)$ as $ b^{\dag}_\bp b_{r(\bp)}^{\dag}$
% and writing $b_r(\bp)$ as $ b_{r(\bp)}b_\bp$, one can not get the correction anti-commutation
% relation between $b_r(\bp)$ and $b_r^\dag(\bp)$ from those for the momentum and spinor
% degrees of freedom.

%\bibitem{footnote-H0}
% If, instead of Eq.(\ref{b-dag-def}), the creation operator $b_r^{\dag}(\bp)$ is defined by
% $ \sqrt{p^0} b_r^{\dag}(\bp)|0\ra = |\bp_{r}\ra $, then, one can verify
% that $ \{ b_r(\bp) , b_s^{\dag}(\bq) \} = \delta_{rs} \delta^{3}(\bp-\bq)$
% and that $H_0$ in Eq.(\ref{H0-elec}) has the ordinary expression
% $ H_0  = \int d^3p \  p^0 b_r^{\dag}(\bp) b^r(\bp)$.


% \bibitem{MS} F.~Mandl and G.~Shaw, {\it Quantum field theory}
% (John Wiley \& Sons, Chichester, 1993).
 %\bibitem{RMP} F.~Wilczek, Rev. Mod. Phys. {\bf 71} (1999) S85;
 %J.~H.~Schwarz and N.~Seiberg, Rev. Mod. Phys. {\bf 71} (1999) S112.
\end{thebibliography}
\end{document}